\documentclass[sigconf, nonacm]{acmart}
\pdfoutput=1

\usepackage{etoolbox}
\usepackage{balance}
\usepackage{standalone}
\usepackage{tikz}
\usepackage{pgfplots}
\usepackage{enumitem}
\usepackage{thm-restate}
\usepackage[noend]{algpseudocode}
\usepackage{algorithm}
\usepackage{subcaption}
\usepackage{multirow}
\usepackage{diagbox}
\usepackage{macros}
\usetikzlibrary{positioning,shadows,arrows,trees,shapes,fit,shapes.geometric,backgrounds,arrows.meta,quotes,angles,calc,positioning,scopes,quotes,angles}
\pgfplotsset{compat=1.16}

\newtoggle{withappendix}
\toggletrue{withappendix}

\allowdisplaybreaks
\setlist{parsep=1mm,leftmargin=1.2em}
\setlist[enumerate,1]{label={\arabic*.}}

\begin{document}

\title{Rewriting the Infinite Chase}

\author{Michael Benedikt}
\affiliation{%
    \institution{Oxford University}
    \streetaddress{Parks Road}
    \city{Oxford}
    \country{United Kingdom}
    \postcode{OX1 3QD}
}
\email{michael.benedikt@cs.ox.ac.uk}

\author{Maxime Buron}
\affiliation{%
    \institution{LIRMM, Inria, Univ. of Montpellier}
    \streetaddress{rue St Priest}
    \city{Montpellier}
    \country{France}
    \postcode{34095}
}
\email{maxime.buron@inria.fr}

\author{Stefano Germano}
\affiliation{%
    \institution{Oxford University}
    \streetaddress{Parks Road}
    \city{Oxford}
    \country{United Kingdom}
    \postcode{OX1 3QD}
}
\email{stefano.germano@cs.ox.ac.uk}

\author{Kevin Kappelmann}
\affiliation{%
    \institution{Technical University of Munich}
    \streetaddress{Boltzmannstra{\ss}e 3}
    \city{Munich}
    \country{Germany}
    \postcode{85748}
}
\email{kevin.kappelmann@tum.de}

\author{Boris Motik}
\affiliation{%
    \institution{Oxford University}
    \streetaddress{Parks Road}
    \city{Oxford}
    \country{United Kingdom}
    \postcode{OX1 3QD}
}
\email{boris.motik@cs.ox.ac.uk}

\begin{abstract}
Guarded tuple-generating dependencies (GTGDs) are a natural extension of
description logics and referential constraints. It has long been known that
queries over GTGDs can be answered by a variant of the \emph{chase}---a
quintessential technique for reasoning with dependencies. However, there has
been little work on concrete algorithms and even less on implementation. To
address this gap, we revisit \emph{Datalog rewriting} approaches to query
answering, where GTGDs are transformed to a Datalog program that entails the
same base facts on each base instance. We show that the rewriting can be seen
as containing ``shortcut'' rules that circumvent certain chase steps, we
present several algorithms that compute the rewriting by simulating specific
types of chase steps, and we discuss important implementation issues. Finally,
we show empirically that our techniques can process complex GTGDs derived from
synthetic and real benchmarks and are thus suitable for practical use.

\end{abstract}

\maketitle



\section{Introduction}

Tuple-generating dependencies (TGDs) are a natural extension of description
logics and referential constraints, and they are extensively used in databases.
For example, they are used in data integration to capture semantic restrictions
on data sources, mapping rules between data sources and the mediated schema,
and constraints on the mediated schema. A fundamental computational problem in
such applications is \emph{query answering under TGDs}: given a query $Q$, a
collection of facts $I$, and a set of TGDs $\Sigma$, find all the answers to
$Q$ that logically follow from $I$ and $\Sigma$. This problem has long been
seen as a key component of a declarative data integration systems
\cite{alonlogicdataint,teenage}, and it also arises in answering querying using
views and accessing data sources with restrictions
\cite{alonviews,candbsigrecord,meier}.

The \emph{chase} is a quintessential technique for reasoning with TGDs. It
essentially performs ``forward reasoning'' by extending a set of given facts
$I$ to a set $I'$ of all facts implied by $I$ and a set of TGDs $\Sigma$. To
answer a query, one can compute $I'$ using the chase and then evaluate the
query in $I'$. Unfortunately, the chase does not necessarily terminate, and in
fact query answering for general TGDs is undecidable. Considerable effort was
devoted to identifying classes of TGDs for which query answering is decidable.
One line of work has focused on TGDs where the chase terminates;
\emph{weakly-acyclic TGDs} \cite{DBLP:journals/tcs/FaginKMP05} are perhaps the
best-known such class. Another line of work focused on \emph{guarded TGDs
(GTGDs)}. GTGDs are interesting since they can capture common constraints used
in data integration, and ontologies expressed in variants of \emph{description
logic (DL)} \cite{dl-handbook-2} can be translated directly into GTGDs.
Example~\ref{ex:motivation} illustrates the use of GTGDs used in a practical
data integration scenario.

\begin{example}\label{ex:motivation}
The IEC Common Information Model (CIM) is an open model for describing power
generation and distribution networks. It is frequently used as a semantic layer
in applications that integrate data about power systems
\cite{DBLP:conf/rr/GahaZLBVV13}. CIM is defined in UML, but its formal
semantics has been provided by a translation into an OWL ontology. The domain
of CIM is described using \emph{classes} and \emph{properties}, which
correspond to unary and binary relations, respectively. Moreover, semantic
relationships between classes and properties are represented as OWL axioms,
many of which can be translated into GTGDs. A significant portion of CIM
describes power distribution equipment using GTGDs such as
\eqref{ex:run:1}--\eqref{ex:run:4}.
\begin{align}
    \kw{ACEquipment}(x) \rightarrow \exists y ~ \kw{hasTerminal}(x,y) \wedge \kw{ACTerminal}(y) \label{ex:run:1} \\
    \kw{ACTerminal}(x) \rightarrow \kw{Terminal}(x)                                             \label{ex:run:2} \\
    \kw{hasTerminal}(x,z) \wedge \kw{Terminal}(z) \rightarrow \kw{Equipment}(x)                 \label{ex:run:3} \\
    \kw{ACTerminal}(x) \rightarrow \exists y ~ \kw{partOf}(x,y) \wedge \kw{ACEquipment}(y)      \label{ex:run:4}
\end{align}
Data integration is then achieved by populating the vocabulary using
\emph{mappings}, which can be seen queries over the data sources that produce a
set of facts called a \emph{base instance}. A key issue in data integration is
dealing with incompleteness of data sources. For example, it is not uncommon
that one data source mentions two switches $\kw{sw}_1$ and $\kw{sw}_2$, while
another data source provides information about connected terminals only for
switch $\kw{sw}_1$.
\begin{align}
    \kw{ACEquipment}(\kw{sw}_1) \quad \kw{ACEquipment}(\kw{sw}_2)               \label{ex:run:fct:1} \\
    \kw{hasTerminal}(\kw{sw}_1,\kw{trm}_1) \quad \kw{ACTerminal}(\kw{trm}_1)    \label{ex:run:fct:2}
\end{align}
GTGDs can be used to complete the data. For example, if a user asks to list all
pieces of equipment known to the system, both $\kw{sw}_1$ and $\kw{sw}_2$ will
be returned, even though the base instance does not explicitly classify either
switch as a piece of equipment.
\end{example}

Even though the chase for GTGDs does not necessarily terminate, query answering
for GTGDs is decidable \cite{datalogpmj}. To prove decidability, one can argue
that the result of a chase is \emph{tree-like}---that is, the facts derived by
the chase can be arranged into a particular kind of tree. Next, one can develop
a \emph{finite representation} of potentially infinite trees. One possibility
is to describe the trees using a \emph{finite tree automaton}, so query
answering can be reduced to checking automaton emptiness. While theoretically
elegant, this method is not amenable to practical use: building the automaton
and the emptiness test are both complex and expensive, and such algorithms
always exhibit worst-case complexity. Alternatively, one can use
\emph{blocking} to identify a tree prefix sufficient for query evaluation.
Blocking is commonly used in description logic reasoning \cite{dl-handbook-2},
and it was later lifted to \emph{guarded logic} \cite{hirschthesis}. However,
blocking was shown to be impractical for query answering: the required tree
prefix can be much larger than the base instance $I$ so, as $I$ grows in size,
the size of the tree prefix becomes unmanageable.

More promising query answering techniques for GTGDs are based on \emph{Datalog
rewriting} \cite{marnette}. The idea was initially proposed by
\citet{marnette}, and it was later extended to broader classes of TGDs
\cite{baget2011walking,gottlobsimkus} and settings \cite{bbcmfcs}. The main
idea is to transform an input set of GTGDs $\Sigma$ into a set $\rew(\Sigma)$
of \emph{Datalog} rules such that $\Sigma$ and $\rew(\Sigma)$ entail the same
base facts on each base instance. Thus, given a base instance $I$, instead of
computing the chase of $I$ and $\Sigma$ (which may not terminate), we compute
the chase $I'$ of $I$ and $\rew(\Sigma)$. Since Datalog rules essentially
correspond to existential-free TGDs, $I'$ is always finite and it can be
computed using optimized Datalog engines. Moreover, $\Sigma$ and $\rew(\Sigma)$
entail the same base facts on $I$, so we can answer any \emph{existential-free}
conjunctive query (i.e., queries where all variables are answer variables) by
evaluating in $I'$. The restriction to existential-free queries is technical:
existentially quantified variables in a query can be matched to objects
introduced by existential quantification, and these are not preserved in a
Datalog rewriting. However, practical queries are typically existential-free
since all query variables are usually answer variables.

\begin{example}
A Datalog program consisting of rules \eqref{ex:run:2}--\eqref{ex:run:3} and
\eqref{ex:rewrite:1} is a rewriting of GTGDs \eqref{ex:run:1}--\eqref{ex:run:4}.
\begin{align}
    \kw{ACEquipment}(x) \rightarrow \kw{Equipment}(x)   \label{ex:rewrite:1}
\end{align}
Rule \eqref{ex:rewrite:1} is a logical consequence of GTGDs
\eqref{ex:run:1}--\eqref{ex:run:3}, and it provides a ``shortcut'' for the
inferences of the other GTGDs.
\end{example}

The advantage of rewriting-based approaches is scalability in the size of the
base instance $I$. Such techniques have been implemented and practically
validated in the context of description logics
\cite{motikdatalog1,motikdatalog2}, but practical algorithms have not yet been
proposed for GTGDs. This raises several theoretical and practical questions.

\myparagraphnoperiod{How to compute the Datalog rules needed for completeness?}
Existing Datalog rewriting algorithms often prove their correctness indirectly.
For example, completeness of a rewriting algorithm for description logics
\cite{motikdatalog1} uses a proof-theoretic argument, which does not provide an
intuition about why the algorithm actually works. Our first contribution is to
\emph{relate Datalog rewriting approaches to the chase}. Towards this goal, we
introduce the \emph{one-pass} variant of the chase, which we use to develop a
general completeness criterion for Datalog rewriting algorithms. This, in turn,
provides us with a better understanding of how rewriting algorithms work, and
it allows us to discover new algorithms in a systematic way.

\myparagraphnoperiod{What does the space of rewriting algorithms look like?}
Computing the rewriting $\rew(\Sigma)$ usually requires extending $\Sigma$ with
certain logical consequences of $\Sigma$. We show that we can select the
relevant consequences using different criteria. Some methods require deriving
TGDs with existential quantifiers in the head, others generate Datalog rules
directly, and yet other methods derive logical implications with function
symbols. We relate all of these methods to the one-pass chase mentioned
earlier, and we provide theoretical worst-case guarantees about their
performance.

\myparagraphnoperiod{How do we ensure scalability of rewriting algorithms?}
Implementations of Datalog rewriting algorithms have thus far been mainly
considered in the setting of description logics
\cite{motikthesis,motikdatalog1}. \emph{To the best of our knowledge, we
provide the first look at optimization and implementation of Datalog rewriting
algorithms for GTGDs}. We achieve scalability by developing and combining
various indexing and redundancy elimination techniques.

\myparagraphnoperiod{How do we evaluate rewriting algorithms?}
We provide a benchmark for GTGD query answering algorithms, and we use it to
evaluate our methods. To the best of our knowledge, this is the first attempt
to evaluate query answering techniques for GTGDs.

\myparagraph{Summary of contributions}
We give an extensive account of Datalog rewriting for GTGDs. In particular, we
develop a theoretical framework that allows us to understand, motivate, and
show completeness of rewriting algorithms. Moreover, we present several
concrete algorithms, establish worst-case complexity bounds, and discuss their
relationships. We complement this theoretical analysis with a discussion of how
to adapt techniques from first-order theorem proving to the setting of GTGDs.
Finally, we empirically evaluate our techniques using an extensive benchmark.
All proofs and the details of one algorithm are given in the
\iftoggle{withappendix}{appendix of this paper}{extended version
\cite{saturation-github}}. Our implementation and a more detailed account of
our experimental results can be found online \cite{saturation-github}.

\section{Related Work}\label{sec:related}

Answering queries via rewriting has been extensively considered in description
logics. For example, queries over ontologies in the DL-Lite family of languages
can be rewritten into first-order queries \cite{dllite}, and fact entailment
for $\mathcal{SHIQ}$ ontologies can be rewritten to disjunctive Datalog
\cite{motikdatalog1}. These techniques provide the foundation for the Ontop
\cite{DBLP:journals/semweb/CalvaneseCKKLRR17} and KAON2 \cite{motikthesis}
systems, respectively.

In the context of TGDs, first-order rewritings were considered in data
integration systems with inclusion and key dependencies
\cite{calioldrewriting}. Datalog rewritings have been considered for GTGDs
\cite{marnette} and their extensions such as frontier-guarded TGDs
\cite{bbcmfcs}, and nearly frontier-guarded and nearly guarded TGDs
\cite{gottlobsimkus}. The focus in these studies was to identify complexity
bounds and characterize expressivity of TGD classes rather than provide
practical algorithms. Existing implements of query answering for TGDs use
first-order rewriting for linear TGDs \cite{nyaya}, chase variants for TGDs
with terminating chase \cite{chasebench}, chase with blocking for warded TGDs
\cite{vadalogvldb}, chase with the magic sets transformation for shy TGDs
\cite{alviano}, and Datalog rewriting for separable and weakly separable TGDs
\cite{drewer}. These TGD classes are all different from GTGDs, and we are
unaware of any attempts to implement and evaluate GTGD rewriting algorithms.

Our algorithms are related to resolution-based decision procedures for variants
of guarded logics
\cite{denivelle,ganzinger99superposition,DBLP:conf/aaai/ZhengS20}. Moreover,
our characterization of Datalog rewritings is related to a chase variant used
to answer queries over data sources with access patterns
\cite{DBLP:journals/lmcs/AmarilliB22}. Finally, a variant of the one-pass chase
from Section~\ref{sec:chaserewrite} was generalized to the broader context of
disjunctive GTGDs \cite{DBLP:journals/corr/abs-1911-03679}.

\section{Preliminaries}\label{sec:preliminaries}

In this section, we recapitulate the well-known definitions and notation that
we use to formalize our technical results.

\myparagraph{TGDs}
Let $\consts$, $\vars$, and $\nulls$ be pairwise disjoint, infinite sets of
\emph{constants}, \emph{variables}, and \emph{labeled nulls}, respectively. A
\emph{term} is a constant, a variable, or a labeled null; moreover, a term is
\emph{ground} if it does not contain a variable. For $\alpha$ a formula or a
set thereof, $\consts(\alpha)$, $\vars(\alpha)$, $\nulls(\alpha)$, and
$\trms(\alpha)$ are the sets of constants, free variables, labeled nulls, and
terms, respectively, in $\alpha$.

A \emph{schema} is a set of relations, each of which is associated with a
nonnegative integer arity. A \emph{fact} is an expression of the form $R(\vec
t)$, where $R$ is an $n$-ary relation and $\vec t$ is a vector of $n$ ground
terms; moreover, $R(\vec t)$ is a \emph{base fact} if $\vec t$ contains only
constants. An \emph{instance} $I$ is a finite set of facts, and $I$ is a
\emph{base instance} if it contains only base facts. An \emph{atom} is an
expression of the form $R(\vec t)$, where $R$ is an $n$-ary relation and $\vec
t$ is a vector of $n$ terms not containing labeled nulls. Thus, each base fact
is an atom. We often treat conjunctions as sets of conjuncts; for example, for
$\gamma$ a conjunction of facts and $I$ an instance, ${\gamma \subseteq I}$
means that each conjunct of $\gamma$ is contained $I$.

A \emph{tuple generating dependency} (TGD) is a first-order formula of the form
${\forall \vec x [\body \rightarrow \exists \vec y ~ \head]}$, where $\body$
and $\head$ are conjunctions of atoms, $\head$ is not empty, the free variables
of $\body$ are $\vec x$, and the free variables of $\head$ are contained in
${\vec x \cup \vec y}$. Conjunction $\body$ is the \emph{body} and formula
$\exists \vec y ~ \head$ is the \emph{head} of the TGD. We often omit $\forall
\vec x$ when writing a TGD. A TGD is \emph{full} if $\vec y$ is empty;
otherwise, the TGD is \emph{non-full}. A TGD is in \emph{head-normal form} if
it is full and its head contains exactly one atom, or it is non-full and each
head atom contains at least one existentially quantified variable. Each TGD can
be easily transformed to an equivalent set of TGDs in head-normal form. A full
TGD in head-normal form is a \emph{Datalog rule}, and a \emph{Datalog program}
is a finite set of Datalog rules. The \emph{head-width} ($\hwidth$) and the
\emph{body-width} ($\bwidth$) of a TGD are the numbers of variables in the head
and body, respectively; these are extended to sets of TGDs by taking the maxima
over all TGDs. The notion of an instance satisfying a TGD is inherited from
first-order logic. A base fact $F$ is \emph{entailed} by an instance $I$ and a
finite set of TGDs $\Sigma$, written ${I, \Sigma \models F}$, if ${F \in I'}$
holds for each instance ${I' \supseteq I}$ that satisfies $\Sigma$.

A \emph{substitution} $\sigma$ is a function that maps finitely many variables
to terms. The domain and the range of $\sigma$ are $\dom(\sigma)$ and
$\rng(\sigma)$, respectively. For $\gamma$ a term, a vector of terms, or a
formula, $\sigma(\gamma)$ is obtained by replacing each free occurrence of a
variable $x$ in $\gamma$ such that ${x \in \dom(\sigma)}$ with $\sigma(x)$.

\myparagraph{Fact Entailment for Guarded TGDs}
Fact entailment for general TGDs is semidecidable, and many variants of the
\emph{chase} can be used to define a (possibly infinite) set of facts that is
homomorphically contained in each modef of a base instance and a set of TGDs.

Fact entailment is decidable for \emph{guarded} TGDs (GTGDs): a TGD ${\forall
\vec x [\body \rightarrow \exists \vec y ~ \head]}$ is guarded if $\body$
contains an atom (called a \emph{guard}) that contains all variables of $\vec
x$. Note that a guard need not be unique in $\body$. Let $\Sigma$ be a finite
set of GTGDs. We say that a set of ground terms $G$ is $\Sigma$-\emph{guarded}
by a fact $R(\vec t)$ if ${G \subseteq \vec t \cup \consts(\Sigma)}$. Moreover,
$G$ is $\Sigma$-\emph{guarded} by a set of facts $I$ if $G$ is $\Sigma$-guarded
by some fact in $I$. Finally, a fact $S(\vec u)$ is $\Sigma$-guarded by a fact
$R(\vec t)$ (respectively a set of facts $I$) if $\vec u$ is $\Sigma$-guarded
by $R(\vec t)$ (respectively $I$).

By adapting the reasoning techniques for guarded logics
\cite{vardirobust,gforig} and referential database constraints
\cite{johnsonklug}, fact entailment for GTGDs can be decided by a chase variant
that works on tree-like structures. A \emph{chase tree} $T$ consists of a
directed tree, one tree vertex that is said to be \emph{recently updated}, and
a function mapping each vertex $v$ in the tree to a finite set of facts $T(v)$.
A chase tree $T$ can be transformed to another chase tree $T'$ in the following
two ways. \begin{itemize} \item One can apply a \emph{chase step} with a GTGD
${\tau = \forall \vec x [\body \rightarrow \exists \vec y ~ \head]}$ in
head-normal form. The precondition is that there exist a vertex $v$ in $T$ and
a substitution $\sigma$ with domain $\vec x$ such that ${\sigma(\body)
\subseteq T(v)}$. The result of the chase step is obtained as follows.
\begin{itemize} \item If $\tau$ is full (and thus $\head$ is a single atom),
then chase tree $T'$ is obtained from $T$ by making $v$ recently updated in
$T'$ and setting ${T'(v) = T(v) \cup \{ \sigma(\head) \}}$.

        \item If $\tau$ is not full, then $\sigma$ is extended to a
        substitution $\sigma'$ that maps each variable in $\vec y$ to a labeled
        null not occurring in $T$, and chase tree $T'$ is obtained from $T$ by
        introducing a fresh child $v'$ of $v$, making $v'$ recently updated in
        $T'$, and setting ${T(v') = \sigma'(\head) \cup \{ F \in T(v) \mid F
        \text{ is $\Sigma$-guarded by } \sigma'(\head) \}}$.
    \end{itemize}

    \item One can apply a \emph{propagation step} from a vertex $v$ to a vertex
    $v'$ in $T$. Chase tree $T'$ is obtained from $T$ by making $v'$ recently
    updated in $T'$ and setting ${T'(v') = T(v') \cup S}$ for some nonempty set
    $S$ satisfying ${S \subseteq \{ F \in T(v) \mid F \text { is
    $\Sigma$-guarded by } T(v') \}}$.
\end{itemize}

A \emph{tree-like chase sequence} for a base instance $I$ and a finite set of
GTGDs $\Sigma$ in head-normal form is a finite sequence of chase trees ${T_0,
\dots, T_n}$ such that $T_0$ contains exactly one \emph{root vertex} $r$ that
is recently updated in $T_0$ and ${T_0(r) = I}$, and each $T_i$ with ${0 < i
\leq n}$ is obtained from $T_{i-1}$ by a chase step with some ${\tau \in
\Sigma}$ or a propagation step. For each vertex $v$ in $T_n$ and each fact ${F
\in T_n(v)}$, this sequence is a \emph{tree-like chase proof of} $F$
\emph{from} $I$ and $\Sigma$. It is well known that ${I, \Sigma \models F}$ if
and only if there exists a tree-like chase proof of $F$ from $I$ and $\Sigma$
(e.g., \cite{datalogpmj}). Example~\ref{ex:one-pass} in
Section~\ref{sec:chaserewrite} illustrates these definitions. One can decide
${I, \Sigma \models F}$ by imposing an upper bound on the size of chase trees
that need to be considered \cite{datalogpmj}.

\myparagraph{Rewriting}
A \emph{Datalog rewriting} of a finite set of TGDs $\Sigma$ is a Datalog
program $\rew(\Sigma)$ such that ${I, \Sigma \models F}$ if and only if ${I,
\rew(\Sigma) \models F}$ for each base instance $I$ and each base fact $F$. If
$\Sigma$ contains GTGDs only, then a Datalog rewriting $\rew(\Sigma)$ is
guaranteed to exist (which is not the case for general TGDs). Thus, we can
reduce fact entailment for GTGDs to Datalog reasoning, which can be solved
using highly optimized Datalog techniques
\cite{motikthesis,ahmetaj2018rewriting}. For example, given a base instance
$I$, we can compute the \emph{materialization} of $\rew(\Sigma)$ on $I$ by
applying the rules of $\rew(\Sigma)$ to $I$ up to a fixpoint. This will compute
precisely all base facts entailed by $\rew(\Sigma)$ (and thus also by $\Sigma$)
on $I$, and it can be done in time polynomial in the size of $I$.

\myparagraph{Encoding Existentials by Function Symbols}
It is sometimes convenient to represent existentially quantified values using
functional terms. In such cases, \emph{we use a slightly modified notions of
terms, atoms, and rules}. It will be clear from the context which definitions
we use in different parts of the paper.

We adjust the notion of a term as either a constant, a variable, or an
expression of the form $f(\vec t)$ where $f$ is an $n$-ary \emph{function
symbol} and $\vec t$ is a vector of $n$ terms. The notions of ground terms,
(base) facts, and (base) instances, and atoms are the same as before, but they
use the modified notion of terms. A \emph{rule} is a first-order implication of
the form ${\forall \vec x [\body \rightarrow H]}$ where $\body$ is a
conjunction of atoms whose free variables are $\vec x$, and $H$ is an atom
whose free variables are contained in $\vec x$; as for TGDs, we often omit
$\forall \vec x$. A rule thus contains no existential quantifiers, but its head
contains exactly one atom that can contain function symbols. Also, a Datalog
rule, a function-free rule, and a full TGD in head-normal form are all
synonyms. Finally, a base fact still contains only constants.

\emph{Skolemization} allow us to replace existential quantifiers in TGDs by
functional terms. Specifically, let ${\tau = \forall \vec x [\body \rightarrow
\exists \vec y ~ \head]}$, and let $\sigma$ be a substitution defined on each
${y \in \vec y}$ as ${\sigma(y) = f_{\tau,y}(\vec x)}$ where $f_{\tau,y}$ is a
fresh $|\vec x|$-ary \emph{Skolem} symbol uniquely associated with $\tau$ and
$y$. Then, the Skolemization of $\tau$ produces rules ${\forall \vec x [\body
\rightarrow \sigma(H)]}$ for each atom ${H \in \head}$. Moreover, the
Skolemization $\Sigma'$ of a finite set of TGDs $\Sigma$ is the union of the
rules obtained by Skolemizing each ${\tau \in \Sigma}$. It is well known that
${I, \Sigma \models F}$ if and only if ${I, \Sigma' \models F}$ for each base
instance $I$ and each base fact $F$.

\myparagraph{Unification}
A \emph{unifier} of atoms ${A_1, \dots, A_n}$ and ${B_1, \dots, B_n}$ is a
substitution $\theta$ such that ${\theta(A_i) = \theta(B_i)}$ for ${1 \leq i
\leq n}$. Such $\theta$ is a \emph{most general unifier} (MGU) if, for each
unifier $\sigma$ of ${A_1, \dots, A_n}$ and ${B_1, \dots, B_n}$, there exists a
substitution $\rho$ such that ${\sigma = \rho \circ \theta}$ (where $\circ$ is
function composition). An MGU is unique up to variable renaming if it exists,
and it can be computed in time ${O(\sum_{i=1}^n |A_i| + |B_i|)}$ where $|A_i|$
and $|B_i|$ are the encoding sizes of $A_i$ and $B_i$
\cite{robinson1965machine,DBLP:journals/jcss/PatersonW78}.

\section{Chase-Based Datalog Rewriting}\label{sec:chaserewrite}

Our objective is to develop rewriting algorithms that can handle complex GTGDs.
Each algorithm will derive Datalog rules that provide ``shortcuts'' in
tree-like chase proofs: instead of introducing a child vertex $v'$ using a
chase step with a non-full GTGD at vertex $v$, performing some inferences in
$v'$, and then propagating a derived fact $F$ back from $v'$ to $v$, these
``shortcuts'' will derive $F$ in one step without having to introduce $v'$. The
main question is how to derive all ``shortcuts'' necessary for completeness
while keeping the number of derivations low. In this section we lay the
technical foundations that will allow us to study different strategies for
deriving ``shortcuts'' in Section~\ref{sec:algorithms}. We show that, instead
of considering arbitrary chase proofs, we can restrict our attention to chase
proofs that are \emph{one-pass} according to Definition~\ref{def:one-pass}.
Then, we identify the parts of such proofs that we need to be able to
circumvent using ``shortcuts''. Finally, we present sufficient conditions that
guarantee completeness of rewriting algorithms. We start by describing formally
the structure of tree-like chase proofs.

\begin{definition}\label{def:one-pass}
    A tree-like chase sequence ${T_0, \dots, T_n}$ for a base instance $I$ and
    a finite set of GTGDs $\Sigma$ in head-normal form is \emph{one-pass} if,
    for each ${0 < i \leq n}$, chase tree $T_i$ is obtained by applying one of
    the following two steps to the recently updated vertex $v$ of $T_{i-1}$:
    \begin{itemize}
        \item a propagation step copying exactly one fact from $v$ to its
        parent, or

        \item a chase step with a GTGD from $\Sigma$ provided that no
        propagation step from $v$ to the parent of $v$ is applicable.
    \end{itemize}
\end{definition}

Thus, each step in a tree-like chase sequence is applied to a ``focused''
vertex; steps with non-full TGDs move the ``focus'' from a parent to a child,
and propagation steps move the ``focus'' in the opposite direction. Moreover,
once a child-to-parent propagation takes place, the child cannot be revisited
in further steps. Theorem~\ref{thm:one-pass-proof-exists} states a key property
about chase proofs for GTGDs: whenever a proof exists, there exists a one-pass
proof too. Example~\ref{ex:one-pass} illustrates important aspects of
Definition~\ref{def:one-pass} and Theorem~\ref{thm:one-pass-proof-exists}.

\begin{restatable}{theorem}{OnePassProofExists}\label{thm:one-pass-proof-exists}
    For each base instance $I$, each finite set of GTGDs $\Sigma$ in
    head-normal form, and each base fact $F$ such that ${I, \Sigma \models F}$,
    there exists a one-pass tree-like chase proof of $F$ from $I$ and
    $\Sigma$.
\end{restatable}

\begin{example}\label{ex:one-pass}
Let ${I = \{ A(a,b) \}}$ and let $\Sigma$ contain GTGDs
\eqref{ex:one-pass:1}--\eqref{ex:one-pass:6}.
\begin{align}
    A(x_1,x_2)                      & \rightarrow \exists y ~ B(x_1,y) \wedge C(x_1,y)              \label{ex:one-pass:1} \\
    C(x_1,x_2)                      & \rightarrow D(x_1,x_2)                                        \label{ex:one-pass:2} \\
    B(x_1,x_2) \wedge D(x_1,x_2)    & \rightarrow E(x_1)                                            \label{ex:one-pass:3} \\
    A(x_1,x_2) \wedge E(x_1)        & \rightarrow \exists y_1, y_2 ~ F(x_1,y_1) \wedge F(y_1,y_2)   \label{ex:one-pass:4} \\
    E(x_1) \wedge F(x_1,x_2)        & \rightarrow G(x_1)                                            \label{ex:one-pass:5} \\
    B(x_1,x_2) \wedge G(x_1)        & \rightarrow H(x_1)                                            \label{ex:one-pass:6}
\end{align}

\begin{figure}[tb]
    \centering
    \begin{tikzpicture}
    \newcommand{\bag}[2]{\node (#1) {$\begin{array}{@{}c@{}} #2 \end{array}$};}
    \tikzset{
        ->,
        >=latex,
        every node/.style={shape=rectangle,draw},
        scope/.style={draw=none},
        lnode/.style={draw=none},
        updated/.style={red,line width=1.7pt},
        pics/drawv/.style args={#1/#2/#3/#4}{
            code = {
                \draw[-,dashed] let \p1=(#1.east),\p2=(#2.west),\p3=(#3.north),\p4=(#4.south) in ({(\x1+\x2)/2},\y3) -- ({(\x1+\x2)/2},\y4);
            }
        },
        pics/drawh/.style args={#1/#2/#3/#4}{
            code = {
                \draw[-,dashed] let \p1=(#1.south),\p2=(#2.north),\p3=(#3.west),\p4=(#4.east) in (\x3,{(\y1+\y2)/2}) -- (\x4,{(\y1+\y2)/2});
            }
        },
        node distance=0.3cm,
        r-0/.pic={\bag{r}{A(a,b)}},
        r-1/.pic={\bag{r}{A(a,b) \\ E(a)}},
        v1-1/.pic={\bag{v1}{B(a,n_1) \\ C(a,n_1)}},
        v1-2/.pic={\bag{v1}{B(a,n_1) \\ C(a,n_1) \\ D(a,n_1)}},
        v1-3/.pic={\bag{v1}{B(a,n_1) \\ C(a,n_1) \\ D(a,n_1) \\ E(a)}},
        v1-4/.pic={\bag{v1}{B(a,n_1) \\ C(a,n_1) \\ D(a,n_1) \\ E(a) \\ G(a)}},
        v1-5/.pic={\bag{v1}{B(a,n_1) \\ C(a,n_1) \\ D(a,n_1) \\ E(a) \\ G(a) \\ H(a)}},
        v2-1/.pic={\bag{v2}{E(a) \\ F(a,n_2) \\ F(n_2,n_3)}},
        v2-2/.pic={\bag{v2}{E(a) \\ F(a,n_2) \\ F(n_2,n_3) \\ G(a)}},
    }
    \node[scope] (T0) {
        \begin{tikzpicture}
            \pic[updated]{r-0};
            \node[lnode,above=0.1cm of r]{$T_0$};
        \end{tikzpicture}
    };
    \node[scope, right=of T0] (T1) {
        \begin{tikzpicture}
            \pic{r-0};
            \pic[below=of r, updated]{v1-1};
            \path (r) edge (v1);
            \node[lnode,above=0.1cm of r]{$T_1$};
        \end{tikzpicture}
    };
    \node[scope, right=of T1] (T2) {
        \begin{tikzpicture}
            \pic{r-0};
            \pic[below=of r, updated]{v1-2};
            \path (r) edge (v1);
            \node[lnode,above=0.1cm of r]{$T_2$};
        \end{tikzpicture}
    };
    \node[scope, right=of T2] (T3) {
        \begin{tikzpicture}
            \pic{r-0};
            \pic[below=of r, updated]{v1-3};
            \path (r) edge (v1);
            \node[lnode,above=0.1cm of r]{$T_3$};
        \end{tikzpicture}
    };
    \node[scope, right=of T3] (T4) {
        \begin{tikzpicture}
            \pic[updated]{r-1};
            \pic[below=of r]{v1-3};
            \path (r) edge (v1);
            \node[lnode,above=0.1cm of r]{$T_4$};
        \end{tikzpicture}
    };
    \node[scope, right=of T4] (T5) {
        \begin{tikzpicture}
            \pic{r-1};
            \pic[below=of r, xshift=-0.75cm]{v1-3};
            \pic[below=of r, xshift=0.75cm, updated]{v2-1};
            \path
                (r) edge (v1)
                (r) edge (v2)
            ;
            \node[lnode,above=0.1cm of r]{$T_5$};
        \end{tikzpicture}
    };
    \node[scope, below=1.8cm of T0, xshift=2cm] (T6) {
        \begin{tikzpicture}
            \pic{r-1};
            \pic[below=of r, xshift=-0.75cm]{v1-3};
            \pic[below=of r, xshift=0.75cm, updated]{v2-2};
            \path
                (r) edge (v1)
                (r) edge (v2)
            ;
            \node[lnode,left=of r]{$T_6$};
        \end{tikzpicture}
    };
    \node[scope, right=of T6] (T7) {
        \begin{tikzpicture}
            \pic{r-1};
            \pic[below=of r, xshift=-0.75cm, updated]{v1-4};
            \pic[below=of r, xshift=0.75cm]{v2-2};
            \path
                (r) edge (v1)
                (r) edge (v2)
            ;
            \node[lnode,left=of r]{$T_7$};
        \end{tikzpicture}
    };
    \node[scope, right=of T7] (T8) {
        \begin{tikzpicture}
            \pic{r-1};
            \pic[below=of r, xshift=-0.75cm, updated]{v1-5};
            \pic[below=of r, xshift=0.75cm]{v2-2};
            \path
                (r) edge (v1)
                (r) edge (v2)
            ;
            \node[lnode,left=of r] {$T_8$};
        \end{tikzpicture}
    };
    \pic{drawh=T4/T8/T0/T5};
    \pic{drawv=T0/T1/T4/T4};
    \pic{drawv=T1/T2/T4/T4};
    \pic{drawv=T2/T3/T4/T4};
    \pic{drawv=T3/T4/T4/T4};
    \pic{drawv=T4/T5/T4/T4};
    \pic{drawv=T6/T7/T8/T8};
    \pic{drawv=T7/T8/T8/T8};
\end{tikzpicture}
    \caption{Tree-Like Chase Sequence for Example~\ref{ex:one-pass}}\label{fig:chase-sequence}
\end{figure}
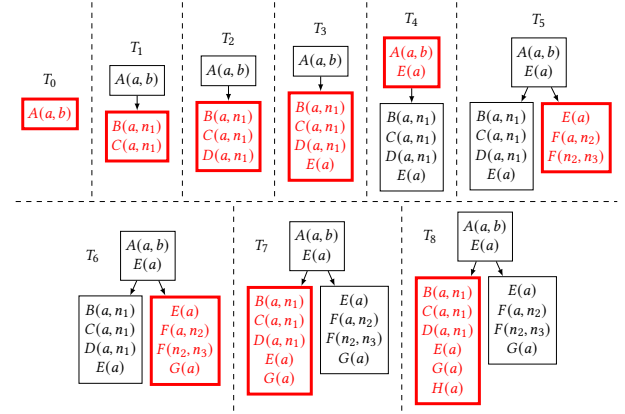

A tree-like chase sequence for $I$ and $\Sigma$ is shown in
Figure~\ref{fig:chase-sequence}, and it provides a proof of the base fact
$H(a)$ from $I$ and $\Sigma$. The recently updated vertex of each chase tree is
shown in red. We denote the root vertex by $r$, and its left and right children
by $v_1$ and $v_2$, respectively. The step producing $T_7$ from $T_6$ does
\emph{not} satisfy the requirements of one-pass chase: it propagates the fact
$G(a)$ from $v_2$ to $v_1$, where the latter is a ``sibling'' of the former.

To obtain a one-pass chase sequence, we could try to ``slow down'' the
propagation of $G(a)$: we first propagate $G(a)$ from $v_2$ to $r$, and then
from $r$ to $v_1$. The former step is allowed in one-pass chase, but the latter
step is not: once we leave the subtree rooted at $v_1$, we are not allowed to
revisit it later. Note, however, that $B(a,n_1)$ and $G(a)$ must occur jointly
in a vertex of a chase tree in order to derive $H(a)$. Moreover, note that no
reordering of chase steps will derive $H(a)$: we must first produce $v_1$ to be
able to derive $v_2$, and we must combine $G(a)$ from $v_2$ and $B(a,n_1)$ from
$v_1$.

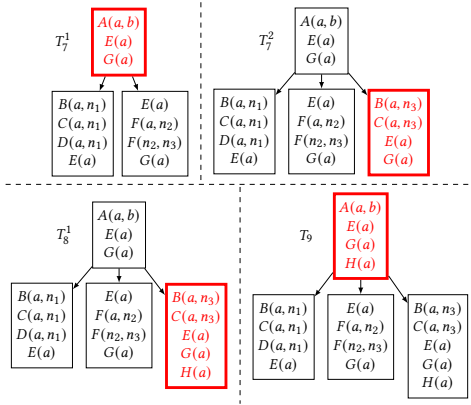
\begin{figure}[tb]
    \centering
    \begin{tikzpicture}
    \newcommand{\bag}[2]{\node (#1) {$\begin{array}{@{}c@{}} #2 \end{array}$};}
    \tikzset{
        ->,
        >=latex,
        every node/.style={shape=rectangle,draw},
        scope/.style={draw=none},
        lnode/.style={draw=none},
        updated/.style={red,line width=1.7pt},
        pics/drawv/.style args={#1/#2/#3/#4}{
            code = {
                \draw[-,dashed] let \p1=(#1.east),\p2=(#2.west),\p3=(#3.north),\p4=(#4.south) in ({(\x1+\x2)/2},\y3) -- ({(\x1+\x2)/2},\y4);
            }
        },
        pics/drawh/.style args={#1/#2/#3/#4}{
            code = {
                \draw[-,dashed] let \p1=(#1.south),\p2=(#2.north),\p3=(#3.west),\p4=(#4.east) in (\x3,{(\y1+\y2)/2}) -- (\x4,{(\y1+\y2)/2});
            }
        },
        node distance=0.3cm,
        r-2/.pic={\bag{r}{A(a,b) \\ E(a) \\ G(a)}},
        r-3/.pic={\bag{r}{A(a,b) \\ E(a) \\ G(a) \\ H(a)}},
        v1-3/.pic={\bag{v1}{B(a,n_1) \\ C(a,n_1) \\ D(a,n_1) \\ E(a)}},
        v2-2/.pic={\bag{v2}{E(a) \\ F(a,n_2) \\ F(n_2,n_3) \\ G(a)}},
        v3-1/.pic={\bag{v3}{B(a,n_3) \\ C(a,n_3) \\ E(a) \\ G(a)}},
        v3-2/.pic={\bag{v3}{B(a,n_3) \\ C(a,n_3) \\ E(a) \\ G(a) \\ H(a)}},
        v3-3/.pic={\bag{v3}{B(a,n_3) \\ C(a,n_3) \\ E(a) \\ G(a) \\ H(a)}},
    }
    \node[scope] (T7-1) {
        \begin{tikzpicture}
            \pic[updated]{r-2};
            \pic[below=of r, xshift=-0.75cm]{v1-3};
            \pic[below=of r, xshift=0.75cm]{v2-2};
            \path
                (r) edge (v1)
                (r) edge (v2)
            ;
            \node[lnode,left=of r]{$T_7^1$};
        \end{tikzpicture}
    };
    \node[scope, right=of T7-1] (T7-2) {
        \begin{tikzpicture}
            \pic{r-2};
            \pic[below=of r, xshift=-1.6cm]{v1-3};
            \pic[below=of r]{v2-2};
            \pic[below=of r, xshift=1.6cm, updated]{v3-1};
            \path
                (r) edge (v1)
                (r) edge (v2)
                (r) edge (v3)
            ;
            \node[lnode,left=of r]{$T_7^2$};
        \end{tikzpicture}
    };
    \node[scope, below=of T7-1] (T8-1) {
        \begin{tikzpicture}
            \pic{r-2};
            \pic[below=of r, xshift=-1.6cm]{v1-3};
            \pic[below=of r]{v2-2};
            \pic[below=of r, xshift=1.6cm, updated]{v3-2};
            \path
                (r) edge (v1)
                (r) edge (v2)
                (r) edge (v3)
            ;
            \node[lnode,left=of r]{$T_8^1$};
        \end{tikzpicture}
    };
    \node[scope, right=of T8-1] (T9) {
        \begin{tikzpicture}
            \pic[updated]{r-3};
            \pic[below=of r, xshift=-1.6cm]{v1-3};
            \pic[below=of r]{v2-2};
            \pic[below=of r, xshift=1.6cm]{v3-2};
            \path
                (r) edge (v1)
                (r) edge (v2)
                (r) edge (v3)
            ;
            \node[lnode,left=of r]{$T_9$};
        \end{tikzpicture}
    };
    \pic{drawh=T7-2/T9/T8-1/T9};
    \pic{drawv=T7-1/T7-2/T7-2/T7-2};
    \pic{drawv=T8-1/T9/T9/T9};
\end{tikzpicture}
    \caption{One-Pass Chase Sequence Obtained from Figure~\ref{fig:chase-sequence}}\label{fig:one-pass}
\end{figure}

The solution, which is used in the proof of
Theorem~\ref{thm:one-pass-proof-exists}, is to replace propagation to the child
by ``regrowing'' the entire subtree. In our example, we replace the steps
producing $T_7$ and $T_8$ with the steps shown in Figure~\ref{fig:one-pass}.
Chase tree $T_7^1$ is obtained from $T_6$ by propagating $G(a)$ from $v_2$ to
$r$. Then, instead of propagating $G(a)$ from $r$ to $v_1$, a new vertex $v_3$
is created in $T_7^2$ by reapplying \eqref{ex:one-pass:1} and fact $G(a)$ is
pushed to $v_3$ as part of the chase step with a non-full GTGD. This allows
$H(a)$ to be derived in vertex $v_3$ of $T_8^1$.

Fact $D(n_3)$ can be derived in vertex $v_3$, but this is not needed to prove
$H(a)$. Moreover, our chase is \emph{oblivious} \cite{datalogpmj}: a non-full
TGD can be applied to the same facts several times, each time introducing a
fresh vertex and fresh labeled nulls. The number of children of a vertex is
thus not naturally bounded, and our objective is not to apply all chase steps
exhaustively to obtain a universal model of $\Sigma$. Instead, we are
interested only in \emph{chase proofs}, which must only contain steps needed to
demonstrate entailment of a specific fact.
\end{example}

One-pass chase proofs are interesting because they can be decomposed into
\emph{loops} as described in Definition~\ref{def:loop}.

\begin{definition}\label{def:loop}
    For ${T_0, \dots, T_n}$ a one-pass tree-like chase sequence for some
    $I$ and $\Sigma$, a \emph{loop at vertex} $v$ \emph{with output fact} $F$
    is a subsequence ${T_i, \dots, T_j}$ with ${0 \leq i < j \leq n}$ such that
    \begin{itemize}
        \item $T_{i+1}$ is obtained by a chase step with a non-full GTGD,

        \item $T_j$ is obtained by a propagation step that copies $F$, and

        \item $v$ is the recently updated vertex of both $T_i$ and $T_j$.
    \end{itemize}
    The \emph{length} of the loop is defined as $j - i$.
\end{definition}

\begin{example}\label{ex:loop}
Subsequence ${T_0, T_1, T_2, T_3, T_4}$ of the chase trees from
Example~\ref{ex:one-pass} is a loop at the root vertex $r$ with output fact
$E(a)$: chase tree $T_1$ is obtained by applying a non-full GTGD to $r$, and
chase tree $T_4$ is obtained by propagating $E(a)$ back to $r$. Analogously,
${T_4, T_5, T_6, T_7^1}$ is another loop at $r$ with output fact $G(a)$.
Finally, ${T_7^1, T_7^2, T_8^1, T_9}$ is a loop at $r$ with output fact $H(a)$.
\end{example}

Thus, a loop is a subsequence of chase steps that move the ``focus'' from a
parent to a child vertex, perform a series of inferences in the child and its
descendants, and finally propagate one fact back to the parent. If non-full
TGDs are applied to the child, then the loop can be recursively decomposed into
further loops at the child. The properties of the one-pass chase ensure that
each loop is finished as soon as a fact is derived in the child that can be
propagated to the parent, and that the vertices introduced in the loop are not
revisited at any later point in the proof. In this way, each loop at vertex $v$
can be seen as taking the set $T_i(v)$ as input and producing the output fact
$F$ that is added to $T_j(v)$. This leads us to the following idea: for each
loop with the input set of facts $T_i(v)$, a rewriting should contain a
``shortcut'' Datalog rule that derives the loop's output.

\begin{example}\label{ex:rewriting}
One can readily check that rules \eqref{ex:rewriting:1}--\eqref{ex:rewriting:3}
provide ``shortcuts'' for the three loops identified in Example~\ref{ex:loop}.
\begin{align}
    A(x_1,x_2)                  & \rightarrow E(x_1) \label{ex:rewriting:1} \\
    A(x_1,x_2) \wedge E(x_1)    & \rightarrow G(x_1) \label{ex:rewriting:2} \\
    A(x_1,x_2) \wedge G(x_1)    & \rightarrow H(x_1) \label{ex:rewriting:3}
\end{align}
Moreover, these are all relevant ``shortcuts'': the union of rules
\eqref{ex:rewriting:1}--\eqref{ex:rewriting:3} and the Datalog rules from
Example~\ref{ex:one-pass}---that is, rules \eqref{ex:one-pass:2},
\eqref{ex:one-pass:3}, \eqref{ex:one-pass:5}, and \eqref{ex:one-pass:6}---is a
rewriting of the set $\Sigma$ from Example~\ref{def:one-pass}.
\end{example}

These ideas are formalized in Proposition~\ref{prop:rewriting-conditions},
which will provide us with a correctness criterion for our algorithms.

\begin{restatable}{proposition}{RewritingConditions}\label{prop:rewriting-conditions}
    A Datalog program $\Sigma'$ is a rewriting of a finite set of GTGDs
    $\Sigma$ in head-normal form if
    \begin{itemize}
        \item $\Sigma'$ is a logical consequence of $\Sigma$,

        \item each Datalog rule of $\Sigma$ is a logical consequence of
        $\Sigma'$, and

        \item for each base instance $I$, each one-pass tree-like chase
        sequence ${T_0, \dots, T_n}$ for $I$ and $\Sigma$, and each loop ${T_i,
        \dots, T_j}$ at the root vertex $r$ with output fact $F$, there exist a
        Datalog rule ${\body \rightarrow H \in \Sigma'}$ and a substitution
        $\sigma$ such that ${\sigma(\body) \subseteq T_i(r)}$ and ${\sigma(H) =
        F}$.
    \end{itemize}
\end{restatable}

Intuitively, the first condition ensures soundness: rewriting $\Sigma'$ should
not derive more facts than $\Sigma$. The second condition ensures that
$\Sigma'$ can mimic direct applications of Datalog rules from $\Sigma$ at the
root vertex $r$. The third condition ensures that $\Sigma'$ can reproduce the
output of each loop at vertex $r$ using a ``shortcut'' Datalog rule.

\section{Rewriting Algorithms}\label{sec:algorithms}

We now consider ways to produce ``shortcut'' Datalog rules satisfying
Proposition \ref{prop:rewriting-conditions}. In
Subsection~\ref{subsec:algorithms:ExbDR} we present the $\ExbDR$ algorithm that
manipulates GTGDs directly, and in Subsections~\ref{subsec:algorithms:SkDR}
and~\ref{subsec:algorithms:HypDR} we present the $\SkDR$ and $\HypDR$
algorithms, respectively, that manipulate rules obtained by Skolemizing the
input GTGDs. All of these algorithms can produce intermediate GTGDs/rules that
are not necessarily Datalog rules. In
\iftoggle{withappendix}{Appendix~\ref{app:FullDR}}{the online appendix
\cite{saturation-github}} we present the $\FullDR$ algorithm that manipulates
GTGDs, but derives only Datalog rules. However, the performance of $\FullDR$
proved to not be competitive, so we do not discuss it any further here.

Each algorithm is defined by an inference rule $\Inf$ that can be applied to
several TGDs/rules to derive additional TGDs/rules. For simplicity, we use the
same name for the rule and the resulting algorithm. Given a set of GTGDs
$\Sigma$, the algorithm applies $\Inf$ to (the Skolemization of) $\Sigma$ as
long as possible and then returns all produced Datalog rules. This process,
however, can derive a large number of TGDs/rules, so it is vital to eliminate
TGDs/rules whenever possible. We next define notions of \emph{redundancy} that
can be used to discard certain TGDs/rules produced by $\Inf$.

\begin{definition}\label{def:redundancy}
    A TGD ${\tau_1 = \forall \vec x_1[\body_1 \rightarrow \exists \vec y_1 ~
    \head_1]}$ is a \emph{syntactic tautology} if it is in head-normal form and
    ${\body_1 \cap \head_1 \neq \emptyset}$. TGD $\tau_1$ \emph{subsumes} a TGD
    ${\tau_2 = \forall \vec x_2[\body_2 \rightarrow \exists \vec y_2~\head_2]}$
    if there exists a substitution $\mu$ such that ${\dom(\mu) = \vec x_1 \cup
    \vec y_1}$, ${\mu(\vec x_1) \subseteq \vec x_2}$, ${\mu(\vec y_1) \subseteq
    \vec y_1 \cup \vec y_2}$, ${\mu(y) \neq \mu(y')}$ for distinct $y$ and $y'$
    in $\vec y_1$, ${\mu(\body_1) \subseteq \body_2}$, and ${\mu(\head_1)
    \supseteq \head_2}$.

    A rule ${\tau_1 = \forall \vec x_1[\body_1 \rightarrow H_1]}$ is a
    \emph{syntactic tautology} if ${H_1 \in \body_1}$. Rule $\tau_1$
    \emph{subsumes} a rule ${\tau_2 = \forall \vec x_2[\body_2 \rightarrow
    H_2]}$ if there exists a substitution $\mu$ such that ${\mu(\body_1)
    \subseteq \body_2}$ and ${\mu(H_1) = H_2}$.

    A TGD/rule $\tau$ is \emph{contained in a set of TGDs/rules $S$ up to
    redundancy} if $\tau$ is a syntactic tautology or some ${\tau' \in S}$
    subsumes $\tau$.
\end{definition}

The following example illustrates Definition~\ref{def:redundancy}.

\begin{example}\label{ex:redundancy}
Rule ${A(x) \wedge B(x) \rightarrow A(x)}$ is a syntactic tautology: applying a
chase step with it cannot produce a new fact. A non-full TGD in head-normal
form cannot be a syntactic tautology since each head atom of such a TGD
contains an existentially quantified variable that does not occur in the TGD
body.

Rule ${\tau_1 = A(f(x_1),f(x_1)) \wedge B(x_1) \rightarrow B(f(x_1))}$ is
subsumed by rule ${\tau_2 = A(x_2,x_3) \rightarrow B(x_2)}$ using substitution
$\mu_1$ that maps both $x_2$ and $x_3$ to $f(x_1)$. If $\tau_1$ derives
$B(f(t))$ in one step from a set of facts $I$ by a substitution $\sigma$ where
${\sigma(x_1) = t}$, then $\tau_2$ also derives $B(f(t))$ from $I$ in one step
by substitution ${\sigma \circ \mu_1}$. Thus, rule $\tau_1$ is not needed when
rule $\tau_2$ is present, so $\tau_1$ can be discarded.

While syntactic tautologies and rule subsumption are standard in first-order
theorem proving \cite{bachmair2001resolution}, subsumption of TGDs is more
involved. TGD ${\tau_3 = A(x_1,x_1) \wedge B(x_1) \rightarrow \exists y_1 ~
C(x_1,y_1)}$ is subsumed by TGD ${\tau_4 = A(x_2,x_3) \rightarrow \exists y_2,
y_3 ~ C(x_2,y_2) \wedge D(x_3, y_3)}$ by substitution $\mu_2$ where
${\mu_2(x_2) = \mu_2(x_3) = x_1}$, ${\mu_2(y_2) = y_1}$, and ${\mu_2(y_3) =
y_3}$. The conditions on substitution $\mu_2$ in
Definition~\ref{def:redundancy} ensure that $y_2$ and $y_3$ are not mapped to
each other or to $x_1$. Thus, as in the previous paragraph, the result of each
chase step with $\tau_3$ and substitutions $\sigma$ and $\sigma'$ can always be
obtained (up to isomorphism) by a chase step with $\tau_4$ and substitutions
${\sigma \circ \mu_2}$ and ${\sigma' \circ \mu_2}$.
\end{example}

In Definition~\ref{def:rewriting-algorithm} we formalize the notion of applying
$\Inf$ exhaustively up to redundancy. The definition, however, does not say how
to actually do it: we discuss this and other issues in
Section~\ref{sec:implementation}.

\begin{definition}\label{def:rewriting-algorithm}
    For $\Inf$ an inference rule and $\Sigma$ a finite set of GTGDs,
    $\Inf(\Sigma)$ is the subset of all Skolem-free Datalog rules of $\Sigma'$,
    where $\Sigma'$ is the smallest set that contains up to redundancy each
    TGD/rule obtained by
    \begin{itemize}
        \item transforming $\Sigma$ into head-normal form if $\Inf$ manipulates
        TGDs or Skolemizing $\Sigma$ if $\Inf$ manipulates rules, and

        \item selecting an adequate number of premises in $\Sigma'$, renaming
        any variables shared by distinct premises, applying $\Inf$ to the
        renamed premises, and transforming the result into head-normal form.
    \end{itemize}
\end{definition}

\subsection{The Existential-Based Rewriting}\label{subsec:algorithms:ExbDR}

As we discussed in Section~\ref{sec:chaserewrite}, each loop ${T_i, \dots,
T_j}$ at vertex $v$ in a one-pass chase sequence can be seen as taking $T_i(v)$
as input and producing one fact included in $T_j(v)$ as output. Let $v'$ be
child of $v$ introduced in $T_{i+1}$. The idea behind the $\ExbDR$ algorithm is
to derive all GTGDs such that, for each $k$ with ${i < k \leq j}$, all facts of
$T_k(v')$ can be derived from the input $T_i(v)$ in one step. The output of the
loop can then also be derived from $T_i(v)$ in one step by full GTGD, so this
GTGD provides us with the desired loop ``shortcut''. Before formalizing this
idea, we slightly adapt the notion of unification.

\begin{definition}
    For $X$ a set of variables, an \emph{$X$-unifier} and an \emph{$X$-MGU}
    $\theta$ of atoms ${A_1, \dots, A_n}$ and ${B_1, \dots, B_n}$ are defined
    as in Section~\ref{sec:preliminaries}, but with the additional requirement
    that ${\theta(x) = x}$ for each ${x \in X}$.
\end{definition}

It is straightforward to see that an $X$-MGU is unique up to the renaming of
variables not contained in $X$, and that it can be computed as usual while
treating variables in $X$ as if they were constants. We are now ready to
formalize the $\ExbDR$ algorithm.

\begin{definition}\label{def:ExbDR}
    The \emph{Existential-Based Datalog Rewriting} inference rule $\ExbDR$
    takes two guarded TGDs
    \begin{align*}
        \tau    & = \forall \vec x [\body \rightarrow \exists \vec y~\head \wedge A_1 \wedge \dots \wedge A_n] \quad \text{with } n \geq 1 \text{ and} \\
        \tau'   & = \forall \vec z [A_1' \wedge \dots \wedge A_n' \wedge \body' \rightarrow H']
    \end{align*}
    and, for $\theta$ a $\vec y$-MGU of ${A_1, \dots, A_n}$ and ${A_1', \dots,
    A_n'}$, , if ${\theta(\vec x) \cap \vec y = \emptyset}$ and
    ${\vars(\theta(\body')) \cap \vec y = \emptyset}$, it derives
    \begin{displaymath}
        \theta(\body) \wedge \theta(\body') \rightarrow \exists \vec y~\theta(\head) \wedge \theta(A_1) \wedge \dots \wedge \theta(A_n) \wedge \theta(H').
    \end{displaymath}
\end{definition}

\begin{example}\label{ex:ExbDR}
Consider again the set $\Sigma$ from Example~\ref{ex:one-pass}. The idea behind
the $\ExbDR$ algorithm is illustrated in Figure~\ref{fig:ExbDR-loop}, which
summarizes the steps of the loop ${T_0, T_1, T_2, T_3, T_4}$ from
Figure~\ref{fig:chase-sequence}. We denote the vertices by $r$ and $v_1$ as in
Example~\ref{ex:one-pass}.

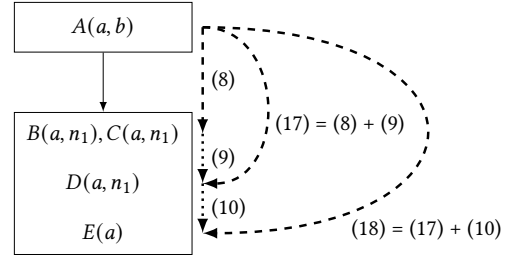
\begin{figure}[tb]
    \centering
    \begin{tikzpicture}
    \tikzstyle{nonfull}=[->,>=latex,dashed,line width=1pt]
    \tikzstyle{full}=[->,>=latex,dotted,line width=1pt]

    \newcommand{\fact}[5]{
        \node       (#1) at (#3,#4)   {#5} ;
        \coordinate (#2) at (1.4,#4)  {} ;
    }

    \fact{A}{nA}{0}{0}{$A(a,b)$}

    \fact{BC}{nBC}{0}{-1.5}{$B(a,n_1), C(a,n_1)$}

    \fact{D}{nD}{0}{-2.2}{$D(a,n_1)$}

    \fact{E}{nE}{0}{-2.9}{$E(a)$}

    \node (r)  at (A) [draw,minimum width=2.5cm,minimum height=0.7cm] {} ;

    \node (v1) at (D) [draw,minimum width=2.5cm,minimum height=2cm] {} ;

    \path[-latex] (r) edge (v1) ;
    
    \draw[nonfull] (nA)  to                                node[right]         {\eqref{ex:one-pass:1}}                                              (nBC) ;
    \draw[full]    (nBC) to                                node[right]         {\eqref{ex:one-pass:2}}                                              (nD)  ;
    \draw[nonfull] (nA)  to[out=0,in=0,min distance=1.2cm] node[below right]   {\eqref{ex:ExbDR:1} = \eqref{ex:one-pass:1} + \eqref{ex:one-pass:2}} (nD)  ;
    \draw[full]    (nD)  to                                node[right]         {\eqref{ex:one-pass:3}}                                              (nE)  ;
    \draw[nonfull] (nA)  to[out=0,in=0,min distance=4.2cm] node[yshift=-1.4cm] {\eqref{ex:ExbDR:2} = \eqref{ex:ExbDR:1} + \eqref{ex:one-pass:3}}    (nE)  ;
    
\end{tikzpicture}
    \caption{Deriving ``shortcuts'' for the loop $T_0$--$T_4$ in $\ExbDR$}\label{fig:ExbDR-loop}
\end{figure}

Fact $A(a,b)$ is the input to the loop, and the first step of the loop derives
$B(a,n_1)$ and $C(a,n_1)$ using GTGD \eqref{ex:one-pass:1}. Next, GTGD
\eqref{ex:one-pass:2} evolves vertex $v_1$ by deriving $D(a,n_1)$. To capture
this, the $\ExbDR$ inference rule combines \eqref{ex:one-pass:1}, the GTGD that
creates $v_1$, with \eqref{ex:one-pass:2}, the GTGD that evolves $v_1$. This
produces GTGD \eqref{ex:ExbDR:1}, which derives all facts of $v_1$ from the
input fact in one step. Vertex $v_1$ is evolved further using GTGD
\eqref{ex:one-pass:3} to derive $E(a)$. To reflect this, the $\ExbDR$ inference
rule combines \eqref{ex:ExbDR:1} and \eqref{ex:one-pass:3} to produce
\eqref{ex:ExbDR:2}, which again derives all facts of $v_1$ from the loop's
input in one step.
\begin{align}
    A(x_1,x_2)  & \rightarrow \exists y ~ B(x_1,y) \wedge C(x_1,y) \wedge D(x_1,y)               \label{ex:ExbDR:1} \\
    A(x_1,x_2)  & \rightarrow \exists y ~ B(x_1,y) \wedge C(x_1,y) \wedge D(x_1,y) \wedge E(x_1) \label{ex:ExbDR:2}
\end{align}
Fact $E(a)$ does not contain the labeled null $n_1$ that is introduced when
creating $v_1$, so it can be propagated to the root vertex $r$ as the output of
the loop. This is reflected in \eqref{ex:ExbDR:2}: atom $E(x_1)$ does not
contain any existential variables. Definition~\ref{def:rewriting-algorithm}
requires each derived GTGD to be brought into head-normal, so
\eqref{ex:ExbDR:2} is broken up into \eqref{ex:ExbDR:1} and
\eqref{ex:rewriting:1}. The latter GTGD is full, and it provides us with the
desired shortcut for the loop.

Next, \eqref{ex:one-pass:5} and atom $F(x_1,y_1)$ of \eqref{ex:one-pass:4}
produce \eqref{ex:ExbDR:3}, and transformation into head-normal form produces
\eqref{ex:one-pass:4} and \eqref{ex:rewriting:2}. Moreover,
\eqref{ex:one-pass:1} and \eqref{ex:one-pass:6} produce \eqref{ex:ExbDR:4}, and
transformation \eqref{ex:ExbDR:4} into head-normal form produces
\eqref{ex:rewriting:3} and \eqref{ex:ExbDR:5}.
\begin{align}
    A(x_1,x_2) \wedge E(x_1)    & \rightarrow \exists y_1, y_2 ~ F(x_1,y_1) \wedge F(y_1,y_2) \wedge G(x_1) \label{ex:ExbDR:3} \\
    A(x_1,x_2) \wedge G(x_1)    & \rightarrow \exists y ~  B(x_1,y) \wedge C(x_1,y) \wedge H(x_1)           \label{ex:ExbDR:4} \\
    A(x_1,x_2) \wedge G(x_1)    & \rightarrow \exists y ~  B(x_1,y) \wedge C(x_1,y)                         \label{ex:ExbDR:5}
\end{align}
GTGD \eqref{ex:ExbDR:5} is subsumed by \eqref{ex:one-pass:1} so it can be
dropped. No further inferences are possible after this, so all derived full
GTGDs are returned as the rewriting of $\Sigma$.
\end{example}

Before proceeding, we present an auxiliary result showing certain key properties
of the $\ExbDR$ inference rule.

\begin{restatable}{proposition}{ExbDRProperties}\label{prop:ExbDR-properties}
    Each application of the $\ExbDR$ inference rule to $\tau$, $\tau'$, and
    $\theta$ as in Definition~\ref{def:ExbDR} satisfies the following
    properties.
    \begin{enumerate}
        \item Some atom $A_i'$ with ${1 \leq i \leq n}$ is a guard in $\tau'$.

        \item For each ${1 \leq i \leq n}$ such that $A_i'$ is a guard of
        $\tau'$, and for $\sigma$ the $\vec y$-MGU of $A_i'$ and the
        corresponding atom $A_i$ such that ${\sigma(\vec x) \cap \vec y =
        \emptyset}$, it is the case that ${\vars\bigl(\sigma(A_j')\bigr) \cap
        \vec y \neq \emptyset}$ for each ${1 \leq j \leq n}$.

        \item The result is a GTGD whose body and head width are at most
        $\bwidth(\Sigma)$ and $\hwidth(\Sigma)$, respectively.
    \end{enumerate}
\end{restatable}

In the second claim of Proposition~\ref{prop:ExbDR-properties}, $\sigma$
unifies only $A_i$ and $A_i'$, whereas $\theta$ unifies all ${A_1, \dots, A_n}$
and ${A_1', \dots, A_n'}$; thus, $\sigma$ and $\theta$ are not necessarily the
same. The third claim is needed to prove termination of $\ExbDR$.

Proposition~\ref{prop:ExbDR-properties} can be used to guide the application of
the $\ExbDR$ inference rule. Consider an attempt to apply the $\ExbDR$
inference rule to two candidate GTGDs ${\tau = \body \rightarrow \exists \vec y
~ \head}$ and ${\tau' = \body' \rightarrow H'}$. The first claim of
Proposition~\ref{prop:ExbDR-properties} tells us that a guard of $\tau'$ will
definitely participate in the inference. Thus, we can choose one such guard
${G' \in \body'}$ of $\tau'$ and try to find a $\vec y$-MGU $\sigma$ of $G'$
and a counterpart atom ${G \in \head}$ from the head of $\tau$. Next, we need
to check whether ${\sigma(\vec x) \cap \vec y = \emptyset}$; if not, there is
no way for ${\theta(\vec x) \cap \vec y = \emptyset}$ to hold so the inference
is not possible. By the second claim of
Proposition~\ref{prop:ExbDR-properties}, all candidates for the atoms
participating in the inference will contain a variable that is mapped by
$\sigma$ to a member of $\vec y$; thus, ${S' = \bigl\{ \sigma(A') \mid A' \in
\body'\wedge \vars(\sigma(A')) \cap \vec y \neq \emptyset \bigr\}}$ is the set
of all relevant side atoms. Note that we apply $\sigma$ to the atoms in $S'$ to
simplify further matching. The next step is to identify the corresponding head
atoms of $\tau$. To achieve this, for each atom ${A' \in S'}$ of the form
$R(t_1, \dots, t_n)$, we identify the set $C[A']$ of candidate counterpart
atoms as the set of atoms of the form ${R(s_1, \dots, s_n) \in \sigma(\head)}$
such that, for each argument position $i$ with ${1 \leq i \leq n}$, if either
${t_i \in \vec y}$ or ${s_i \in \vec y}$, then ${t_i = s_i}$. Finally, we
consider each possible combination $S$ of such candidates, and we try to find
an MGU $\theta$ of sets $S$ and $S'$. If unification succeeds, we derive the
corresponding GTGD.

\begin{restatable}{theorem}{ExbDRCorrectnessComplexity}\label{thm:ExbDR-correctness-complexity}
    Program $\ExbDR(\Sigma)$ is a Datalog rewriting of a finite set of GTGDs
    $\Sigma$. Moreover, the rewriting can be computed in time ${O(b^{r^d \cdot
    (w_b + c)^{d a} \cdot r^d \cdot (w_h + c)^{d a}})}$ for $r$ the number of
    relations in $\Sigma$, $a$ the maximum relation arity in $\Sigma$, ${w_b =
    \bwidth(\Sigma)}$, ${w_h = \hwidth(\Sigma)}$, ${c = |\consts(\Sigma)|}$,
    and some $b$ and $d$.
\end{restatable}

Program $\ExbDR(\Sigma)$ can thus be large in the worst case. In
Section~\ref{sec:experiments} we show empirically that rewritings are suitable
for practical use. From a theoretical point of view, checking fact entailment
via $\ExbDR(\Sigma)$ is worst-case optimal. To see why, let $r$, $a$, and $c$
be as in Theorem~\ref{thm:ExbDR-correctness-complexity}, and consider a base
instance $I$ with $c'$ constants. The fixpoint of $\ExbDR(\Sigma)$ on $I$
contains at most ${r (c+c')^a}$ facts, and it can be computed in time ${O(r
(c+c')^a \cdot |\ExbDR(\Sigma)|)}$: each rule ${\tau \in \ExbDR(\Sigma)}$ is
guarded so we can apply a chase step with $\tau$ by matching a guard and then
checking the remaining body atoms. Hence, we can compute $\ExbDR(\Sigma)$ and
find its fixpoint in \textsc{2ExpTime}, in \textsc{ExpTime} if the relation
arity is fixed, and in \textsc{PTime} if $\Sigma$ is fixed (i.e., if we
consider \emph{data complexity}). These results match the lower bounds for
checking fact entailment for GTGDs \cite{datalogpmj}.

\subsection{Using Skolemization}\label{subsec:algorithms:SkDR}

The $\ExbDR$ algorithm exhibits two drawbacks. First, each application of the
$\ExbDR$ inference rule potentially introduces a head atom, so the rule heads
can get very long. Second, each inference requires matching a subset of body
atoms of $\tau'$ to a subset of the head atoms of $\tau$; despite the
optimizations outlined after Proposition~\ref{prop:ExbDR-properties}, this can
be costly, particularly when rule heads are long.

We would ideally derive GTGDs with a single head atom and unify just one body
atom of $\tau'$ with the head atom of $\tau$, but this does not seem possible
if we stick to manipulating GTGDs. For example, atoms $C(y)$ and $D(y)$ of GTGD
\eqref{ex:ExbDR:1} refer to the same labeled null (represented by variable
$y$), and this information would be lost if we split \eqref{ex:ExbDR:1} into
two GTGDs. We thus need a way to refer to the same existentially quantified
object in different logical formulas. This can be achieved by replacing
existentially quantified variables by Skolem terms, which in turns gives rise
to the $\SkDR$ algorithm from Definition~\ref{def:SkDR}. Before presenting the
algorithm, in Definition~\ref{def:guarded-rule} we generalize the notion of
guardedness to rules.

\begin{definition}\label{def:guarded-rule}
    Rule ${\forall \vec x [\body \rightarrow H]}$ is \emph{guarded} if each
    function symbol in the rule is a Skolem symbol, the body $\body$ contains a
    Skolem-free atom ${A \in \body}$ such that ${\vars(A) = \vec x}$, and each
    Skolem term in the rule is of the form $f(\vec t)$ where
    ${\vars\bigl(f(\vec t)\bigr) = \vec x}$ and $\vec t$ is function-free.
\end{definition}

\begin{definition}\label{def:SkDR}
    The \emph{Skolem Datalog Rewriting} inference rule $\SkDR$ takes two
    guarded rules
    \begin{displaymath}
        \tau = \body \rightarrow H \qquad \text{and} \qquad \tau' = A' \wedge \body' \rightarrow H'
    \end{displaymath}
    such that
    \begin{itemize}
        \item $\body$ is Skolem-free and $H$ contains a Skolem symbol, and

        \item $A'$ contains a Skolem symbol, or $\tau'$ is Skolem-free and $A'$
        contains all variables of $\tau'$,
    \end{itemize}
    and, for $\theta$ an MGU of $H$ and $A'$, it derives
    \begin{displaymath}
        \theta(\body) \wedge \theta(\body') \rightarrow \theta(H').
    \end{displaymath}
\end{definition}

\begin{example}\label{ex:SkDR}
Skolemizing GTGDs \eqref{ex:one-pass:1} and \eqref{ex:one-pass:4} produces
rules \eqref{ex:SkDR:1}--\eqref{ex:SkDR:2}, and
\eqref{ex:SkDR:3}--\eqref{ex:SkDR:4}, respectively.
\begin{align}
    A(x_1,x_2)                  & \rightarrow B(x_1,f(x_1,x_2))         \label{ex:SkDR:1} \\
    A(x_1,x_2)                  & \rightarrow C(x_1,f(x_1,x_2))         \label{ex:SkDR:2} \\
    A(x_1,x_2) \wedge E(x_1)    & \rightarrow F(x_1,g(x_1,x_2))         \label{ex:SkDR:3} \\
    A(x_1,x_2) \wedge E(x_1)    & \rightarrow F(g(x_1,x_2),h(x_1,x_2))  \label{ex:SkDR:4}
\end{align}
Intuitively, rules \eqref{ex:SkDR:1} and \eqref{ex:SkDR:2} jointly represent
the facts introduced by the non-full GTGD \eqref{ex:one-pass:1}: functional
term $f(x_1,x_2)$ allows both rules to ``talk'' about the same labeled nulls.
This allows the $\SkDR$ inference rule to simulate the $\ExbDR$ inference rule
while unifying just pairs of atoms. In particular, $\SkDR$ combines
\eqref{ex:SkDR:1} and \eqref{ex:one-pass:3} to obtain \eqref{ex:SkDR:5}; it
combines \eqref{ex:SkDR:2} and \eqref{ex:one-pass:2} to obtain
\eqref{ex:SkDR:6}; and it combines \eqref{ex:SkDR:5} and \eqref{ex:SkDR:6} to
obtain the ``shortcut'' rule \eqref{ex:rewriting:1}.
\begin{align}
    A(x_1,x_2) \wedge D(x_1,f(x_1,x_2)) & \rightarrow E(x_1)            \label{ex:SkDR:5} \\
    A(x_1,x_2)                          & \rightarrow D(x_1,f(x_1,x_2)) \label{ex:SkDR:6}
\end{align}
The rules with Skolem-free bodies derived in this way allow us to reconstruct
derivations in one step analogously to Example~\ref{ex:ExbDR}, and the rules
with Skolem symbols in body atoms capture the intermediate derivation steps.
For example, rules \eqref{ex:SkDR:5} and \eqref{ex:SkDR:7} capture the result
of matching the first and the second body atom, respectively, of rule
\eqref{ex:one-pass:3} to facts produced by rules \eqref{ex:SkDR:1} and
\eqref{ex:SkDR:6}, respectively. To complete the rewriting, $\SkDR$ combines
\eqref{ex:SkDR:3} with \eqref{ex:one-pass:5} to obtain \eqref{ex:rewriting:2},
and it combines \eqref{ex:SkDR:1} with \eqref{ex:one-pass:6} to derive
\eqref{ex:rewriting:3}.

However, $\SkDR$ also combines \eqref{ex:one-pass:3} and \eqref{ex:SkDR:6} into
\eqref{ex:SkDR:7}, which with \eqref{ex:SkDR:1} derives \eqref{ex:rewriting:1}
the second time. These inferences are superfluous: they just process the two
body atoms of \eqref{ex:one-pass:3} in a different order. Also, $\SkDR$
combines \eqref{ex:one-pass:5} and \eqref{ex:SkDR:4} into rule
\eqref{ex:SkDR:8}, which is a ``dead-end'' in that it does not further
contribute to a Datalog rule.
\begin{align}
    A(x_1,x_2) \wedge B(x_1,f(x_1,x_2))             & \rightarrow E(x_1)        \label{ex:SkDR:7} \\
    A(x_1,x_2) \wedge E(x_1) \wedge E(g(x_1,x_2))   & \rightarrow G(g(x_1,x_2)) \label{ex:SkDR:8}
\end{align}
Our $\HypDR$ algorithm in Subsection~\ref{subsec:algorithms:HypDR} can avoid these
overheads, but at the expense of using more than two rules at a time.
\end{example}

Proposition~\ref{prop:SkDR-properties} and
Theorem~\ref{thm:SkDR-correctness-complexity} capture the relevant properties
of the $\SkDR$ algorithm.

\begin{restatable}{proposition}{SkDRProperties}\label{prop:SkDR-properties}
    Each application of the $\SkDR$ inference rule to rules $\tau$ and $\tau'$
    as in Definition~\ref{def:SkDR} produces a guarded rule.
\end{restatable}

\begin{restatable}{theorem}{SkDRCorrectnessComplexity}\label{thm:SkDR-correctness-complexity}
    Program $\SkDR(\Sigma)$ is a Datalog rewriting of a finite set of GTGDs
    $\Sigma$. Moreover, the rewriting can be computed in time ${O(b^{r^d \cdot
    (e + w_b + c)^{d a}})}$ for $r$ the number of relations in $\Sigma$, $a$
    the maximum relation arity in $\Sigma$, $e$ the number of existential
    quantifiers in $\Sigma$, ${w_b = \bwidth(\Sigma)}$, ${c =
    |\consts(\Sigma)|}$, and some $b$ and $d$.
\end{restatable}

It is natural to wonder whether $\SkDR$ is guaranteed to be more efficient than
$\ExbDR$. We next show that neither algorithm is generally better: there exist
families of inputs on which $\SkDR$ performs exponentially more inferences than
$\ExbDR$, and vice versa.

\begin{proposition}\label{prop:ExbDR-worse-SkDR}
    There exists a family ${\{ \Sigma_n \}_{n \in \ints}}$ of finite sets of
    GTGDs such that the number of GTGDs derived by $\ExbDR$ is $O(2^n)$ times
    larger than the number of rules derived by $\SkDR$ on each $\Sigma_n$.
\end{proposition}

\begin{proof}\let\qed\relax
For each ${n \in \ints}$, let $\Sigma_n$ contain the following GTGDs.
\begin{align}
    A(x)                            & \rightarrow \exists \vec y ~ B_1(x,y_1) \wedge \dots \wedge B_n(x,y_n)    \label{eq:ExpDR:1} \\
    B_i(x_1,x_2) \wedge C_i(x_1)    & \rightarrow D_i(x_1,x_2) \text{ for } 1 \leq i \leq  n                    \label{eq:ExpDR:2}
\end{align}
On such $\Sigma_n$, $\ExbDR$ derives a GTGD of the form \eqref{eq:ExpDR:3} for
each subset ${\{ k_1, \dots, k_m \} \subseteq \{ 1, \dots, n \}}$, and there
are $2^n$ such TGDs. In contrast, the Skolemization of \eqref{eq:ExpDR:1}
consists of $n$ rules shown in equation~\eqref{eq:ExpDR:4}, so $\SkDR$ derives
just $n$ rules shown in equation~\eqref{eq:ExpDR:5}.
\begin{align}
    A(x) \wedge \bigwedge_{i=1}^m C_{k_i}(x)    & \rightarrow \exists \vec y ~ \bigwedge_{i=1}^n B_i(x,y_i) \wedge \bigwedge_{i=1}^m D_{k_i}(x,y_{k_i}) \label{eq:ExpDR:3} \\
    A(x)                                        & \rightarrow B_i(x,f_i(x)) \text{ for } 1 \leq i \leq  n                                               \label{eq:ExpDR:4} \\
    A(x) \wedge C_i(x)                          & \rightarrow D_i(x,f_i(x)) \text{ for } 1 \leq i \leq  n \rlap{\qquad\qquad\qedsymbol}                 \label{eq:ExpDR:5}
\end{align}
\end{proof}

\begin{proposition}\label{prop:SkDR-worse-ExbDR}
    There exists a family ${\{ \Sigma_n \}_{n \in \ints}}$ of finite sets of
    GTGDs such that the number of rules derived by $\SkDR$ is $O(2^n)$ times
    larger than the number of TGDs derived by $\ExbDR$ on each $\Sigma_n$.
\end{proposition}

\begin{proof}\let\qed\relax
For each ${n \in \ints}$, let $\Sigma_n$ contain the following GTGDs.
\begin{align}
    A(x) \rightarrow \exists y ~ B_1(x,y) \wedge \dots \wedge B_n(x,y)  \label{eq:SkDR:1} \\
    B_1(x_1,x_2) \wedge \dots \wedge B_n(x_1,x_2) \rightarrow C(x_1)    \label{eq:SkDR:2}
\end{align}
On such $\Sigma_n$, $\ExbDR$ derives just GTGD \eqref{eq:SkDR:3} in one step.
In contrast, the Skolemization of \eqref{eq:SkDR:1} consists of $n$ rules of
the form \eqref{eq:SkDR:4} for each ${1 \leq i \leq n}$. Thus, $\SkDR$ combines
these with \eqref{eq:SkDR:2} to derive $2^n-1$ rules of the form
\eqref{eq:SkDR:5}, one for each subset ${\{ k_1, \dots, k_m \} \subsetneq \{ 1,
\dots, n \}}$.
\begin{align}
    & A(x) \rightarrow C(x)                                                                                     \label{eq:SkDR:3} \\
    & A(x) \rightarrow B_i(x,f(x))                                                                              \label{eq:SkDR:4} \\
    & A(x) \wedge B_{k_1}(x,f(x)) \wedge \dots \wedge B_{k_m}(x,f(x)) \rightarrow C(x) \rlap{\;\;\qedsymbol}    \label{eq:SkDR:5}
\end{align}
\end{proof}

\subsection{Combining Several $\SkDR$ Steps into One}\label{subsec:algorithms:HypDR}

The $\SkDR$ algorithm can produce many rules with Skolem symbols in the body,
which is the main reason for Proposition~\ref{prop:SkDR-worse-ExbDR}. We next
present the $\HypDR$ algorithm, which uses the \emph{hyperresolution} inference
rule as a kind of ``macro'' to combine several $\SkDR$ steps into one. We show
that this can be beneficial for several reasons.

\begin{definition}\label{def:HypDR}
    The \emph{Hyperresolution Rewriting} inference rule $\HypDR$ takes guarded
    rules
    \begin{align*}
        \tau_1  & = \body_1 \rightarrow H_1 \quad \dots \quad \tau_n = \body_n \rightarrow H_n \text{ and}\\
        \tau'   & = A_1' \wedge \dots \wedge A_n' \wedge \body' \rightarrow H'
    \end{align*}
    such that
    \begin{itemize}
        \item for each $i$ with ${1 \leq i \leq n}$, conjunction $\body_i$ is
        Skolem-free and atom $H_i$ contains a Skolem symbol, and

        \item rule $\tau'$ is Skolem-free,
    \end{itemize}
    and, for $\theta$ an MGU of ${H_1, \dots, H_n}$ and ${A_1', \dots, A_n'}$,
    if conjunction $\theta(\body')$ is Skolem-free, it derives
    \begin{displaymath}
        \theta(\body_1) \wedge \dots \wedge \theta(\body_n) \wedge \theta(\body') \rightarrow \theta(H').
    \end{displaymath}
\end{definition}

\begin{example}\label{ex:HypDR}
The $\HypDR$ inference rule simulates chase steps in the child vertex of a loop
analogously to $\ExbDR$: all body atoms matching a fact introduced in the child
vertex are resolved in one step. We can see two benefits of this on our running
example.

First, $\HypDR$ derives \eqref{ex:SkDR:6} from \eqref{ex:SkDR:2} and
\eqref{ex:one-pass:2}, and it derives \eqref{ex:rewriting:1} from
\eqref{ex:one-pass:3}, \eqref{ex:SkDR:1}, and \eqref{ex:SkDR:6}. Rule
\eqref{ex:rewriting:1} is derived just once, and without intermediate rules
\eqref{ex:SkDR:5} and \eqref{ex:SkDR:7}. In other words, the $\HypDR$ inference
rule does not resolve the body atoms of a rule in every possible order. As
Proposition~\ref{prop:SkDR-worse-HypDR} below shows, this can reduce the number
of derived rules by an exponential factor.

Second, $\HypDR$ derives only rules with Skolem-free bodies, and thus does not
derive the ``dead-end'' rule \eqref{ex:SkDR:8}. In other words, all
consequences of $\HypDR$ derive in one step one fact in the child vertex of a
loop from the loop's input $T_i(v)$.

The downside of $\HypDR$ is that more than two rules can participate in an
inference. This requires more complex unification and selection of candidates
that can participate in an inference.
\end{example}

Proposition~\ref{prop:HypDR-properties} and
Theorem~\ref{thm:HypDR-correctness-complexity} capture the properties of
$\HypDR$, and Proposition~\ref{prop:SkDR-worse-HypDR} compares it to $\SkDR$.

\begin{restatable}{proposition}{HypDRProperties}\label{prop:HypDR-properties}
    Each application of the $\HypDR$ inference rule to rules ${\tau_1, \dots,
    \tau_n}$ and $\tau'$ as in Definition~\ref{def:HypDR} produces a guarded
    rule.
\end{restatable}

\begin{restatable}{theorem}{HypDRCorrectnessComplexity}\label{thm:HypDR-correctness-complexity}
    Program $\HypDR(\Sigma)$ is a Datalog rewriting of a finite set of GTGDs
    $\Sigma$. Moreover, the rewriting can be computed in time time ${O(b^{r^d
    \cdot (e + w_b + c)^{d a}})}$ for $r$ the number of relations in $\Sigma$,
    $a$ the maximum relation arity in $\Sigma$, $e$ the number of existential
    quantifiers in $\Sigma$, ${w_b = \bwidth(\Sigma)}$, ${c =
    |\consts(\Sigma)|}$, and some $b$ and $d$.
\end{restatable}

\begin{proposition}\label{prop:SkDR-worse-HypDR}
    There exists a family ${\{ \Sigma_n \}_{n \in \ints}}$ of finite sets of
    GTGDs such that $\SkDR$ derives $O(2^n)$ more rules than $\HypDR$ on each
    $\Sigma_n$.
\end{proposition}

\begin{proof}
For each ${n \in \ints}$, let $\Sigma_n$ contain the following GTGDs.
\begin{align}
    A(x) \rightarrow \exists y ~ B(x,y)                                                 \label{eq:HypDR:1} \\
    B(x_1,x_2) \wedge C_i(x_1) \rightarrow D_i(x_1,x_2) \text{ for } 1 \leq i \leq  n   \label{eq:HypDR:2} \\
    D_1(x_1,x_2) \wedge \dots \wedge D_n(x_1,x_2) \rightarrow E(x_1)                    \label{eq:HypDR:3}
\end{align}
Skolemizing \eqref{eq:HypDR:1} produces \eqref{eq:HypDR:4}. Thus, $\SkDR$
combines \eqref{eq:HypDR:4} with each \eqref{eq:HypDR:2} to derive each
\eqref{eq:HypDR:5}, and it uses \eqref{eq:HypDR:5} and \eqref{eq:HypDR:3} to
derive $2^n-1$ rules of the form \eqref{eq:HypDR:6} for each set of indexes $I$
satisfying ${\emptyset \subsetneq I \subseteq \{ 1, \dots, n \}}$; note that
none of these rules are redundant.
\begin{align}
    A(x) \rightarrow B(x,f(x))                                                                                                  \label{eq:HypDR:4} \\
    A(x) \wedge C_i(x) \rightarrow D_i(x,f(x)) \text{ for } 1 \leq i \leq  n                                                    \label{eq:HypDR:5} \\
    A(x) \wedge  \bigwedge_{i \in I} C_i(x) \wedge \bigwedge_{j \in \{ 1, \dots, n \} \setminus I} D_j(x,f(x)) \rightarrow E(x) \label{eq:HypDR:6}
\end{align}
In contrast, $\HypDR$ derives each \eqref{eq:HypDR:5} just like $\SkDR$, and it
combines in one step \eqref{eq:HypDR:3} and all \eqref{eq:HypDR:5} to derive
\eqref{eq:HypDR:6} for ${I = \{ 1, \dots, n \}}$.
\end{proof}

\section{Implementation and Optimizations}\label{sec:implementation}

In this section, we discuss numerous issues that have to be addressed to make
the computation of a rewriting practical.

\myparagraph{Computing $\Inf(\Sigma)$ in Practice}
Definition~\ref{def:rewriting-algorithm} does not specify how to compute the
set $\Sigma'$, and redundancy elimination makes this question nontrivial. When
$\Inf$ derives a TGD/rule $\tau$, we can apply subsumption in two ways. First,
we can discard $\tau$ if $\tau$ is subsumed by a previously derived TGD/rule;
this is known as \emph{forward subsumption}. Second, if $\tau$ is not
discarded, we can discard each previously derived TGD/rule that is subsumed by
$\tau$; this is known as \emph{backward subsumption}. The set of derived
TGD/rules can thus grow and shrink, so the application of $\Inf$ has to be
carefully structured to ensure that all inferences are performed eventually.

We address this problem by a variant of the \emph{Otter loop}
\cite{DBLP:journals/jar/McCuneW97} used in first-order theorem provers. The
pseudo-code is shown in Algorithm~\ref{alg:rewriting}. The algorithm maintains
two sets of TGDs/rules: the \emph{worked-off} set $\workedOffSet$ contain
TGDs/rules that have been processed by $\Inf$, and the \emph{unprocessed} set
$\unprocessedSet$ contains TGDs/rules that are still to be processed. Set
$\workedOffSet$ is initially empty (line~\ref{alg:rewriting:init:W}), and set
$\unprocessedSet$ is initialized to the head-normal form of $\Sigma$ if $\Inf$
manipulates TGDs, or to the Skolemization of $\Sigma$ if $\Inf$ manipulates
rules. The algorithm then processes each ${\tau \in \unprocessedSet}$ until
$\unprocessedSet$ becomes empty
(lines~\ref{alg:rewriting:loop:start}--\ref{alg:rewriting:loop:end}). It is
generally beneficial to process shorter TGDs/rules first as that improves
chances of redundancy elimination. After moving $\tau$ to $\workedOffSet$
(line~\ref{alg:rewriting:move:W}), the algorithm applies $\Inf$ to $\tau$ and
$\workedOffSet$ and transforms the results into head-normal form
(line~\ref{alg:rewriting:evolve}). The algorithm discards each resulting
${\tau' \in \evolvedSet}$ that is a syntactic tautology or is forward-subsumed
by an element of ${\workedOffSet \cup \unprocessedSet}$
(line~\ref{alg:rewriting:forward}). If $\tau'$ is not discarded, the algorithm
applies backward subsumption to $\tau'$, $\workedOffSet$, and $\unprocessedSet$
(line~\ref{alg:rewriting:backward}) and adds $\tau'$ to $\unprocessedSet$
(line~\ref{alg:rewriting:add:taup}). When all TGDs/rules are processed, the
algorithm returns all Skolem-free Datalog rules from $\workedOffSet$
(line~\ref{alg:rewriting:return}). The result of applying $\Inf$ to TGDs/rules
in $\workedOffSet$ is thus contained in ${\workedOffSet \cup \unprocessedSet}$
up to redundancy at all times so, upon algorithm's termination, set
$\workedOffSet$ satisfies the condition on $\Sigma'$ from
Definition~\ref{def:rewriting-algorithm}.

\begin{algorithm}[tb]
\caption{Computing $\Inf(\Sigma)$ for $\Sigma$ a finite set of GTGDs}\label{alg:rewriting}
\algrenewcommand\algorithmicindent{0.3cm}
\begin{algorithmic}[1]
    \State $\workedOffSet = \emptyset$                                                                      \label{alg:rewriting:init:W}
    \State $\unprocessedSet = \text{the head-normal form or the Skolemization of } \Sigma$                  \label{alg:rewriting:init:U}
    \While{$\unprocessedSet \neq \emptyset$}                                                                \label{alg:rewriting:loop:start}
        \State Choose some $\tau \in \unprocessedSet$ and remove it from $\unprocessedSet$
        \State $\workedOffSet = \workedOffSet \cup \{ \tau \}$                                              \label{alg:rewriting:move:W}
        \State Let $\evolvedSet$ be the result of applying $\Inf$ to $\tau$ and a subset of $\workedOffSet$ \label{alg:rewriting:evolve}
        \Statex \qquad\; and transforming the result into head-normal form
        \For{\textbf{each} $\tau' \in \evolvedSet$}                                                         \label{alg:rewriting:taup:start}
            \If{$\tau'$ is not contained in $\workedOffSet \cup \unprocessedSet$ up to redundancy}          \label{alg:rewriting:forward}
                \State Remove from $\workedOffSet$ and $\unprocessedSet$ each $\tau''$ subsumed by $\tau'$  \label{alg:rewriting:backward}
                \State $\unprocessedSet = \unprocessedSet \cup \{ \tau' \}$                                 \label{alg:rewriting:add:taup}
            \EndIf
        \EndFor                                                                                             \label{alg:rewriting:taup:end}
    \EndWhile                                                                                               \label{alg:rewriting:loop:end}
    \State \textbf{return} $\{ \tau \in \workedOffSet \mid \tau \text{ is a Skolem-free Datalog rule} \}$   \label{alg:rewriting:return}
\end{algorithmic}
\end{algorithm}

\myparagraph{Checking Subsumption}
Checking whether TGD/rule $\tau_1$ subsumes $\tau_2$ is \textsc{NP}-complete
\cite{DBLP:conf/cade/KapurN86}, and the main difficulty is in matching the
variables of $\tau_1$ to the variables of $\tau_2$. Thus, we use an approximate
check in our implementation. First, we normalize each TGD to use fixed
variables ${x_1, x_2, \dots}$ and ${y_1, y_2, \dots}$: we sort the body and
head atoms by their relations using an arbitrary, but fixed ordering and
breaking ties arbitrarily, and then we rename all variables so that the
$i^{th}$ distinct occurrence of a universally (respectively existentially)
quantified variable from left to right is $x_i$ (respectively $y_i$). To see
whether ${\tau_1 = \body_1 \rightarrow \exists \vec y ~ \head_1}$ subsumes
${\tau_2 = \body_2 \rightarrow \exists \vec y ~ \head_2}$, we determine whether
${\body_1 \subseteq \body_2}$ and ${\head_1 \supseteq \head_2}$ holds, which
requires only polynomial time. We use a similar approximation for rules.
Variable normalization ensures termination, and using a modified subsumption
check does not affect the correctness of the rewriting: set $\workedOffSet$ may
contain more TGDs/rules than strictly necessary, but these are all logical
consequences of (the Skolemization of) $\Sigma$.

\myparagraph{Subsumption Indexing}
Sets $\workedOffSet$ and $\unprocessedSet$ can be large, so we use a variant of
\emph{feature vector indexing} \cite{schulz} to retrieve subsumption candidates
in ${\workedOffSet \cup \unprocessedSet}$. For simplicity, we consider only
TGDs in the following discussion, but rules can be handled analogously. Note
that a TGD $\tau_1$ can subsume TGD $\tau_2$ only if the set of relations
occurring in the body of $\tau_1$ (respectively the head of $\tau_2$) is a
subset of the set of relations occurring in the body of $\tau_2$ (respectively
the head of $\tau_1$). Thus, we can reduce the problem of retrieving
subsumption candidates to the problem of, given a domain set $D$, a set $N$ of
subsets of $D$, a subset ${S \subseteq D}$, and ${{\bowtie} \in \{ {\subseteq},
{\supseteq} \}}$, retrieving each ${S' \in N}$ satisfying ${S' \bowtie S}$. The
set-trie data structure \cite{DBLP:conf/IEEEares/Savnik13} can address this
problem. The idea is to order $D$ in an arbitrary, yet fixed way, so that we
can treat each subset of $N$ as a word over $D$. We then index $N$ by
constructing a trie over the words representing the elements of $N$. Finally,
we retrieve all ${S' \in N}$ satisfying ${S' \bowtie S}$ by traversing the
trie, where the ordering on $D$ allows us to considerably reduce the number of
vertices we visit during the traversal.

A minor issue is that retrieving TGDs that subsume a given TGD requires both
subset and superset testing for body and head relations, respectively, and vice
versa for retrieval of subsumed TGDs. To address this, we introduce a distinct
symbol $R^b$ and $R^h$ for each relation $R$ occurring in $\Sigma$, and we
represent each TGD $\tau$ as a \emph{feature vector} $F_\tau$ of these symbols
corresponding to the body and head of $\tau$. Moreover, we combine in the
obvious way the subset and superset retrieval algorithms. For example, when
searching for a TGD ${\tau' \in \workedOffSet \cup \unprocessedSet}$ that
subsumes a given TGD $\tau$, we use the subset retrieval for the symbols $R^b$
and the superset retrieval for symbols $R^h$. Finally, we order these symbols
by the decreasing frequency of the order of the symbols' occurrence in the set
$\Sigma$ of input TGDs, and moreover we order each $R^b$ before all $R^h$.

\myparagraph{Relation Clustering}
We observed that the subsumption indexes can easily get very large, so index
traversal can become a considerable source of overhead. To reduce the index
size, we group the symbols $R^b$ and $R^h$ into clusters $C^b$ and $C^h$,
respectively. Then, the feature vector $F_\tau$ associated with each TGD $\tau$
consists of all clusters $C^b$ and $C^h$ that contain a relation occurring in
the body and head, respectively, of $\tau$. We adapt the trie traversal
algorithms in the obvious way to take into account this change. The number of
clusters is computed using the average numbers of symbols and atoms in the
input TGDs, and clusters are computed with the aim of balancing the number of
TGDs stored in each leaf vertex.

\myparagraph{Unification Indexing}
We construct indexes over $\workedOffSet$ that allow us to quickly identify
TGDs/rules that can participate in an inference with some $\tau$. For TGDs, we
maintain a hash table that maps each relation $R$ to a set of TGDs containing
$R$ in the body, and another hash table that does the same but for TGD heads.
To index rules, we use a variant of a \emph{path indexing} \cite{stickel89}:
each atom in a rule is represented as a sequence of relations and function
symbols occurring in the atom, and such sequences are entered into two tries
(one for body and one for head atoms). Then, given rule $\tau$, we consider
each body and head atom $A$ of $\tau$, we convert $A$ into the corresponding
sequence, and we use the sequence to query the relevant trie for all candidates
participating in an inference with $\tau$ on $A$.

\myparagraph{Cheap Lookahead Optimization}
Consider an application of the $\ExbDR$ inference rule to GTGDs $\tau$ and
$\tau'$ as in Definition~\ref{def:ExbDR}, producing a GTGD $\tau''$ where
${\vars(\theta(H')) \cap \vec y \neq \emptyset}$ and the relation of $H'$ does
not occur in the body of a GTGD in $\Sigma$. In each one-pass chase sequence
for some base instance and $\Sigma$, no GTGD of $\Sigma$ can be applied to a
fact obtained by instantiating $\theta(H')$, so deriving this fact is
redundant. Consequently, we can drop such $\tau''$ as soon as we derive it in
line~\ref{alg:rewriting:evolve}. Analogously, when the $\SkDR$ inference rule
is applied to rules $\tau$ and $\tau'$ as in Definition~\ref{def:SkDR}, we can
drop the resulting rule if $\theta(H')$ is not full and it contains a relation
not occurring in the body of a GTGD in $\Sigma$.

\section{Experimental Evaluation}\label{sec:experiments}

We implemented a system that can produce a Datalog rewriting of a set of GTGDs
using our algorithms, and we conducted an empirical evaluation using a
comprehensive collection of 428 synthetic and realistic inputs. Our objectives
were to show that our algorithms can indeed rewrite complex GTGDs, and that the
rewriting can be successfully processed by modern Datalog systems. In
Subsection~\ref{subsec:experiments:setting} we describe the test setting. Then,
in Subsection~\ref{subsec:experiments:ontologies} we discuss the rewriting
experiments with GTGDs obtained from ontologies, and in
Subsection~\ref{subsec:experiments:end-to-end} we validate the usefulness of
the rewriting approach end-to-end. Finally, in
Subsection~\ref{subsec:experiments:higher-arity} we discuss rewriting GTGDs of
higher arity. Due to the very large number of inputs, we can only summarize our
results in this paper; however, our complete evaluation results are available
online \cite{saturation-github}.

\subsection{Input GTGDs, Competitors, \& Test Setting}\label{subsec:experiments:setting}

Before discussing our results, we next describe our test setting.

\myparagraph{Input GTGDs}
We are unaware of any publicly available sets of GTGDs that we could readily
use in our evaluation, so we derived the input GTGDs for our evaluation from
the ontologies in the Oxford Ontology Library \cite{library}. At the time of
writing, this library contained 787 ontologies, each assigned a unique
five-digit identifier. After removing closely-related ontology variants, we
were left with 428 core ontologies. We loaded each ontology using the parser
from the Graal system~\cite{graal}, discarded axioms that cannot be translated
into GTGDs, and converted the remaining axioms into GTGDs. We used the standard
translation of description logics into first-order logic \cite{dl-handbook-2},
where each class corresponds to a unary relation, and each property corresponds
to a binary relation. We thus obtained 428 sets of input GTGDs with properties
shown in Table~\ref{tab:inputs}.

\begin{table}[tb]
\caption{Input GTGDs at a Glance}\label{tab:inputs}
\newcommand{\mc}[2]{\multicolumn{#1}{c}{#2}}
\newcommand{\mcl}[2]{\multicolumn{#1}{c|}{#2}}
\begin{tabular}{r|r@{\;\;}r@{\;\;}r@{\;\;}r|r@{\;\;}r@{\;\;}r@{\;\;}r}
    \hline
    \mcl{1}{Inputs} & \mcl{4}{\# Full TGDs}                 & \mc{4}{\# Non-Full TGDs} \\
                    & Min   & Max       & Avg       & Med   & Min   & Max       & Avg       & Med \\
    \hline
    428             & 1     & 171,905   & 11,030    & 789   & 2     & 156,743   & 5,255     & 283 \\
    \hline
\end{tabular}
\end{table}

To evaluate our algorithms on TGDs containing relations of arity higher than
two, we devised a way to ``blow up'' relation arity. Given a set of GTGDs and a
blowup factor $b$, our method proceeds as follows. First, in each atom of each
GTGD, it replaces each variable argument with $b$ fresh variables uniquely
associated with the variable; for example, for ${b = 2}$, atom $A(x,y)$ is
transformed into atom ${A(x_1,x_2,y_1,y_2)}$. Next, the method randomly
introduces fresh head and body atoms over the newly introduced variables; in
doing so, it ensures that the new atoms do not introduce patterns that would
prevent application of the $\ExbDR$ inference rule.

\myparagraph{Competitors}
We compared the $\ExbDR$, $\SkDR$, and $\HypDR$ algorithms, implemented as
described in Section~\ref{sec:implementation}. As noted in
Section~\ref{sec:related}, no existing system we are aware of implements a
Datalog rewriting algorithm for GTGDs. However, the KAON2
system~\cite{kaon2,motikdatalog1,motikthesis} can rewrite GTGDs obtained from
OWL ontologies, so we used KAON2 as a baseline in our experiments with
OWL-based GTGDs. We made sure that all inputs to KAON2 and our algorithms
include only GTGDs that all methods can process.

\myparagraph{Test Setting}
We conducted all experiments on a laptop with an Intel Core i5-6500 CPU {@}
3.20~GHz and 16~GB of RAM, running Ubuntu 20.04.4 LTS and Java 11.0.15. In each
test run, we loaded a set of TGDs, measured the wall-clock time required to
compute the rewriting of a set of GTGDs, and saved the produced Datalog
rewriting. We used a timeout of ten minutes for each test run.

\subsection{Experiments with GTGDs from Ontologies}\label{subsec:experiments:ontologies}

We computed the Datalog rewriting of GTGDs obtained from OWL ontologies using
our three algorithms and KAON2. Figure~\ref{tab:results:ontology} shows the
number of inputs that each algorithm processed in a given time, provides
information about the inputs and outputs of each system, and compares the
performance among systems. The input size for $\ExbDR$ is the number of GTGDs
after transforming the input into head-normal form, and for $\SkDR$ and
$\HypDR$ it is the number of rules after Skolemization. Input size is not
available for KAON2 since this system reads an OWL ontology and transforms it
into GTGDs internally. The output size is the number of Datalog rules in the
rewriting. Finally, the blowup is the ratio of the output and the input sizes.
Each input GTGD contained at most seven body atoms. Out of 428 inputs, 349 were
processed within the ten minute limit by our three systems, and 334 inputs were
processed by all four systems. Moreover, 32 inputs, each containing between
20,270 and 221,648 GTGD, were not processed by any system.

\begin{figure}[tb]
    \includestandalone[width=.35\textwidth]{figures/ontology-times} \\[2ex]
    {
        \small
        \newcommand{\tc}[1]{\multicolumn{2}{|l|}{#1}}
        \newcommand{\Times}{\multirow{4}{*}{\rotatebox[origin=c]{90}{Time (s)}}}
        \begin{tabular}{|l|l|rrrr|}
            \hline
            \tc{}                           & $\ExbDR$  & $\SkDR$   & $\HypDR$  & KAON2   \\
            \hline
            \tc{\# of Processed Inputs}     & 367       & 377       & 382       & 362     \\
            \tc{Max.\ Processed Input Size} & 185,515   & 324,092   & 324,092   & N/A     \\
            \tc{Max.\ Output Size}          & 196,594   & 124,846   & 124,846   & 61,964  \\
            \tc{Max.\ Size Blowup}          & 8.95      & 8.85      & 8.85      & N/A     \\
            \tc{Max.\ Body Atoms in Output} & 7         & 6         & 6         & 4       \\
            \tc{\# Blowup $\geq$ 1.5}       & 26        & 14        & 16        & N/A     \\
            \hline
            \Times  & Min.\                 & 0.05      & 0.05      & 0.04      & 0.21    \\
                    & Max.\                 & 582.18    & 584.79    & 404.34    & 547.53  \\
                    & Avg.\                 & 23.23     & 14.34     & 6.38      & 18.66   \\
                    & Med.\                 & 0.82      & 0.52      & 0.55      & 0.49    \\
            \hline
        \end{tabular}
    }\\[2ex]
    {
        \footnotesize
        \begin{tabular}{|c|c@{\;}c@{\;}c@{\;}c||c@{\;}c@{\;}c@{\;}c|}
            \multicolumn{1}{@{}c@{}}{}          & \multicolumn{4}{c}{$\mathit{time}(Y) / \mathit{time}(X) \geq 10$}           & \multicolumn{3}{c}{$X$ and $Y$ both fail} \\
            \hline
            \diagbox[width=1.25cm]{$X$}{$Y$}    & $\ExbDR$  & $\SkDR$   & $\HypDR$      & KAON2 & $\ExbDR$  & $\SkDR$   & $\HypDR$  & KAON2 \\
            \hline
            $\ExbDR$                            &           & 19        & 0             & 19    & 61        &           &           &       \\
            $\SkDR$                             & 37        &           & 0             & 26    & 33        & 51        &           &       \\
            $\HypDR$                            & 37        & 12        &               & 31    & 35        & 43        & 46        &       \\
            KAON2                               & 35        & 15        & 0             &       & 37        & 47        & 46        & 66    \\
            \hline
        \end{tabular}
    }
    \caption{Results for TGDs Derived from Ontologies}\label{tab:results:ontology}
\end{figure}

\myparagraph{Discussion}
As one can see in Figure~\ref{tab:results:ontology}, all algorithms were able
to compute the rewriting of large inputs containing 100k+ GTGDs. Moreover, for
the vast majority of inputs that were successfully processed, the size of the
rewriting and the number of body atoms in the rewriting are typically of the
same order of magnitude as the input. Hence, the worst-case exponential blowup
from Theorems~\ref{thm:ExbDR-correctness-complexity},
\ref{thm:SkDR-correctness-complexity},
and~\ref{thm:HypDR-correctness-complexity} does not appear in practice: the
size of the rewriting seems to be determined primarily by the input size.

\myparagraph{Relative Performance}
No system can be identified as the best in general, but $\HypDR$ seems to offer
the best performance on average. The algorithm was able to process most inputs;
it was at least 35\% faster than the other systems on the slowest input; it was
never slower by an order of magnitude; there were only 14 inputs that could be
processed by some other algorithm but not $\HypDR$; and the output of $\HypDR$
does not differ significantly from the output of $\SkDR$. This is in line with
our motivation for $\HypDR$ outlined in Example~\ref{ex:HypDR}. Specifically,
$\HypDR$ derives rules with just one head atom, but it does not derive
intermediate rules with functional body atoms. The main source of overhead in
$\HypDR$ seems to be more complex selection of rules participating in an
inference.

\myparagraph{Impact of Subsumption}
All algorithms spend a considerable portion of their running time checking
TGD/rule subsumption, so it is natural to wonder whether this overhead is
justified. To answer this question, we ran our three approaches using a
modification of Algorithm~\ref{alg:rewriting}: we replaced the check for
containment up to redundancy in line~\ref{alg:rewriting:forward} with just
checking ${\tau' \not\in \workedOffSet \cup \unprocessedSet}$, and we removed
line~\ref{alg:rewriting:backward}. Note that our normalization of variables
described in Section~\ref{sec:implementation} still guarantees termination.
This change significantly increased the number of derivations: the numbers of
derived TGDs/rules increased on average by a factor of 104, 185, and 103 on
$\ExbDR$, $\SkDR$, and $\HypDR$, respectively. Interestingly, this increase did
not affect the performance uniformly. While $\SkDR$ was able to process 12
inputs an order of magnitude faster, $\ExbDR$ and $\HypDR$ timed out on 72 and
17 additional inputs, respectively. This, we believe, is due to how different
inference rules select inference candidates. The $\SkDR$ rule is applied to
just pairs of rules, and candidate pairs can be efficiently retrieved using
unification indexes. In contrast, $\ExbDR$ requires matching several head atoms
with as many body atoms, which makes developing a precise index for candidate
pair retrieval difficult; thus, as the number of derived TGDs increases, the
number of false candidates retrieved from the index increases as well. Finally,
$\HypDR$ can be applied to an arbitrary number of rules, so selecting inference
candidates clearly becomes more difficult as the number of derived rules
increases.

\myparagraph{Impact of Structural Transformation}
KAON2 uses \emph{structural transformation} \cite{Baaz+:HandbookAR:normal:2001}
to simplify ontology axioms before translating them into GTGDs. For example,
axiom ${A \sqsubseteq \exists B.\exists C.D}$ is transformed into ${A
\sqsubseteq \exists B.X}$ and ${X \sqsubseteq \exists C.D}$ for $X$ a fresh
class. The resulting axioms have simpler structure, which is often beneficial
to performance. To see how this transformation affects our algorithms, we reran
our experiments while transforming the input axioms in the same way as in
KAON2. This indeed improved the performance of $\SkDR$ by one order of
magnitude on 22 ontologies, and it did not hurt the performance of $\HypDR$.
The main challenge is to generalize this transformation to arbitrary GTGDs:
whereas description logic axioms exhibit syntactic nesting that lends itself
naturally to this transformation, it is less clear how to systematically apply
this transformation to TGDs, where heads and bodies consist of ``flat''
conjunctions. We leave this question for future work.

\subsection{End-to-End Experiments}\label{subsec:experiments:end-to-end}

To validate our approach end-to-end, we selected ten inputs where $\ExbDR$
produced the largest rewritings. For each of these, we generated a large base
instance using WatDiv~\cite{DBLP:conf/semweb/AlucHOD14}, and we computed the
fixpoint of the rewriting and the instance using the RDFox \cite{rdfox} Datalog
system v5.4. Table~\ref{tab:fact-entailment} summarizes our results.

\begin{table}[tb]
\caption{Computing the Fixpoint of the Rewriting}\label{tab:fact-entailment}
\small
\begin{tabular}{l|rrrr}
    \hline
    Ont.\ ID    & \# Rules  & \# Input Facts    & \# Output Facts   & Time (s) \\
    \hline
    00387       & 63,422    & 4,403,105         & 51,439,424        & 53       \\
    00448       & 67,986    & 5,510,444         & 107,235,697       & 110      \\
    00470       & 75,146    & 10,532,943        & 141,396,446       & 242      \\
    00471       & 78,977    & 11,077,423        & 128,954,126       & 253      \\
    00472       & 75,146    & 10,533,008        & 141,396,576       & 279      \\
    00473       & 78,977    & 11,077,459        & 128,954,198       & 291      \\
    00573       & 113,959   & 9,197,254         & 155,118,592       & 206      \\
    00682       & 68,461    & 5,183,460         & 105,431,952       & 101      \\
    00684       & 81,553    & 6,057,017         & 66,981,628        & 109      \\
    00686       & 124,846   & 10,402,324        & 166,366,039       & 238      \\
    \hline
\end{tabular}
\end{table}

All programs used in this experiment are at least several orders of magnitude
larger than what is usually encountered in practical applications of Datalog,
but RDFox nevertheless computed the fixpoint of all rewritings in a few
minutes. Moreover, although the fixpoints seem to be an order of magnitude
larger than the base instance, this is not a problem for highly optimized
systems such as RDFox. Hence, checking fact entailment via rewritings produced
by our algorithms is feasible in practice.

\subsection{GTGDs With Relations of Higher Arity}\label{subsec:experiments:higher-arity}

Finally, we computed the rewriting of GTGDs obtained by blowing up relation
arity as described in Subsection~\ref{subsec:experiments:setting} using a
blowup factor of five. We did not use KAON2 since this system supports
relations of arity at most two. Figure~\ref{tab:results:higher-arity}
summarizes our results. Out of 428 inputs, 187 were processed within the ten
minute limit by our three systems, and 128 inputs were not processed by any
system.

\begin{figure}[tb]
    \includestandalone[width=.35\textwidth]{figures/blowup-times} \\[2ex]
    {
        \small
        \newcommand{\tc}[1]{\multicolumn{2}{|l|}{#1}}
        \newcommand{\Times}{\multirow{4}{*}{\rotatebox[origin=c]{90}{Time (s)}}}
        \begin{tabular}{|l|l|rrr|}
            \hline
            \tc{}                           & $\ExbDR$  & $\SkDR$   & $\HypDR$  \\
            \hline
            \tc{\# of Processed Inputs}     & 274       & 238       & 199       \\
            \tc{Max.\ Processed Input Size} & 69,046    & 182,569   & 38,362    \\
            \tc{Max.\ Output Size}          & 58,749    & 171,832   & 38,335    \\
            \tc{Max.\ Size Blowup}          & 9.00      & 5.84      & 5.84      \\
            \tc{\# Blowup $\geq$ 1.5}       & 26        & 5         & 3         \\
            \hline
            \Times  & Min.\                 & 0.06      & 0.05      & 0.04      \\
                    & Max.\                 & 591.82    & 504.49    & 557.75    \\
                    & Avg.\                 & 26.70     & 38.39     & 17.05     \\
                    & Med.\                 & 0.61      & 1.65      & 1.72      \\
            \hline
        \end{tabular}
    }\\[2ex]
    {
        \footnotesize
        \begin{tabular}{|c|c@{\;\;}c@{\;\;}c||c@{\;\;}c@{\;\;}c|}
            \multicolumn{1}{@{}c@{}}{}          & \multicolumn{3}{c}{$\mathit{time}(Y) / \mathit{time}(X) \geq 10$}           & \multicolumn{3}{c}{$X$ and $Y$ both fail} \\
            \hline
            \diagbox[width=1.25cm]{$X$}{$Y$}    & $\ExbDR$  & $\SkDR$   & $\HypDR$      &  $\ExbDR$  & $\SkDR$   & $\HypDR$ \\
            \hline
            $\ExbDR$                            &           & 61        & 87            & 154        &           &          \\
            $\SkDR$                             & 11        &           & 21            & 128        & 190       &          \\
            $\HypDR$                            & 6         & 4         &               & 148        & 184       & 229      \\
            \hline
        \end{tabular}
    }
    \caption{Results for TGDs with Higher-Arity Relations}\label{tab:results:higher-arity}
\end{figure}

While $\HypDR$ performed best on GTGDs derived from ontologies,
Figure~\ref{tab:results:higher-arity} shows it to be worst-performing on
higher-arity GTGDs: it successfully processed only 199 inputs within the ten
minute timeout, whereas $\SkDR$ and $\ExbDR$ processed 238 and 274 inputs,
respectively. This is mainly due to additional body atoms introduced by our
``blowup'' method: these increase the number of rules participating in an
application of the $\HypDR$ inference rule, which makes selecting the
participating rules harder.

This experiment proved to be more challenging, as most problems discussed in
Section~\ref{sec:implementation} became harder. For example, in $\ExbDR$,
higher arity of atoms increases the likelihood that an atom retrieved through a
unification index does not unify with a given atom, and that the atoms of the
selected GTGDs cannot be successfully matched. Subsumption indexing is also
more difficult for similar reasons. However, the inputs used in this experiment
consist of a large numbers of GTGDs with relations of arity ten, so they can be
seen as a kind of a ``stress test''. Our algorithms were able to process more
than half of such inputs, which leads us to believe that they can also handle
more well-behaved GTGDs used in practice.

\section{Conclusion}\label{sec:conclusion}

We presented several algorithms for rewriting a finite set of guarded TGDs into
a Datalog program that entails the same base facts on each base instance. Our
algorithms are based on a new framework that establishes a close connection
between Datalog rewritings and a particular style of the chase. In future, we
plan to generalize our framework to wider classes of TGDs, such as
frontier-guarded TGDs, as well as provide rewritings for conjunctive queries
under certain answer semantics. Moreover, we shall investigate whether the
extension of our framework to disjunctive guarded TGDs
\cite{DBLP:journals/corr/abs-1911-03679} can be used to obtain practical
algorithms for rewriting disjunctive guarded TGDs into disjunctive Datalog
programs.

\begin{acks}
This work was funded by the EPSRC grants OASIS (EP/S032347/1), QUINTON
(EP/T022124/1), UK FIRES (EP/S019111/1), AnaLOG (EP/P025943/1), and Concur
(EP/V050869/1). For the purpose of Open Access, the author has applied a CC BY
public copyright licence to any Author Accepted Manuscript (AAM) version
arising from this submission.
\end{acks}

\balance

\bibliographystyle{ACM-Reference-Format}
\bibliography{paper}

\iftoggle{withappendix}{
    \newpage
    \appendix
    \onecolumn

    \section{Proofs for Section~\ref{sec:chaserewrite}: One-pass Chase Proofs}

In Section~\ref{sec:chaserewrite} we introduced the notion of a one-pass chase
proof, which allows us to establish a completeness criterion for saturations
that is tied to the chase. We provide details of the proofs in this appendix.

\subsection{Proof of Theorem \ref{thm:one-pass-proof-exists}: Existence of One-Pass Chase Proofs}

\OnePassProofExists*

Throughout this section, we fix an arbitrary base instance $I$ and a finite set
of GTGDs $\Sigma$. It is known that ${I, \Sigma \models F}$ if and only if
there exists a tree-like chase proof of $F$ from $I$ and $\Sigma$. We next
prove Theorem~\ref{thm:one-pass-proof-exists} by showing that each such proof
can be transformed to a one-pass chase proof of $F$ from $I$ and $\Sigma$. This
argument was developed jointly with Antoine Amarilli, and it is related to
proofs by \citet{DBLP:journals/lmcs/AmarilliB22} and
\citet{DBLP:journals/corr/abs-1911-03679}; however, note that
Definition~\ref{def:one-pass} imposes slightly stronger conditions on one-pass
chase sequences than related definitions in those works.

Towards our goal, we first state two basic properties of tree-like chase
sequences. The first claim is a variation of the well-known fact that any chase
tree produced for GTGDs represents a tree decomposition \cite{cali2013taming}.
The second claim captures the idea that, as the chase progresses, facts may be
added within a vertex, but this will not produced new guarded sets of terms.

\begin{lemma}\label{lem:guarded-preservation}
    Let ${T_0, \dots, T_n}$ be an arbitrary tree-like chase sequence for
    $I$ and $\Sigma$.
    \begin{enumerate}
        \item For each ${0 \leq i \leq n}$, all vertices $v_1$ and $v_2$ in
        $T_i$, each set $G$ of ground terms that is $\Sigma$-guarded by
        $T_i(v_1)$ and by $T_i(v_2)$, and each vertex $v_3$ on the unique path
        in $T_i$ between $v_1$ and $v_2$, set $G$ is $\Sigma$-guarded by
        $T_i(v_3)$.

        \item For each ${0 \leq i \leq n}$, each vertex $v$ in $T_i$, each set
        $G$ of ground terms that is $\Sigma$-guarded by $T_i(v)$, and each ${0
        \leq j \leq i}$ such that $T_j$ contains $v$, set $G$ is
        $\Sigma$-guarded by $T_j(v)$.
    \end{enumerate}
\end{lemma}

\begin{proof}[Proof of Claim 1.]
The proof is by induction on $i$ with ${0 \leq i \leq n}$. For ${i = 0}$, chase
tree $T_0$ contains just one vertex so the claim holds trivially. Now assume
that the property holds for some ${0 \leq i < n}$ and consider ways in which
$T_{i+1}$ can be derived from $T_i$. First, $T_{i+1}$ can be obtained by
applying a chase step to $T_i$ at vertex $v$ with some GTGD ${\tau \in
\Sigma}$. Let $v_1$ be the recently updated vertex of $T_{i+1}$. Thus, $v_1$ is
either $v$ or a fresh child of $v$. Moreover, consider each fact $R(\vec t)$
derived by the step, each set of ground terms ${G \subseteq \vec t}$, each
vertex $v_2$ such that $G$ is $\Sigma$-guarded by $T_{i+1}(v_2)$, and each
vertex $v_3$ on the unique path in $T_{i+1}$ between $v_1$ and $v_2$. If $G$
contains a labeled null that is freshly introduced in $T_{i+1}$, the claim
holds trivially because $v_2$ and $v_3$ are necessarily the same as $v_1$.
Otherwise, $\tau$ is guarded, so $T_i(v)$ contains a fact $S(\vec u)$ such that
${G \subseteq \vec u}$. But then, $G$ is $\Sigma$-guarded by $T_i(v_3)$ by the
induction assumption. Moreover, ${T_i(v_3) \subseteq T_{i+1}(v_3)}$ ensures
that $G$ is $\Sigma$-guarded by $T_{i+1}(v_3)$, as required. Second, $T_{i+1}$
can be obtained by applying a propagation step to $T_i$, but then the property
clearly holds.
\end{proof}

\begin{proof}[Proof of Claim 2.]
The proof is by induction on $i$ with ${0 \leq i \leq n}$. The base case for
${i = 0}$ is trivial. For the induction step, assume that the property holds
for some $i$. If $T_{i+1}$ is obtained from $T_i$ by a chase step with a
non-full GTGD, then the claim clearly holds for $T_{i+1}$ because the step
introduces a fresh vertex that does not occur in any $T_j$ with ${0 \leq j \leq
i}$. Otherwise, $T_{i+1}$ is obtained by extending some $T_i(v)$, so consider
an arbitrary fact ${F \in T_{i+1}(v) \setminus T_i(v)}$. Clearly, $F$ is
$\Sigma$-guarded by $T_i(v)$: if the step involves a full GTGD, then a body
atom of the GTGD is matched to a fact ${F' \in T_i(v)}$ such that $F$ is
$\Sigma$-guarded by $F'$; moreover, if the step involves propagation, then by
definition there exists a fact ${F' \in T_i(v)}$ such that $F$ is
$\Sigma$-guarded by $F'$. Thus, each set of ground terms $G$ that is
$\Sigma$-guarded by $T_{i+1}(v)$ is also $\Sigma$-guarded by ${F' \in T_i(v)}$,
so the claim holds.
\end{proof}

In the rest of the proof, we show how to convert an arbitrary tree-like chase
proof into a one-pass one through a series of transformations. Before
proceeding, we next describe formally the types of chase sequence that we
consider in our transformations.

\begin{definition}~

    \begin{itemize}
        \item A chase sequence is \emph{local} if each propagation step in the
        sequence copies just one fact to either the parent or a child vertex.

        \item A chase sequence is \emph{rootward} if each propagation step in
        the sequence copies just one fact from a child to its parent.

        \item A chase sequence is \emph{almost one-pass} if it is rootward and
        each chase or propagation step is applied to the recently updated
        vertex or an ancestor thereof, and a chase step is applied only if a
        propagation step is not applicable to the recently updated vertex or an
        ancestor thereof.
    \end{itemize}
\end{definition}

Note that facts can still be copied from a parent to a child in a rootward
chase sequence, but this can be done only in chase steps with non-full GTGDs
that introduce a child. Furthermore, the use of ``almost'' in the ``almost
one-pass'' reflects the caveat that, in an almost one-pass chase sequence, a
step can be applied to an ancestor of the recently updated vertex, thus
``jumping rootward'' in the tree, whereas such steps are forbidden in a
one-pass chase sequence.

We capture formally the relationship between the chase sequences produced by
our transformations using the notion introduced in Definition~\ref{def:subset}.

\begin{definition}\label{def:subset}
    A chase tree $T$ is a \emph{subset} of a chase tree $T'$, written ${T
    \subseteq T'}$, if the tree of $T$ is a subtree of $T'$ (i.e., the root of
    $T$ is the root of $T'$, and whenever vertex $v$ is a parent of vertex $v'$
    in $T$, then $v$ is a parent of $v'$ in $T'$), and ${T(v) \subseteq T'(v)}$
    holds for each vertex $v$ of $T$.
\end{definition}

We are now ready to present our transformations, which we capture in a series
of lemmas. We next summarize the main intuitions.
\begin{itemize}
    \item In Lemma~\ref{lem:to-local}, we show that an arbitrary chase sequence
    can be transformed into a local chase sequence by ``slowing down''
    propagation steps so that facts are copied only between vertices that are
    adjacent in a chase tree.

    \item In Lemma~\ref{lem:to-rootward}, we show that each local chase
    sequence can be transformed into a rootward chase sequence. Intuitively,
    instead of propagating a fact from a parent to a child, we ``regrow'' a
    clone of the relevant child and the entire subtree underneath. The relevant
    fact is then copied as part of the chase step with the non-full GTGD that
    ``regrows'' the child's clone.

    \item In Lemma~\ref{lem:to-almost-one-pass}, we show that each rootward
    chase sequence can be transformed to an almost one-pass chase sequence. The
    main difficulty arises due to the fact that steps in a rootward chase
    sequence can be applied to arbitrary vertices. We address this problem by
    shuffling and regrowing parts of the chase trees.

    \item Finally, in Lemma~\ref{def:to-one-pass}, we show that each almost
    one-pass chase proof can be transformed to a one-pass chase proof by
    pruning irrelevant parts of the chase sequence.
\end{itemize}

\begin{lemma}\label{lem:to-local}
    For each tree-like chase sequence ${T_0, \dots, T_n}$ for $I$ and $\Sigma$,
    there exists a local tree-like chase sequence ${\Tbar_0, \dots, \Tbar_m}$
    for $I$ and $\Sigma$ such that ${T_n \subseteq \Tbar_m}$.
\end{lemma}

\begin{proof}
Each propagation step in ${T_0, \dots, T_n}$ that copies more than one fact can
clearly be ``expanded'' into several steps, each copying just one fact.
Moreover, due to Claim 1 of Lemma~\ref{lem:guarded-preservation}, each
propagation step that copies a fact $F$ between vertices $v$ and $v'$ that are
further apart can be ``expanded'' into several steps that propagate $F$ to all
vertices on the unique path between $v$ and $v'$.
\end{proof}

\begin{lemma}\label{lem:to-rootward}
    For each local tree-like chase sequence ${T_0, \dots, T_n}$ for $I$ and
    $\Sigma$, there exists a rootward tree-like chase sequence ${\Tbar_0,
    \dots, \Tbar_m}$ for $I$ and $\Sigma$ such that
    \begin{enumerate}[wide=0.2cm, label=(S\arabic*)]
        \item\label{lem:to-rootward:last} ${T_n \subseteq \Tbar_m}$, and

        \item\label{lem:to-rootward:tau} for each vertex $v$ in $T_n$ that is
        introduced by a chase step with a non-full GTGD ${\tau \in \Sigma}$ and
        substitutions $\sigma$ and $\sigma'$, vertex $v$ is introduced into
        some $\Tbar_k$ with ${0 \leq k \leq m}$ by a chase step with the same
        $\tau$, $\sigma$, and $\sigma'$.
    \end{enumerate}
\end{lemma}

\begin{proof}
Let ${T_0, \dots, T_n}$ be an arbitrary local tree-like chase sequence for $I$
and $\Sigma$. We prove the claim by induction on ${0 \leq i \leq n}$. The
induction base ${i=0}$ holds trivially. For the induction step, we assume that
the claim holds for some $i$ with ${0 \leq i < n}$. By the inductive
assumption, there exists a rootward chase sequence ${\Tbar_0, \dots, \Tbar_j}$
for $I$ such that ${T_i \subseteq \Tbar_j}$ and property
\ref{lem:to-rootward:tau} holds. Let $v$ be the vertex of $T_i$ to which a
chase or propagation step is applied to derive $T_{i+1}$. By
Definition~\ref{def:subset}, chase tree $\Tbar_j$ contains vertex $v$ and
${T_i(v) \subseteq \Tbar_j(v)}$ holds. We now consider ways in which $T_{i+1}$
can be derived from $T_i$.

Assume that $T_{i+1}$ is obtained from $T_i$ by a chase step with non-full TGD
${\tau \in \Sigma}$, and let $v'$ be the child of $v$ introduced by the step.
Without loss of generality, we can choose $v'$ and the fresh labeled nulls such
that they do not occur in $\Tbar_j$. Now let $\Tbar_{j+1}$ be obtained from
$\Tbar_j$ by adding $v'$ as a child of $v$ and setting ${\Tbar_{j+1}(v') =
T_{i+1}(v')}$. Clearly, ${\Tbar_0, \dots, \Tbar_j, \Tbar_{j+1}}$ is a rootward
chase sequence such that ${T_{i+1} \subseteq \Tbar_{j+1}}$ and property
\ref{lem:to-rootward:tau} hold, as required.

Assume that $T_{i+1}$ is obtained from $T_i$ by a chase step with a full TGD
${\tau \in \Sigma}$ deriving a fact $F$, or by a rootward propagation step that
copies a fact $F$ from $T_i(v)$ to the parent of $v$. Let $v'$ be the recently
updated vertex of $T_{i+1}$. Chase tree $\Tbar_j$ clearly contains $v'$. If ${F
\in \Tbar_j(v')}$, then sequence ${\Tbar_0, \dots, \Tbar_j}$ satisfies the
inductive property. If $T_{i+1}$ is obtained from $T_i$ by a propagation step,
then $F$ is $\Sigma$-guarded by $T_i(v')$. But then, ${T_i(v') \subseteq
\Tbar_j(v')}$ ensures that $F$ is also $\Sigma$-guarded by $\Tbar_j(v')$ and
thus the propagation step is applicable to vertices $v$ and $v'$ in $\Tbar_j$.
Now let $\Tbar_{j+1}$ to be the same as $\Tbar_j$ but with ${\Tbar_{j+1}(v') =
\Tbar_j(v') \cup \{ F \}}$ and with $v'$ being the recently updated vertex.
Clearly, ${\Tbar_0, \dots, \Tbar_j, \Tbar_{j+1}}$ is a rootward chase sequence
satisfying ${T_{i+1} \subseteq \Tbar_{j+1}}$, as required. Moreover, property
\ref{lem:to-rootward:tau} holds by the induction hypothesis.

The only remaining case is when $T_{i+1}$ is obtained from $T_i$ by applying a
propagation step that copies one fact $F$ to a child $v'$ of $v$. By
Definition~\ref{def:subset}, chase tree $\Tbar_j$ contains vertex $v'$ and
${T_i(v') \subseteq \Tbar_j(v')}$ holds. Sequence ${\Tbar_0, \dots, \Tbar_j}$
satisfies the inductive property if ${F \in \Tbar_j(v')}$ holds, so we next
assume ${F \not\in \Tbar_j(v')}$. We next show that we can simulate propagation
by ``replaying'' the chase steps that generate $v'$ and all of its descendants.
Towards this goal, let $T_k$ be the chase tree in the original sequence where
$v'$ is first introduced by applying a chase step with the non-full GTGD ${\tau
= \body \rightarrow \exists \vec y ~ \head \in \Sigma}$, and let $\sigma$ and
$\sigma'$ be substitutions used in the step. By the inductive property
\ref{lem:to-rootward:tau}, there exists $\ell_0$ with ${0 < \ell_0 \leq j}$
such that $v'$ is introduced in $\Tbar_{\ell_0}$ as the result of applying a
chase step with the same non-full TGD $\tau$ and substitutions $\sigma$ and
$\sigma'$. Finally, let ${\Tbar_{\ell_0}, \dots, \Tbar_{\ell_m}}$ be the
subsequence of ${\Tbar_0, \dots, \Tbar_j}$ consisting of precisely those chase
trees that were obtained by applying a chase or a propagation step to $v'$ or a
descendant of $v'$. In other words, the chase steps producing ${\Tbar_{\ell_0},
\dots, \Tbar_{\ell_m}}$ are exactly the steps that we need to ``replay'' to
simulate the propagation of $F$ from $v$ to $v'$.

Our objective is to ``replay'' the steps producing ${\Tbar_{\ell_0}, \dots,
\Tbar_{\ell_m}}$ so that they introduce exactly the same vertices and labeled
nulls, which is needed because property \ref{lem:to-rootward:last} talks about
exact containment of the final chase trees of the two sequences (rather than
containment up to isomorphism). A technical issue is that these vertices and
labeled nulls already occur in the sequence ${\Tbar_0, \dots, \Tbar_j}$; thus,
if we extended this sequence directly, we could not ``reapply'' the chase steps
with non-full GTGDs, which by definition introduce fresh vertices and labeled
nulls. To get around this, we first perform the following renaming step. Let
$N$ be the set of labeled nulls introduced by the chase steps with non-full
TGDs in subsequence ${\Tbar_{\ell_0}, \dots, \Tbar_{\ell_m}}$, and let $W$ be
the set of introduced vertices (thus, $W$ contains $v'$ and all of its
descendants). Moreover, let ${\Ubar_0, \dots, \Ubar_j}$ be the chase sequence
obtained by uniformly replacing in ${\Tbar_0, \dots, \Tbar_j}$ each labeled
null in $N$ with a distinct, fresh labeled null, and by uniformly replacing
each vertex ${w \in W}$ by a fresh vertex.

We next describe the chase trees that will be produced by ``replaying'' the
steps producing the subsequence ${\Tbar_{\ell_0}, \dots, \Tbar_{\ell_m}}$.
Intuitively, we must ``graft'' the results of these steps onto $\Ubar_j$: for
$v'$ or a descendant of $v'$ we take the results of the chase steps in the
subsequence, and for each other vertex we copy the content from $\Ubar_j$.
Formally, let ${\Vbar_0, \dots, \Vbar_m}$ be the sequence obtained from the
subsequence ${\Tbar_{\ell_0}, \dots, \Tbar_{\ell_m}}$ using the following steps.
\begin{enumerate}[wide=0.2cm, label=(R\arabic*)]
    \item\label{step:to-rootward:other-vertices}
    For each ${0 \leq p \leq m}$ and each vertex $w$ in $\Ubar_j$ such
    that $w$ is neither $v'$ nor a descendant of $v'$ in $\Ubar_j$, we set
    ${\Vbar_p(w) = \Ubar_j(w)}$.

    \item\label{step:to-rootward:vprime-descendants}
    For each ${0 \leq p \leq m}$ and each vertex $w$ that occurs in
    $\Tbar_{\ell_p}$ such that $w$ is $v'$ or a descendant of $v'$ in
    $\Tbar_{\ell_p}$, we set ${\Vbar_p(w) = \Tbar_{\ell_p}(w)}$.

    \item\label{step:to-rootward:propagate-0}
    We add to $\Vbar_0(v')$ each fact ${G \in \Ubar_j(v)}$ that is
    $\Sigma$-guarded by $\sigma'(\head)$.

    \item\label{step:to-rootward:propagate-p}
    We analogously extend each $\Vbar_p$ with ${1 \leq p \leq m}$ to ensure
    that each chase step with a non-full GTGD correctly propagates all relevant
    facts to a child.
\end{enumerate}

We now argue that ${\Ubar_0, \dots, \Ubar_j, \Vbar_0, \dots, \Vbar_m}$ is a
rootward chase sequence that satisfies properties \ref{lem:to-rootward:last}
and \ref{lem:to-rootward:tau}. Towards this goal, we make the following
observations.
\begin{itemize}
    \item Sequence ${\Ubar_0, \dots, \Ubar_j}$ is a rootward chase sequence
    produced by the same steps as ${\Tbar_0, \dots, \Tbar_j}$, but with the
    vertices in $W$ and labeled nulls in $N$ uniformly renamed. Also, due to
    step \ref{step:to-rootward:propagate-p}, ${\Vbar_0, \dots, \Vbar_m}$ is a
    rootward chase sequence produced by the same steps as ${\Tbar_{\ell_0},
    \dots, \Tbar_{\ell_m}}$.

    \item Chase tree $\Vbar_0$ coincides with $\Ubar_j$ on each vertex that is
    not $v'$ or a descendant of $v'$. Moreover, $\Ubar_j$ does not contain a
    labeled null in $N$, and it does not contain $v'$ or a descendant of $v'$;
    thus, $\Vbar_0$ can be seen as the result of applying to $\Ubar_j$ a chase
    step with the non-full GTGD $\tau$ and substitutions $\sigma$ and $\sigma'$
    that introduces vertex $v'$ as a child of $v$.

    \item We now show that property \ref{lem:to-rootward:tau} is
    satisfied---that is, that ${T_{i+1} \subseteq \Vbar_m}$ holds. Towards this
    goal, consider an arbitrary vertex $w$ occurring in $T_{i+1}$; by the
    induction assumption, we have ${T_i(w) \subseteq \Tbar_j(w)}$. If $w$ is
    neither $v'$ nor a descendant thereof, then neither $w$ nor a labeled null
    occurring in $T_i(w)$ was renamed in $\Ubar_j$, so we have ${\Tbar_j(w) =
    \Ubar_j(w) = \Vbar_m(w)}$, where the last equality is ensured by step
    \ref{step:to-rootward:other-vertices}; thus, ${T_{i+1}(w) = T_i(w)
    \subseteq \Vbar_m(w)}$ holds, as required. Now assume that $w$ is $v'$ or a
    descendant thereof. Then, ${\Tbar_j(w) = \Tbar_{\ell_m}(w)}$ holds by the
    fact that $\Tbar_{\ell_m}$ is the last place in ${\Tbar_0, \dots, \Tbar_j}$
    where $v'$ or a descendant of $v'$ was modified, and ${\Tbar_{\ell_m}(w) =
    \Vbar_m(w)}$ holds by step \ref{step:to-rootward:vprime-descendants};
    putting it all together, we have ${T_i(w) \subseteq \Vbar_m(w)}$. Now if
    $w$ is not $v'$ (i.e., $w$ is a descendant of $v'$), then ${T_{i+i}(w) =
    T_i(w) \subseteq \Vbar_m(w)}$ holds, as required. We finally consider the
    case when $w$ is $v'$, so ${T_{i+1}(v') = T_i(v') \cup \{ F \}}$. Since the
    propagation step is applicable to $T_i$, fact $F$ is $\Sigma$-guarded by
    $T_i(v')$. By Claim~2 of Lemma~\ref{lem:guarded-preservation}, fact $F$ is
    also $\Sigma$-guarded by $T_k(v')$. Finally, by the definition of a chase
    step with a non-full TGD, fact $F$ is $\Sigma$-guarded by $\sigma'(\head)$.
    But then, step \ref{step:to-rootward:propagate-0} ensures ${F \in
    \Vbar_0(v') \subseteq \Vbar_m(v')}$. Consequently, ${T_{i+i}(v') \subseteq
    \Vbar_m(v')}$ holds, as required.

    \item We now show that property \ref{lem:to-rootward:tau} is satisfied. To
    this end, consider an arbitrary vertex $w$ in $T_{i+1}$ introduced by a
    chase step with a non-full GTGD $\tau$ and substitutions $\sigma$ and
    $\sigma'$. If $w$ is not $v'$ or a descendant thereof, then the labeled
    nulls introduced by the chase step are not renamed in ${\Ubar_0, \dots,
    \Ubar_j}$, so the claim holds by the induction assumption. Otherwise, the
    chase steps producing ${\Vbar_0, \dots, \Vbar_m}$ are exactly the same as
    the chase steps producing ${\Tbar_{\ell_0}, \dots, \Tbar_{\ell_m}}$, so the
    claim holds by the induction assumption too. \qedhere
\end{itemize}
\end{proof}

\begin{lemma}\label{lem:to-almost-one-pass}
    For each rootward tree-like chase sequence ${T_0, \dots, T_n}$ for $I$ and
    $\Sigma$, there exists an almost one-pass chase sequence ${\Tbar_0, \dots,
    \Tbar_m}$ for $I$ and $\Sigma$ such that ${T_n \subseteq \Tbar_m}$.
\end{lemma}

\begin{proof}
Let ${T_0, \dots, T_n}$ be an arbitrary rootward tree-like chase sequence for
$I$ and $\Sigma$. The induction base ${i=0}$ holds trivially. For the induction
step, we assume that the claim holds for some $i$ with ${0 \leq i < n}$. By the
inductive assumption, there exists an almost one-pass chase sequence ${\Tbar_0,
\dots, \Tbar_j}$ for $I$ and $\Sigma$ such that ${T_i \subseteq \Tbar_j}$
holds. Now assume that $T_{i+1}$ is obtained by applying a chase or a
propagation step to some vertex $v$ of $T_i$, and let $k$ be the maximal number
such that ${0 \leq k \leq j}$ and $v$ is recently updated in $\Tbar_k$. Such
$k$ clearly exists since $v$ occurs in $\Tbar_j$, and ${T_i(v) \subseteq
\Tbar_k(v)}$ holds because $k$ is maximal. We now consider ways in which
$T_{i+1}$ can be derived from $T_i$.

Assume that $T_{i+1}$ is obtained from $T_i$ by a chase step with non-full GTGD
${\tau \in \Sigma}$ and substitutions $\sigma$ and $\sigma'$, and let $v'$ be
the child of $v$ introduced by the step. Without loss of generality, we can
choose $v'$ and the fresh labeled nulls such that they do not occur in
$\Tbar_j$. We shall now ``move'' this chase step so that it is performed
immediately after $\Tbar_k$. Towards this goal, we describe the chase trees
that are obtained by this move. For each $p$ with ${k \leq p \leq j}$, let
$\Ubar_p$ be the chase tree obtained from $\Tbar_p$ by adding vertex $v'$ and
letting ${\Ubar_p(v') = T_{i+1}(v')}$. We now argue that ${\Tbar_0, \dots,
\Tbar_k, \Ubar_k, \dots, \Ubar_j}$ is an almost one-pass chase sequence
satisfying the conditions of the lemma.
\begin{itemize}
    \item Chase tree $\Ubar_k$ can be seen as obtained from $\Tbar_k$ by a
    chase step with $\tau$ and substitutions $\sigma$ and $\sigma'$. Moreover,
    for each $p$ with ${k \leq p < j}$, chase tree $\Ubar_{p+1}$ is obtained
    from $\Ubar_p$ in the same way as $\Tbar_{p+1}$ is obtained from $\Tbar_p$.
    Thus, all preconditions of all chase steps are satisfied.

    \item Chase tree $\Ubar_k$ is obtained from $\Tbar_k$ by applying the chase
    step to the recently updated vertex $v$ of $\Tbar_k$. Moreover, if ${k <
    j}$, then $\Tbar_{k+1}$ is obtained from $\Tbar_k$ by applying a step to
    $v$ or an ancestor of $v$, and so $\Ubar_{k+1}$ is obtained from $\Ubar_k$
    by applying a step to an ancestor of the recently updated vertex of
    $\Ubar_k$. Thus, the sequence is almost one-pass.

    \item The construction clearly satisfies ${T_{i+1} \subseteq \Ubar_j}$.
\end{itemize}

In the rest of this proof we consider the case when $T_{i+1}$ is obtained from
$T_i$ by a chase step with a full GTGD ${\tau \in \Sigma}$ deriving a fact $F$,
or by a propagation step that copies a fact $F$ from $T_i(v)$ to the parent of
$v$. Let $v'$ be the recently updated vertex of $T_{i+1}$. Chase tree $\Tbar_j$
clearly contains $v'$. If ${F \in \Tbar_k(v')}$, then sequence ${\Tbar_0,
\dots, \Tbar_j}$ satisfies the inductive property, so we next assume that ${F
\not\in \Tbar_k(v')}$ holds. We shall now transform ${\Tbar_0, \dots, \Tbar_j}$
so that this step is applied immediately after $\Tbar_k$, and fact $F$ is
propagated towards the root as far as possible. Since this will move the
recently updated vertex towards the root, we will then ``reapply'' all relevant
steps from ${\Tbar_0, \dots, \Tbar_j}$ to ``regrow'' the relevant part of the
sequence. In each case, we specify the structure of the chase trees and discuss
the steps that produce these trees.

Let $\Ubar_0$ be obtained from $\Tbar_k$ by adding $F$ to $\Tbar_k(v)$. We
argue that $\Ubar_0$ can be seen as being obtained from $\Tbar_k$ by the same
step that produces $T_{i+1}$ from $T_i$.
\begin{itemize}
    \item If $T_{i+1}$ is obtained from $T_i$ by a chase step with a full GTGD,
    then ${F \not\in \Tbar_k(v')}$ holds ensures that the same step is
    applicable to $\Tbar_k(v')$ (where $v' = v$).

    \item If $T_{i+1}$ is obtained from $T_i$ by a propagation step, then $F$
    is $\Sigma$-guarded by $T_i(v')$. But then, ${T_i(v') \subseteq
    \Tbar_j(v')}$ ensures that $F$ is also $\Sigma$-guarded by $\Tbar_j(v')$,
    and Claim 2 of Lemma~\ref{lem:guarded-preservation} ensures that $F$ is
    $\Sigma$-guarded by $\Tbar_k(v')$. Thus, the propagation step is applicable
    to vertices $v$ and $v'$ in $\Tbar_k$.
\end{itemize}
Moreover, let ${\Ubar_1, \dots, \Ubar_s}$ be the chase trees obtained by
propagating $F$ starting from $\Ubar_0$ towards the root using local steps as
long as possible. Clearly, ${\Tbar_0, \dots, \Tbar_k, \Ubar_0, \Ubar_1, \dots,
\Ubar_s}$ is a correctly formed almost one-pass chase sequence. Let $v''$ be
the recently updated vertex of $\Ubar_s$,

We cannot simply append the step producing $T_{k+1}$ after $\Ubar_s$ because
this step might not be applicable to $v''$ or an ancestor thereof. Thus, to
obtain the chase sequence satisfying the claim of the lemma, we shall find a
place in sequence ${\Tbar_0, \dots, \Tbar_j}$ where vertex $v''$ is introduced,
and we shall ``replay'' all steps from that point onwards. In doing so, we
shall use chase steps that introduce the same vertices and labeled nulls, so we
will first need to rename these in the sequence ${\Tbar_0, \dots, \Tbar_k,
\Ubar_0, \Ubar_1, \dots, \Ubar_s}$.

Let $\ell$ be the smallest integer such that $\Tbar_\ell$ contains $v''$.
Clearly, ${\ell \leq k}$ holds. Now let $N$ be the set of labeled nulls
introduced by applying a chase step to $v''$ or a descendant thereof, and let
$W$ the the set of descendants of $v''$ in the sequence ${\Tbar_0, \dots,
\Tbar_j}$. Moreover, let ${\Tbar_0', \dots, \Tbar_k', \Ubar_0', \Ubar_1',
\dots, \Ubar_s'}$ be the chase sequence obtained by uniformly replacing in
${\Tbar_0, \dots, \Tbar_k, \Ubar_0, \Ubar_1, \dots, \Ubar_s}$ each labeled null
in $N$ with a distinct, fresh labeled null, and by uniformly replacing each
vertex ${w \in W}$ by a fresh vertex.

We now transform chase trees ${\Tbar_{\ell+1}, \dots, \Tbar_j}$ into chase
tress ${\Vbar_{\ell+1}, \dots, \Vbar_j}$ that reflect the result of
``replaying'' after $\Ubar_s'$ the steps producing the former sequence.
Intuitively, each $\Vbar_p$ is a ``union'' of $\Ubar_s$ and $\Tbar_p$.
Formally, for each $p$ with ${\ell < p \leq j}$, we define $\Vbar_p$ as follows.
\begin{enumerate}[wide=0.2cm, label=(S\arabic*)]
    \item\label{step:to-almost-one-pass:vertices}
    The chase tree $\Vbar_p$ contains the union of the vertices of $\Ubar_s'$
    and $\Tbar_p$.

    \item\label{step:to-almost-one-pass:just-one}
    For each vertex $w$ occurring only in $\Ubar_s'$ (resp.\ $\Tbar_p$), we
    define ${\Vbar_p(w) = \Ubar_s'(w)}$ (resp.\ ${\Vbar_p(w) = \Tbar_p(w)}$).

    \item\label{step:to-almost-one-pass:both}
    For each vertex $w$ occurring in both $\Ubar_s'$ and $\Tbar_p$, we define
    ${\Vbar_p(w) = \Ubar_s'(w) \cup \Tbar_p(w)}$.

    \item\label{step:to-almost-one-pass:propagate}
    If $\Tbar_p$ is obtained by applying to a vertex $w$ of $\Tbar_{p-1}$ a
    chase step with a non-full GTGD ${\tau = \forall \vec x [\body \rightarrow
    \exists \vec y ~ \head]}$ and substitutions $\sigma$ and $\sigma'$, then,
    for $w'$ the child of $w$ introduced by the step, we extend $\Vbar_p(w')$
    with each fact ${G \in \Vbar_{p-1}(w)}$ that is $\Sigma$-guarded by
    $\sigma'(\head)$.
\end{enumerate}

We now argue that ${\Tbar_0', \dots, \Tbar_k', \Ubar_0', \Ubar_1', \dots,
\Ubar_s', \Vbar_{\ell+1}, \dots, \Vbar_j}$ contains an almost one-pass chase
sequence for $\Sigma$ and $I$ that satisfies the conditions of this lemma.
\begin{itemize}
    \item Sequence ${\Tbar_0', \dots, \Tbar_k', \Ubar_0', \Ubar_1', \dots,
    \Ubar_s'}$ is clearly a valid almost one-pass chase sequence.

    \item For ${\ell \leq p < j}$, either $\Vbar_{p+1}$ is obtained from
    $\Vbar_p$ (or $\Ubar_s'$ in case $p = \ell$) by the same step that produces
    $\Tbar_{p+1}$ from $\Tbar_p$, or the step is not applicable. In the latter
    case, we can simply drop such $\Vbar_{p+1}$ from the sequence. By dropping
    all such $\Vbar_{p+1}$, we clearly obtain a valid almost one-pass chase
    sequence.

    \item We have ${T_i \subseteq \Tbar_j}$ by the induction assumption, and
    steps
    \ref{step:to-almost-one-pass:vertices}--\ref{step:to-almost-one-pass:both}
    clearly ensure ${T_i \subseteq \Vbar_j}$. Moreover, $T_{i+1}$ differs from
    $T_i$ only in vertex $v'$, where ${T_{i+1}(v') = T(v') \cup \{ F \}}$
    holds. Our construction, however, clearly ensures ${F \in \Ubar_s'(v'')}$,
    and step \ref{step:to-almost-one-pass:propagate} ensures that $F$ is
    propagated in each chase step with a non-full GTGD introducing a vertex on
    the unique path from $v''$ to $v'$. Thus, ${T_{i+1} \subseteq \Vbar_j}$
    holds. \qedhere
\end{itemize}
\end{proof}

\begin{lemma}\label{def:to-one-pass}
    For each base fact $F$ and each almost one-pass tree-like chase proof of
    $F$ from $I$ and $\Sigma$, there exists a one-pass tree-like chase proof of
    $F$ from $I$ and $\Sigma$.
\end{lemma}

\begin{proof}
Consider an arbitrary base fact $F$ and an arbitrary almost one-pass tree-like
chase proof ${T_0, \dots, T_n}$ of $F$ from $I$ and $\Sigma$. Since $F$ is a
base fact, without loss of generality we can assume that $F$ occurs in the
facts of the root vertex. Now let $T_i$ be the first chase tree that contains
$F$ in the root, and let $W$ be the set containing each non-root vertex $v$
occurring in any of the chase trees such that no propagation step is applied to
$v$. We transform this proof to a one-pass proof as follows. First, we delete
each $T_j$ with ${i < j \leq n}$. Next, we delete in each remaining $T_j$ each
vertex ${v \in W}$ and each descendant of $v$. Finally, we delete each
remaining $T_j$ that is equal to $T_{j+1}$. After this transformation, every
vertex has a propagation step applied to it. It is straightforward to see that
the result is a one-pass tree-like chase sequence. Moreover, since $F$ occurs
in the root, the sequence is a tree-like chase proof of $F$ from $I$ and
$\Sigma$.
\end{proof}

\subsection{Proof of Proposition \ref{prop:rewriting-conditions}: Rewriting Criterion Using One-pass Chase Proofs}

\RewritingConditions*

\begin{proof}
Let $\Sigma$ and $\Sigma'$ be as specified in the proposition, let $I$ be an
arbitrary base instance, and let $F$ be an arbitrary base fact. Since $\Sigma'$
is a logical consequence of $\Sigma$, it is clear that ${I, \Sigma' \models F}$
implies ${I, \Sigma \models F}$. Thus, we assume that ${I, \Sigma \models F}$
holds, and we prove that ${I, \Sigma' \models F}$ holds as well. By
Theorem~\ref{thm:one-pass-proof-exists}, there exists a one-pass tree-like
chase proof ${T_0, \dots, T_n}$ of $F$ from $I$ and $\Sigma$. Without loss of
generality, we can assume that $F$ is produced in the last step of the proof,
and so the recently updated vertex of $T_n$ is root vertex $r$. Let ${i_0 <
\dots < i_m}$ be exactly the indexes between $0$ and $n$ such that the recently
updated vertex of $T_{i_j}$ is $r$. We next construct a tree-like chase
sequence ${\Tbar_0, \dots, \Tbar_k}$ for $I$ and $\Sigma'$ such that ${T_n(r)
\subseteq \Tbar_k(r)}$. To formalize our inductive construction of this chase
sequence, we shall also construct a sequence of indexes ${\ell_0, \dots,
\ell_m}$ such that ${\ell_m = k}$ and, for each $j$ with ${0 \leq j \leq m}$,
we have ${T_{i_j}(r) \subseteq \Tbar_{\ell_j}(r)}$; in other words, each index
$\ell_j$ helps us establish the inductive property by relating $T_{i_j}$ and
$\Tbar_{\ell_j}$. For the base case, ${i_0 = 0}$ holds by the definition of a
tree-like chase proof; thus, we set ${\Tbar_0 = T_0}$ and ${\ell_0 = 0}$, and
the required property clearly holds. For the inductive step, we consider
arbitrary ${0 < j \leq m}$ such that the claim holds for $j-1$, and assume that
the sequence constructed thus far is ${\Tbar_{\ell_0}, \dots,
\Tbar_{\ell_{j-1}}}$. We have the following two cases.
\begin{itemize}
    \item The recently updated vertex of $T_{i_j-1}$ is $r$. Thus, ${i_j-1 =
    i_{j-1}}$, and $T_{i_j}$ is obtained from $T_{i_{j-1}}$ by a chase step
    with a full GTGD ${\body \rightarrow H \in \Sigma}$ producing a fact ${G
    \in T_{i_j}(r)}$. The second condition of the proposition ensures that
    ${\body \rightarrow H}$ is a logical consequence of $\Sigma'$, so $G$ can
    be derived from $\Tbar_{\ell_1}(r)$ and the Datalog rules of $\Sigma'$
    using $\wp$ steps. We then define ${\ell_j = \ell_{j-1} + \wp}$, and we
    append the corresponding steps to obtain the sequence ${\Tbar_0, \dots,
    \Tbar_{\ell_{j-1}}, \dots, \Tbar_{\ell_j}}$.

    \item Otherwise, ${T_{i_{j-1}}, \dots, T_{i_j}}$ is a loop at the root
    vertex $r$ with some output fact ${G \in T_{i_j}(r)}$. The third condition
    of the proposition ensures that there exists a Datalog rule ${\body
    \rightarrow H \in \Sigma'}$ and a substitution $\sigma$ such that
    ${\sigma(\body) \subseteq T_i(r)}$ and ${\sigma(H) = G}$. We define
    ${\ell_j = \ell_{j-1} + 1}$, and we define $\Tbar_{\ell_j}$ as the the
    chase tree containing just the root vertex $r$ such that
    ${\Tbar_{\ell_j}(r) = \Tbar_{\ell_{j-1}}(r) \cup \{ G \}}$; thus,
    $\Tbar_{\ell_j}$ is obtained from $\Tbar_{\ell_{j-1}}$ by applying the
    Datalog rule ${\body \rightarrow H \in \Sigma'}$ to the root vertex $r$.
    Moreover, ${T_{i_j}(r) \subseteq \Tbar_{\ell_j}(r)}$ clearly holds, as
    required. \qedhere
\end{itemize}
\end{proof}

\subsection{Proof of Proposition \ref{prop:loop-shape}: Properties of One-Pass Chase Proofs}

We finally prove a property that will be needed in the proofs in
Appendices~\ref{app:ExbDR}--\ref{app:FullDR}. This property intuitively ensures
that, as soon as a fact $F$ is derived in the child vertex of a loop such that
$F$ does not contain any null values introduced by the child, the loop is
completed and $F$ is propagated to the parent vertex.

\begin{proposition}\label{prop:loop-shape}
    For each loop ${T_i, \dots, T_j}$ at a vertex $v$ in a one-pass
    tree-like chase proof for some $I$ and $\Sigma$, for ${i < k \leq j}$, and
    for $v'$ the vertex introduced in $T_{i+1}$, set ${\trms\bigl(T_k(v')\bigr)
    \setminus \nulls\bigl(T_{i+1}(v')\bigr)}$ is $\Sigma$-guarded by $T_i(v)$.
\end{proposition}

\begin{proof}
Consider an arbitrary loop ${T_i, \dots, T_j}$ at a vertex $v$ in a one-pass
tree-like chase proof for some $I$ and $\Sigma$, and let $v'$ be the child of
$v$ introduced by the chase step producing $T_{i+1}$. We prove the claim by
induction on $k$ with ${i < k \leq j}$. For the induction base ${k = i+1}$, the
definition of a chase step with a non-full GTGD clearly ensures this claim for
$T_{i+1}(v')$. For the induction step, consider an arbitrary $k$ such that the
claim holds. Our claim holds trivially if ${T_{k+1}(v') = T_k(v')}$, so we
assume that ${T_{k+1}(v') \setminus T_k(v')}$ contains exactly one fact $F$,
which can be derived in one of the following two ways.
\begin{itemize}
    \item Assume $F$ is obtained by a propagation step to vertex $v'$. Then,
    $F$ is $\Sigma$-guarded by $T_k(v')$, so ${\trms(F) \subseteq
    \trms\bigl(T_k(v')\bigr) \cup \consts(\Sigma)}$ holds.
    
    \item Assume $F$ is obtained by applying a full GTGD ${\tau \in \Sigma}$ to
    $T_k(v')$ using a substitution $\sigma$. Then, $\tau$ contains a guard atom
    $A$ in the body such that ${\sigma(A) \subseteq T_k(v')}$; moreover, the
    head $\tau$ contains all variables of $A$, and so we have ${\trms(F)
    \subseteq \trms\bigl(T_k(v')\bigr) \cup \consts(\Sigma)}$.
\end{itemize}
Either way, we have ${\trms\bigl(T_{k+1}(v')\bigr) \subseteq
\trms\bigl(T_k(v')\bigr) \cup \consts(\Sigma)}$. By the induction assumption,
set ${\trms\bigl(T_k(v')\bigr) \setminus \nulls(T_{i+1}(v'))}$ is
$\Sigma$-guarded by $T_i(v)$, and so set ${\trms\bigl(T_{k_1}(v')\bigr)
\setminus \nulls(T_{i+1}(v'))}$ is also $\Sigma$-guarded by $T_i(v)$, as
required.
\end{proof}

    \section{Proofs for $\ExbDR$}\label{app:ExbDR}

\subsection{Proof of Proposition~\ref{prop:ExbDR-properties}: Properties of $\ExbDR$}

\ExbDRProperties*

\begin{proof}[Proof of Claim 1]
Let $G$ be a guard for $\tau'$. For the sake of a contradiction, assume that
$G$ is not one of the atoms ${A_1', \dots, A_n'}$---that is, ${G \in \body'}$.
Since ${n \geq 1}$, atom $A_1'$ in the body of $\tau'$ is matched to $A_1$ in
the head of $\tau$. Since $\tau$ is in head-normal form, $A_1$ contains at
least one variable ${y \in \vec y}$. Moreover, the conditions of the $\ExbDR$
inference rule ensure ${\theta(y) = y}$. Since $y$ does not occur in $A_1'$ and
$\theta$ unifies $A_1'$ and $A_1$, atom $A_1'$ contains at some position a
variable $z$ such that ${\theta(z) = y}$. Since $G$ is a guard for $\tau'$,
variable $z$ occurs in $G$. Therefore, we have ${\vars(\theta(G)) \cap \vec y
\neq \emptyset}$, which contradicts the requirement ${\vars(\theta(\body'))
\cap \vec y = \emptyset}$ of the $\ExbDR$ inference rule.
\end{proof}

\begin{proof}[Proof of Claim 2]
Consider arbitrary $i$ such that ${1 \leq i \leq n}$ and $A_i'$ is a guard of
$\tau'$, and let $\sigma$ be an MGU of $A_i'$ and the corresponding atom $A_i$
of $\tau$. Since $\theta$ is a unifier of $A_i'$ and $A_i$ as well as of other
pairs of atoms, there clearly exists a substitution $\rho$ such that ${\theta =
\rho \circ \sigma}$. Now consider an arbitrary $A_j'$ with ${1 \leq j \leq n}$
in $\tau'$. Substitution $\theta$ matches $A_j'$ to the corresponding atom
$A_j$ in the head of $\tau$. Since TGD $\tau$ is in head-normal form, atom
$A_j$ contains at least one variable ${y \in \vec y}$. Since ${\theta(y) = y}$,
we necessarily have ${y \in \vars(\theta(A_j'))}$. Consequently, atom $A_j'$
contains some variable $z$ such that ${\theta(z) = y}$. Since $A_i'$ is a guard
for $\tau'$, variable $z$ occurs in $A_i'$. Now assume for the sake of a
contradiction that ${\sigma(z) \neq y}$. Then $\sigma(z)=\sigma(x)$ for some
$x\in\vars(A_i)$ and $y=\theta(z)=\rho(\sigma(z))=\rho(\sigma(x))=\theta(x)$.
However, this contradicts the requirement ${\theta(\vec x) \cap \vec y =
\emptyset}$ of the $\ExbDR$ inference rule.
\end{proof}

\begin{proof}[Proof of Claim 3]
By Claim 1, there exists $i$ with ${1 \leq i \leq n}$ such that atom $A_i'$ is
a guard for $\tau'$. Thus, ${\vars(\body') \cup \vars(H') \subseteq
\vars(A_i')}$. The $\ExbDR$ inference rule ensures ${\vars(\theta(\body')) \cap
\vec y = \emptyset}$, which in turn ensures ${\vars(\theta(\body')) \subseteq
\vars(\theta(A_i')) \setminus \vec y}$. Now let $G$ be a guard for $\tau$. We
clearly have ${\vars(\theta(\body)) \subseteq \vars(\theta(G))}$. Moreover,
${\theta(y_i) = y_i}$ and ${\theta(x) \cap \vec y = \emptyset}$ ensure
${\vars(\theta(\head)) \cup \vars(\theta(A_i)) \subseteq \vars(\theta(G)) \cup
\vec y}$. Thus, $\theta(G)$ is a guard for the TGD produced by the $\ExbDR$
inference rule. Finally, since $G$ contains all variables of $\tau$, the widths
of the resulting TGD and $\tau$ are equal.
\end{proof}

\subsection{Proof of Theorem~\ref{thm:ExbDR-correctness-complexity}: Correctness and Complexity of $\ExbDR$}

\ExbDRCorrectnessComplexity*

\begin{proof}[Proof of Correctness]
Let $\Sigma'$ be the set $\Sigma$ closed under the $\ExbDR$ inferences rule as
specified in Definition \ref{def:rewriting-algorithm}. It is straightforward to
see that $\Sigma'$ is a logical consequence of $\Sigma$, so $\ExbDR(\Sigma)$ is
also a logical consequence of $\Sigma$. Moreover, $\ExbDR(\Sigma)$ contains
each full GTGD of $\Sigma$ up to redundancy, so each full GTGD of $\Sigma$ is
logically entailed by $\ExbDR(\Sigma)$. We next consider an arbitrary base
instance $I$ and a one-pass tree-like chase sequence for $I$ and $\Sigma$, and
we show the following property:
\begin{quote}
    ($\blacklozenge$) for each loop ${T_i, \dots, T_j}$ in the sequence at some
    vertex $v$ with output fact $F$, there exist a full GTGD ${\body
    \rightarrow H \in \Sigma'}$ and a substitution $\sigma$ such that
    ${\sigma(\body) \subseteq T_i(v)}$ and ${F = \sigma(H)}$.
\end{quote}
Since $\ExbDR(\Sigma)$ contains all full TGDs of $\Sigma'$ and this property
holds for the root vertex $r$, Proposition~\ref{prop:rewriting-conditions}
ensures that $\ExbDR(\Sigma)$ is a rewriting of $\Sigma$.

Our proof is by induction on the length of the loop. The base case and the
inductive step have the same structure, so we consider them jointly. Thus,
consider an arbitrary loop ${T_i, T_{i+1}, \dots, T_{j-1}, T_j}$ at vertex $v$
in the sequence, and assume that the claim holds for all shorter loops in the
sequence. By the definition of the loop, chase tree $T_{i+1}$ is obtained from
$T_i$ by applying a chase step to some non-full TGD ${\forall \vec x[\body_0
\rightarrow \exists \vec y ~ \head_0] \in \Sigma}$. Let $\sigma_0$ and
$\sigma_0'$ be the substitutions used in this chase step, let ${N =
\nulls(\rng(\sigma_0')) \setminus \nulls(\rng(\sigma_0))}$, let $v'$ be the
child of $v$ introduced in $T_{i+1}$, and let ${S \subseteq T_i(v)}$ be the
facts that are copied to $T_{i+1}(v')$ because they are $\Sigma$-guarded by
$\sigma_0'(\head_0)$. Thus, we have ${T_{i+1}(v') = S \cup
\sigma_0'(\head_0)}$. By Proposition~\ref{prop:loop-shape} and the fact that a
chase step is applied only if propagation to the parent is not applicable, the
output fact of the loop is added to $T_{j-1}(v')$ in step $j-1$, and in $T_j$
this fact is propagated back to $T_j(v)$. In other words, for each $k$ with ${i
< k < j-1}$, each fact in ${T_k(v') \setminus S}$ contains at least one labeled
null from $N$, or the fact would be $\Sigma$-guarded by $T_i(v)$ and thus
propagated back to vertex $v$. We show that, in the loop ${T_i, T_{i+1}, \dots,
T_{j-1}, T_j}$ fixed above, the following property holds for each $k$ with ${i
< k < j-1}$:
\begin{quote}
    ($\lozenge$) there exist a GTGD ${\forall \vec x[\body \rightarrow \exists
    \vec y ~ \head] \in \Sigma'}$, a substitution $\sigma$ such that
    ${\sigma(\body) \subseteq T_i(v)}$, and a substitution $\sigma'$ that
    extends $\sigma$ by mapping $\vec y$ to fresh labeled nulls such that
    ${T_k(v') \subseteq S \cup \sigma'(\head)}$.
\end{quote}

We prove ($\lozenge$) by induction on $k$. We have already proved the base case
${k = i+1}$ above. For the inductive step, assume that ($\lozenge$) holds for
some $k$, so there exists a GTGD ${\forall \vec x[\body \rightarrow \exists
\vec y ~ \head] \in \Sigma'}$ and substitutions $\sigma$ and $\sigma'$
satisfying ($\lozenge$) for $k$. Now consider $T_{k+1}$. Property ($\lozenge$)
holds by the inductive hypothesis if $T_{k+1}(v') = T_k(v')$---that is, if the
step involves a descendant of $v'$. Otherwise, ${T_{k+1}(v') = T_k(v') \cup \{
G \}}$ where fact $G$ is obtained in one of the following two ways.
\begin{itemize}
    \item A full GTGD in $\Sigma$ derives $G$ from $T_k(v')$. Set $\Sigma'$
    contains this GTGD up to redundancy, so by Definition~\ref{def:redundancy}
    there exist a full GTGD ${\body'' \rightarrow H' \in \Sigma'}$ and a
    substitution $\rho$ such that ${\rho(\body'') \subseteq T_k(v')}$ and
    ${\rho(H') = G}$.

    \item Fact $G$ is the output of a loop at vertex $v'$. But then, this loop
    is shorter than ${T_i, \dots, T_j}$ so, by property ($\blacklozenge$),
    there exists a full GTGD ${\body'' \rightarrow H' \in \Sigma'}$ and a
    substitution $\rho$ such that ${\rho(\body'') \subseteq T_k(v')}$ and
    ${\rho(H') = G}$.
\end{itemize}
Since $\head$ is in head-normal form, each atom in $\sigma'(\head)$ contains at
least one labeled null of $N$. Now let ${A_1', \dots, A_n'}$ be the atoms of
$\body''$ that are matched to the atoms in $\sigma'(\head)$. Atom $\rho(H')$
contains at least one labeled null of $N$, so ${n \geq 1}$. Thus, we can assume
that ${\body'' \rightarrow H'}$ is of the form ${A_1' \wedge \dots \wedge A_n'
\wedge \body' \rightarrow H'}$ where ${\{ \rho(A_1'), \dots, \rho(A_n') \}
\subseteq \sigma'(\head)}$ and ${\rho(\body') \subseteq S}$. Also, since
${\body'' \rightarrow H'}$ is guarded, at least one of $A_i'$ is a guard for
${\body'' \rightarrow H'}$. Let ${A_1, \dots, A_n}$ be the atoms of $\head$
such that ${\sigma'(A_i) = \rho(A_i')}$ for ${1 \leq i \leq n}$. Since
$\sigma'$ maps each ${y \in \vec y}$ to a distinct labeled null that does not
occur in $T_i$, we have ${\sigma'(\vec x) \cap \sigma'(\vec y) = \emptyset}$.
Thus, there exists a $\vec y$-MGU $\theta$ of ${A_1, \dots, A_n}$ and ${A_1',
\dots, A_n'}$ satisfying ${\theta(\vec x) \cap \vec y = \emptyset}$.
Conjunction $\rho(\body')$ does not contain a labeled null of $N$, so
${\vars(\theta(\body')) \cap \vec y = \emptyset}$ holds. Thus, the
preconditions of the $\ExbDR$ inference rule are satisfied for ${\forall \vec
x[\body \rightarrow \exists \vec y ~ \head]}$ and ${A_1' \wedge \dots \wedge
A_n' \wedge \body' \rightarrow H'}$, so the $\ExbDR$ rule derives ${\tau =
\theta(\body) \wedge \theta(\body') \rightarrow \exists \vec y ~ \theta(\head)
\wedge \theta(H')}$. Moreover, some $A_i'$ is a guard so all variables of
${A_1' \wedge \dots \wedge A_n' \wedge \body' \rightarrow H'}$ participate in
unification, and thus we can extend $\sigma$ and $\sigma'$ to substitutions
$\zeta$ and $\zeta'$, respectively, covering these variables such that
${\zeta(\theta(\body)) \cup \zeta(\theta(\body')) \subseteq T_i(v)}$ and
${T_{k+1}(v') \subseteq S \cup \zeta'(\theta(\head)) \cup \zeta'(\theta(H'))}$.
Set $\Sigma'$ contains $\tau$ up to redundancy. Since ${G \not\in T_k(v')}$,
GTGD $\tau$ is not a syntactic tautology, so there exists a GTGD ${\forall \vec
x_1[\body_1 \rightarrow \exists \vec y_1 ~ \head_1] \in \Sigma'}$ and
substitution $\mu$ such that ${\dom(\mu) = \vec x_1 \cup \vec y_1}$, ${\mu(\vec
x_1) \subseteq \vec x_2}$, ${\mu(\vec y_1) \subseteq \vec y_1 \cup \vec y_2}$
and ${\mu(y) \neq \mu(y')}$ for distinct $y$ and $y'$ in $\vec y_1$, and
${\mu(\body_1) \subseteq \theta(\body) \wedge \theta(\body')}$ and
${\mu(\head_1) \supseteq \theta(\head) \wedge \theta(H')}$. Now let $\sigma_1$
be the substitution defined as ${\sigma_1(x) = \zeta(\mu(x))}$ on each ${x \in
\vec x}$, and let $\sigma_1'$ be the extension of $\sigma_1$ to $\vec y_1$ such
that ${\sigma_1'(y) = \zeta'(\mu(y))}$ for each ${y \in \vec y}$. Clearly,
${\sigma_1(\body_1) \subseteq T_k(v')}$ and ${T_{k+1}(v') \subseteq S \cup
\sigma_1'(\head_1)}$ hold, so property ($\lozenge$) is satisfied.

To complete the proof, consider now the derivation of $T_{j-1}$. By property
($\lozenge$), there exists a GTGD ${\forall \vec x[\body \rightarrow \exists
\vec y ~ \head] \in \Sigma'}$ and substitutions $\sigma$ and $\sigma'$ such
that ${\sigma(\body) \subseteq T_i(v)}$ and ${T_{j-2}(v') = S \cup
\sigma'(\head)}$. Then, as above, $\Sigma'$ contains a full TGD of the form
${A_1' \wedge \dots \wedge A_n' \wedge \body' \rightarrow H'}$ that satisfies
${\rho(A_1') \cup \dots \cup \rho(A_n') \subseteq \sigma'(\head)}$ and
${\rho(\body') \subseteq S}$ for some substitution $\rho$. A minor difference
is that $\rho(H')$ does not contain a labeled null introduced by
$\sigma'(\head_0)$, so ${n = 0}$ is possible; however, in such a case, this TGD
immediately satisfies property ($\blacklozenge$). Moreover, if ${n > 0}$, then
${\body \rightarrow \exists \vec y ~ \head}$ can again be resolved with ${A_1'
\wedge \dots \wedge A_n' \wedge \body' \rightarrow H'}$ to produce
\begin{displaymath}
    \theta(\body) \wedge \theta(\body') \rightarrow \exists \vec y ~ \theta(\head) \wedge \theta(H') \in \Sigma'
\end{displaymath}
satisfying ${\vars(\theta(H')) \cap \vec y = \emptyset}$. This TGD is
transformed into head-normal form by Definition~\ref{def:rewriting-algorithm},
so ${\forall \vec x[\theta(\body) \wedge \theta(\body') \rightarrow
\theta(H')]}$ is contained in $\Sigma'$ up to redundancy. But then, $\Sigma'$
contains a full GTGD that satisfies property ($\blacklozenge$) by the same
argument as above.
\end{proof}

\begin{proof}[Proof of Complexity]
Fix $\Sigma$, $r$, $w_b$, $w_h$, $c$, and $a$ as stated in the theorem. The
number of different body atoms of arity $a$ constructed using $r$ relations,
$w_b$ variables, and $c$ constants is clearly bounded by ${\ell_b = r \cdot
(w_b + c)^a}$. Moreover, by the third claim of
Proposition~\ref{prop:ExbDR-properties}, the number of variables in the head of
each TGD is bounded by $w_h$, so the number of head atoms is bounded by
${\ell_h = r \cdot (w_h + c)^a}$. The body (resp.\ head) of each GTGD
corresponds to a subset of these atoms, so number of different GTGDs up to
variable renaming is bounded by ${\wp = 2^{\ell_b} \cdot 2^{\ell_h}}$. Thus,
the $\ExbDR$ inference rule needs to be applied to at most ${\wp^2 =
2^{2(\ell_b + \ell_h)}}$ pairs of GTGDs. For each such pair, one might need to
consider each possible way to match the $\ell_b$ body atoms of $\tau'$ to
$\ell_h$ head atoms of $\tau$, and there are at most ${(\ell_h)^{\ell_b} \leq
2^{\ell_b \cdot \ell_h}}$ of these. Consequently, unifier $\theta$ may need to
be computed at most ${2^{2(\ell_b + \ell_h)} \cdot 2^{\ell_b \cdot \ell_h} \leq
2^{5 \cdot \ell_b \cdot \ell_h} = 32^{\ell_b \cdot \ell_h}}$ times. To check
whether TGD ${\tau_1 = \forall \vec x_1[\body_1 \rightarrow \vec y_1 ~
\head_1]}$ is subsumed by ${\tau_2 = \forall \vec x_2[\body_2 \rightarrow \vec
y_2 ~ \head_2]}$, we can proceed as follows. First, we consider all possible
ways to match an atom of $\body_1$ to an atom of $\body_2$; since both
conjunctions contain at most $\ell_b$ atoms, there are at most
${\ell_b{}^{\ell_b} \leq 2^{\ell_b{}^2}}$ such matchings. Second, we
analogously consider each of at most $2^{(\ell_h)^2}$ ways to match an atom of
$\head_2$ to an atom of $\head_1$. Once all atoms have been matched, we try to
find a substitution $\mu$ satisfying Definition~\ref{def:redundancy} in linear
time. Thus, a subsumption check for pairs of TGDs takes at most
${2^{\ell_b{}^2} \cdot 2^{\ell_h{}^2}}$ steps. Finally, unification of atoms
requires time that is linear in $a$, and all other steps require linear time
too.
\end{proof}

    \section{Proofs for $\SkDR$}\label{app:SkDR}

\subsection{Proof of Proposition~\ref{prop:SkDR-properties}: Properties of $\SkDR$}

We reuse results by \citet{denivelle} about unification of atoms in guarded
rules. The \emph{variable depth} \cite[Definition~3]{denivelle} of an atom is
defined as $-1$ if the atom is ground, or as the maximum number of nested
function symbols that contain a variable of the atom. Moreover, an atom is
\emph{weakly covering} \cite[Definition~6]{denivelle} if each nonground
functional subterm of the atom contains all variables of the atom. Finally,
\citet[Theorem 1]{denivelle} says that, for $\theta$ an MGU of weakly covering
atoms $A$ and $B$, atom ${C = \theta(A) = \theta(B)}$ is also weakly covering,
the variable depth of $C$ is bounded by the variable depth of $A$ and $B$, and
the number of variables of $C$ is bounded by the number of variables of $A$ and
$B$ too.

\SkDRProperties*

\begin{proof}
Consider arbitrary rules ${\tau = \body \rightarrow H}$ and ${\tau' = A' \wedge
\body' \rightarrow H'}$ and an MGU $\theta$ of $H$ and $A'$ satisfying the
preconditions of the $\SkDR$ inference rule. Atom $H$ thus contains a Skolem
symbol, and rule $\tau$ is guarded; consequently, atom $H$ is weakly covering,
it contains a term of the form $f(\vec t)$ where $\vec t$ consists of constants
and all variables of the rule, and the variable depth of $H$ is at most one.
The corresponding atom $A'$ can be of the following two forms.
\begin{itemize}
    \item Atom $A'$ is Skolem-free. But then, $A'$ contains all variables of
    $\tau'$, and it is clearly weakly covering. By
    \citet[Theorem~1]{denivelle}, atom $\theta(A')$ is weakly covering and has
    variable depth at most one; consequently, each atom in rule ${\theta(A')
    \wedge \theta(\body') \rightarrow \theta(H')}$ is weakly covering and has
    variable depth at most one. Moreover, the variable depth of $\theta(H)$ is
    also at most one, which can be only if $\theta$ maps each variable in $H$
    to another variable or a constant. Thus, each atom in rule ${\theta(\body)
    \rightarrow \theta(H)}$ is weakly covering and has variable depth at most
    one; moreover, $\theta(\body)$ contains an atom that contains all variables
    of the rule. But then, rule ${\theta(\body) \wedge \theta(\body')
    \rightarrow \theta(H')}$ is guarded, as required.

    \item Atom $A'$ contains a Skolem symbol. But then, $A'$ is weakly covering
    by the definition of guarded rules, and its variable depth is at most one.
    By \citet[Theorem~1]{denivelle}, atom ${\theta(H) = \theta(A')}$ is weakly
    covering and has variable depth at most one, which can be the case only if
    $\theta$ maps all variables to other variables or constants. Consequently,
    rules ${\theta(\body) \rightarrow \theta(H)}$ and ${\theta(A') \wedge
    \theta(\body') \rightarrow \theta(H')}$ are both guarded. But then, rule
    ${\theta(\body) \wedge \theta(\body') \rightarrow \theta(H')}$ is guarded,
    as required. \qedhere
\end{itemize}
\end{proof}

\subsection{Proof of Theorem~\ref{thm:SkDR-correctness-complexity}: Correctness and Complexity of $\SkDR$}

\SkDRCorrectnessComplexity*

\begin{proof}[Proof of Correctness]
Let $\Sigma$ be an arbitrary finite set of GTGDs, and let $\Sigma'$ be the set
of rules obtained from $\Sigma$ as specified in
Definition~\ref{def:rewriting-algorithm}. It is straightforward to see that
$\Sigma'$ is a logical consequence of the Skolemization of $\Sigma$, so
$\SkDR(\Sigma)$ is also a logical consequence of $\Sigma$. Moreover,
$\SkDR(\Sigma)$ contains each full TGD of $\Sigma$ up to redundancy, so each
full TGD of $\Sigma$ is logically entailed by $\SkDR(\Sigma)$. We next consider
an arbitrary base instance $I$ and a one-pass tree-like chase sequence for $I$
and $\Sigma$, and we show the following property:
\begin{quote}
    ($\blacklozenge$) for each loop ${T_i, \dots, T_j}$ in the sequence at some
    vertex $v$ with output fact $F$, there exist a Skolem-free rule ${\body
    \rightarrow H \in \Sigma'}$ and a substitution $\sigma$ such that
    ${\sigma(\body) \subseteq T_i(v)}$ and ${F = \sigma(H)}$.
\end{quote}
Since $\SkDR(\Sigma)$ contains all Skolem-free rules of $\Sigma'$ and this
property holds for the root vertex $r$,
Proposition~\ref{prop:rewriting-conditions} ensures that $\SkDR(\Sigma)$ is a
rewriting of $\Sigma$.

Our proof is by induction in the length of the loop. The base case and the
inductive step have the same structure, so we consider them jointly. Thus,
consider an arbitrary loop ${T_i, T_{i+1}, \dots, T_{j-1}, T_j}$ at vertex $v$
in the sequence, and assume that the claim holds for all shorter loops in the
sequence. By the definition of a loop, chase tree $T_{i+1}$ is obtained from
$T_i$ by applying a chase step to some non-full GTGD ${\tau \in \Sigma}$ and
substitution $\gamma$. Let $v'$ be the child of $v$ introduced in $T_{i+1}$,
let ${S \subseteq T_i(v)}$ be the facts that are copied to $T_{i+1}(v')$
because they are $\Sigma$-guarded by the instantiated head of $\tau$, let ${N =
\{ n_1, \dots, n_m \}}$ be the set of labeled nulls introduced in the chase
step for the existentially quantified variables ${y_1, \dots, y_m}$ of $\tau$,
let $\nu$ be a function that maps each labeled null $n_i$ to the ground term
$f_i(\gamma(\vec x))$ where $f_i$ is the symbol used in the Skolemization of
$y_i$. For $U$ a set of facts, let $\nu(U)$ be the result of replacing each
occurrence of a labeled null ${n \in \dom(\nu)}$ in $U$ with $\nu(n)$ and
eliminating any duplicate facts in the result. Clearly, the inverse function
$\nu^-$ is well-defined, and we define $\nu^-(U)$ for $U$ a set of facts in the
obvious way. By Proposition~\ref{prop:loop-shape} and the fact that propagation
is applied eagerly, the output fact of the loop is added to $T_{j-1}(v')$ in
step $j-1$, and in $T_j$ this fact is propagated back to $T_j(v)$. In other
words, for each $k$ with ${i < k < j-1}$, each fact in ${T_k(v') \setminus S}$
contains at least one labeled null from $N$, or the fact would be
$\Sigma$-guarded by $T_i(v)$ and would thus be propagated back to vertex $v$.
We now show that the following property holds for each $k$ with ${i < k \leq
j-1}$:
\begin{quote}
    ($\lozenge$) for each fact ${G \in T_k(v') \setminus S}$, there exist a
    rule ${\body \rightarrow H \in \Sigma'}$ and a substitution $\sigma$ such
    that $\body$ is Skolem-free, ${\sigma(\body) \subseteq \nu(T_i(v))}$, and
    ${\sigma(H) = \nu(G)}$.
\end{quote}
Property ($\lozenge$) implies ($\blacklozenge$): fact $F$ does not contain a
labeled null from $N$, so the rule ${\body \rightarrow H \in \Sigma'}$ whose
existence is implied by ($\lozenge$) for ${k = j - 1}$ is actually a
Skolem-free rule that satisfies ($\blacklozenge$).

We next prove property ($\lozenge$) by a nested induction on $k$. For the base
case ${k = i + 1}$, property ($\lozenge$) holds due to the fact that $\Sigma'$
contains the rules obtained by Skolemizing GTGD $\tau$. For the inductive step,
assume that ($\lozenge$) holds for some $k$ and consider the possible ways to
obtain $T_{k+1}$ from $T_k$. Property ($\lozenge$) holds by the inductive
hypothesis if $T_{k+1}(v') = T_k(v')$---that is, if the step involves a
descendant of $v'$. Otherwise, ${T_{k+1}(v') = T_k(v') \cup \{ G \}}$ where
fact $G$ is obtained in one of the following two ways.
\begin{itemize}
    \item A full TGD in $\Sigma$ derives $G$ from $T_k(v')$. Set $\Sigma'$
    contains this TGD up to redundancy, so by Definition~\ref{def:redundancy}
    there exist a Skolem-free rule ${\body'' \rightarrow H' \in \Sigma'}$ and a
    substitution $\sigma'$ such that ${\sigma'(\body'') \subseteq
    \nu(T_k(v'))}$ and ${\sigma'(H') = \nu(G)}$.

    \item Fact $G$ is the output of a loop at vertex $v'$. But then, this loop
    is shorter than ${T_i, \dots, T_j}$ so, by property ($\blacklozenge$),
    there exist a Skolem-free rule ${\body'' \rightarrow H' \in \Sigma'}$ and a
    substitution $\sigma'$ such that ${\sigma'(\body'') \subseteq
    \nu(T_k(v'))}$ and ${\sigma'(H') = \nu(G)}$.
\end{itemize}
Now let ${W = \{ B' \in \body'' \mid \sigma'(B') \not\in S \}}$. We next show
that set $\Sigma'$ contains up to redundancy the result of ``resolving away''
each atom ${B' \in W}$. A slight complication arises due to the fact that the
$\SkDR$ inference rule considers only two rules at a time, and that the result
of each inference is contained in $\Sigma'$ up to redundancy. Thus, we will
achieve our goal by showing that the $\SkDR$ inference rule can be applied up
to ${n = |W|}$ times. Our proof is by induction on ${1 \leq \ell \leq n}$.
Towards this goal, we shall define $n$ rules ${\body_\ell'' \rightarrow
H_\ell'}$, substitutions $\sigma_\ell'$, and sets of atoms ${W = W_0 \supsetneq
\dots \supsetneq W_n}$ for $\ell$ with ${0 \leq \ell \leq n}$ satisfying the
following invariant:
\begin{quote}
    ($\ast$) ${\sigma_\ell'(\body_\ell'') \subseteq S \cup \{ \sigma'(B') \mid B' \in W_\ell \}}$ and ${\sigma_\ell'(H_\ell') = \sigma'(H')}$.
\end{quote}
For ${\ell = n}$, we have ${W_n = \emptyset}$, and so property ($\ast$) implies
property ($\lozenge$), as required. Our construction proceeds as follows.

For the base case ${\ell = 0}$, property ($\ast$) clearly holds for ${\body_0''
= \body''}$, ${\sigma_0' = \sigma'}$, and let ${W_0 = W}$. For the induction
step, assume that ($\ast$) holds for some ${0 \leq \ell < n}$, so
${\body_\ell'' \rightarrow H_\ell'}$, $\sigma_\ell'$, and $W_\ell$ satisfying
($\ast$) have been defined. First, assume that there exists ${B' \in W_\ell}$
such that ${\sigma'(B') \not\in \sigma_\ell'(\body_\ell')}$. Then, property
($\ast$) clearly holds for ${\body_{\ell+1}'' = \body_\ell''}$, ${H_{\ell+1}' =
H_\ell'}$, ${\sigma_{\ell+1}' = \sigma_\ell'}$, and ${W_{\ell+1} = W_\ell
\setminus \{ B' \}}$. Otherwise, we consider the following possibilities.
\begin{itemize}
    \item If rule ${\body_\ell'' \rightarrow H_\ell'}$ is Skolem-free, the rule
    is of the form ${A_\ell' \wedge \body_\ell' \rightarrow H_\ell'}$ where
    $A_\ell'$ contains all variables of the rule.

    \item Otherwise, rule ${\body_\ell'' \rightarrow H_\ell'}$ is of the form
    ${A_\ell' \wedge \body_\ell' \rightarrow H_\ell'}$ where atom $A_\ell'$
    contains a Skolem symbol, in which case this atom contains all variables of
    the rule.
\end{itemize}
Either way, there exists ${B' \in W_\ell}$ such that ${\sigma_\ell'(A_\ell') =
\sigma'(B')}$ where ${\sigma'(B') \in T_k(v') \setminus S}$. Thus, by property
($\lozenge$), these exist a rule ${\body_\ell \rightarrow H_\ell \in \Sigma'}$
and a substitution $\sigma_\ell$ such that $\body_\ell$ is Skolem-free,
${\sigma_\ell(\body_\ell) \subseteq \nu(T_i(v))}$, and ${\sigma_\ell(H_\ell) =
\sigma'(B')}$; the last observation ensures that $H_\ell$ contains a Skolem
symbol. Moreover, there exists an MGU $\theta_\ell$ of $H_\ell$ and $A_\ell'$,
so the $\SkDR$ inference rule is applicable to ${\body_\ell \rightarrow
H_\ell}$ and ${A_\ell' \wedge \body_\ell' \rightarrow H_\ell'}$, and $\Sigma'$
contains rule ${\theta_\ell(\body_\ell) \wedge \theta_\ell(\body_\ell')
\rightarrow \theta_\ell(H_\ell)}$ up to redundancy. Now let ${\zeta_\ell =
(\sigma_\ell \cup \sigma_\ell') \circ \theta}$ be the composition of
${\sigma_\ell \cup \sigma_\ell'}$ and $\theta$; note that substitution
${\sigma_\ell \cup \sigma_\ell'}$ is correctly defined because rules
${\body_\ell \rightarrow H_\ell}$ and ${A_\ell' \wedge \body_\ell' \rightarrow
H_\ell'}$ do not share variables. Moreover, let ${W_{\ell+1} = W_\ell \setminus
\{ B' \}}$. We clearly have ${\zeta_\ell(\theta_\ell(\body_\ell)) \subseteq
S}$, ${\zeta_\ell(\theta_\ell(\body_\ell')) \subseteq S \cup \{ \sigma'(C')
\mid C' \in W_{\ell+1} \}}$, and ${\zeta_\ell(\theta_\ell(H)) = \sigma'(H')}$.
Since ${G \not\in T_k(v')}$, rule ${\theta_\ell(\body_\ell) \wedge
\theta_\ell(\body_\ell') \rightarrow \theta_\ell(H_\ell)}$ is not a syntactic
tautology. Thus, by Definition~\ref{def:redundancy}, there exist a rule
${\body_{\ell+1}'' \rightarrow H_{\ell+1}' \in \Sigma'}$ and substitution
$\mu_{\ell+1}$ such that ${\mu_{\ell+1}(\body_{\ell+1}'') \subseteq
\theta_\ell(\body_\ell) \cup \theta_\ell(\body_\ell')}$ and
${\mu_{\ell+1}(H_{\ell+1}') = \theta_\ell(H_\ell')}$. Now let
$\sigma_{\ell+1}'$ be the substitution defined on each variable $x$ in
${\body_{\ell+1}'' \rightarrow H_{\ell+1}'}$ such that ${\sigma_{\ell+1}'(x) =
\zeta_\ell(\mu_{\ell+1}(x))}$. Then, property ($\ast$) clearly holds for
${\body_{\ell+1}'' \rightarrow H_{\ell+1}'}$, $\sigma_{\ell+1}'$, and
$W_{\ell+1}$, as required.
\end{proof}

\begin{proof}[Proof of Complexity]
Fix $\Sigma$, $r$, $w_b$, $e$, $c$, and $a$ as stated in the theorem.
Skolemizing a GTGD ${\forall \vec x[\body \rightarrow \exists \vec y ~ \head]}$
produces guarded rules in which each atom is of the form ${R(t_1, \dots, t_n)}$
such that each $t_i$ is a constant, a variable from $\vec x$, or a term of the
form $f(\vec x)$ where $f$ is a Skolem symbol. Moreover, each atom obtained
from ${R(t_1, \dots, t_n)}$ by the $\SkDR$ inference rule is obtained by
replacing a variable in $\vec x$ with another variable or a constant. Thus,
atom ${R(t_1, \dots, t_n)}$ cannot contain more than $|\vec x|$ variables.
Since the number of different symbols obtained by Skolemization is clearly
bounded by $e$, the number of different atoms of such form is bounded by ${\ell
= r \cdot (w_b + e + c)^a}$. The body of each guarded rule corresponds to a
subset of these atoms, so the number of different rules up to variable
remaining is bounded by ${2^\ell \cdot \ell \leq 2^\ell \cdot 2^\ell =
2^{2\ell} = \wp}$. By Definition~\ref{def:rewriting-algorithm}, the result of
applying the $\SkDR$ inference rule is retained in set $\Sigma'$ only if the
set does not contain a variable renaming of the result. Thus, the $\SkDR$
inference rule needs to be applied to at most ${\wp^2 = 2^{4\ell}}$ pairs of
rules. For each pair, one might need to unify at most $\ell$ body atoms of one
rule with the head atom of the other rule, so the unifier $\theta$ may need to
be computed at most ${\wp^2 \cdot \ell \leq \wp^2 \cdot 2^\ell = 2^{5\ell} =
32^\ell}$ times. We can check subsumption between a pair of rules analogously
to Theorem~\ref{thm:ExbDR-correctness-complexity}: for each of at most
$2^{\ell^2}$ ways to match the body atoms of one rule to the body atoms of
another rule, we try to find a substitution $\mu$ satisfying
Definition~\ref{def:redundancy}. Finally, unification of atoms requires time
that is linear in $a$, and all other steps require linear time too.
\end{proof}

    \section{Proofs for $\HypDR$}\label{app:HypDR}

\subsection{Proof of Proposition~\ref{prop:HypDR-properties}: Properties of $\HypDR$}

\HypDRProperties*

\begin{proof}
Consider arbitrary rules ${\tau_i = \body_i \rightarrow H_i}$ with ${1 \leq i
\leq n}$ such that $\body_i$ is Skolem-free and $H_i$ contains a Skolem symbol,
a Skolem-free rule ${\tau' = A_1' \wedge \dots \wedge A_n' \wedge \body'
\rightarrow H'}$, and an MGU $\theta$ of ${H_1, \dots, H_n}$ and ${A_1', \dotsm
A_n'}$ satisfying the preconditions of the $\HypDR$ inference rule. Rule
$\tau_1$ contains a term with a Skolem symbol in the head, and this term is
unified with a variable, say $x$, occurring in a Skolem-free body atom $A_1'$
of rule $\tau'$. Moreover, rule $\tau'$ is guarded, so the body of the rule
contains a Skolem-free atom $G$ that contains all variables of the rule; thus,
$G$ also contains $x$. Since $\theta(x)$ contains a Skolem symbol, $\theta(G)$
contains a Skolem symbol too. However, $\theta(\body')$ is Skolem-free, so $G$
must be one of the atoms ${A_1', \dots, A_n'}$ from the body of rule $\tau'$
that are participating in the $\HypDR$ inference rule. But then, we can show
that the result of the inference is guarded analogously to the proof of
Proposition~\ref{prop:SkDR-properties}.
\end{proof}

\subsection{Proof of Theorem~\ref{thm:HypDR-correctness-complexity}: Correctness and Complexity of $\HypDR$}

\HypDRCorrectnessComplexity*

\begin{proof}[Proof of Correctness]
The correctness proof for $\HypDR$ is almost identical to the correctness proof
in Theorem~\ref{thm:SkDR-correctness-complexity}, so we outline just the
difference. In particular, we wish to prove properties ($\blacklozenge$) and
($\lozenge$) exactly as stated in Theorem~\ref{thm:SkDR-correctness-complexity}
using the same proof structure. In the proof of property ($\lozenge$), we
establish existence of a Skolem-free rule ${\body'' \rightarrow H' \in
\Sigma'}$ and a substitution $\sigma'$ such that ${\sigma'(\body'') \subseteq
\nu(T_k(v'))}$ and ${\sigma'(H') = \nu(G)}$ in exactly the same way. The
difference to the proof of Theorem~\ref{thm:SkDR-correctness-complexity} is
that we ``resolve away'' all relevant body atoms of $\body''$ in one step. To
this end, let ${A_1', \dots, A_n'}$ be precisely the atoms of $\body''$ such
that ${\sigma'(A_i') \not\in S}$ for each ${1 \leq i \leq n}$. Thus, we can
assume that the rule is of the form ${A_1' \wedge \dots \wedge A_n' \wedge
\body' \rightarrow H'}$, and ${\sigma'(\body') \subseteq S}$ clearly holds. By
property ($\lozenge$), for each ${1 \leq \ell \leq n}$, there exist a rule
${\body_\ell \rightarrow H_\ell \in \Sigma'}$ and substitution $\sigma_\ell$
such that $\body_\ell$ is Skolem-free and ${\sigma_\ell(H_\ell) =
\sigma'(A_\ell')}$; the last observation ensures that $H_\ell$ contains a
Skolem symbol. Finally, there exists an MGU $\theta$ of ${H_1, \dots, H_n}$ and
${A_1', \dots, A_n'}$. Since ${\sigma'(\body') \subseteq S}$, conjunction
$\theta(\body')$ is Skolem-free. Thus, the $\HypDR$ inference rule is
applicable to ${\body_1 \rightarrow H_1, \dots, \body_n \rightarrow H_n}$ and
${A_1' \wedge \dots \wedge A_n' \wedge \body' \rightarrow H'}$, so set
$\Sigma'$ contains rule ${\theta(\body_1) \wedge \dots \wedge \theta(\body_n)
\wedge \theta(\body') \rightarrow \theta(H')}$ up to redundancy. Since no
premises share variables, substitution ${\sigma_1 \cup \dots \cup \sigma_n \cup
\sigma'}$ is correctly defined, so let $\zeta$ be the composition of ${\sigma_1
\cup \dots \cup \sigma_n \cup \sigma'}$ and $\theta$. Clearly, we have
${\zeta(\theta(\body_1)) \cup \dots \cup \zeta(\theta(\body_n)) \cup
\zeta(\theta(\body')) \subseteq S}$ and ${\zeta(\theta(H')) = \sigma'(H') =
\nu(G)}$. Since ${G \not\in T_k(v')}$, rule ${\theta(\body_1) \wedge \dots
\wedge \theta(\body_n) \wedge \theta(\body') \rightarrow \theta(H')}$ is not a
syntactic tautology so, by Definition~\ref{def:redundancy}, there exist a rule
${\body \rightarrow H \in \Sigma'}$ and substitution $\mu$ such that
${\mu(\body) \subseteq \theta(\body_1) \cup \dots \cup \theta(\body_n) \cup
\theta(\body')}$ and ${\mu(H) = \theta(H')}$. Let $\sigma$ be the substitution
defined on each variable $x$ in ${\body \rightarrow H}$ such that ${\sigma(x) =
\zeta(\mu(x))}$. Then, ${\sigma(\body) \subseteq S}$ and ${\sigma(H) =
\sigma'(H') = \nu(G)}$, as required for property ($\lozenge$).
\end{proof}

\begin{proof}[Proof of Complexity]
Fix $\Sigma$, $r$, $w_b$, $e$, $c$, and $a$ as stated in the theorem. In the
same way as in the complexity proof of
Theorem~\ref{thm:SkDR-correctness-complexity}, the number of different atoms
can be bounded by ${\ell = r \cdot (w_b + e + c)^a}$, and the number of
different rules can be bounded by ${\wp = 2^{2\ell}}$. Now we can apply the
$\HypDR$ inference rule as follows: we choose one of the $\wp$ rules that plays
the role of $\tau'$ and then, for each of the at most $\ell$ body atoms in
$\tau'$, we select one of the $\wp$ rules that play the role of rules $\tau_i$.
Hence, there are at most ${\wp \cdot \wp^\ell = \wp^{\ell+1}}$ different
applications of the $\HypDR$ inference rule. Thus, we may need to compute the
unifier $\theta$ at most ${(2^{2\ell})^{\ell+1} = 2^{2\ell^2+2\ell} \leq
2^{3\ell^2}}$ times. Finally, the times needed for subsumption checking,
unification, and all other steps can be bounded analogously as in the
complexity proof of Theorem~\ref{thm:SkDR-correctness-complexity}.
\end{proof}

    \section{The $\FullDR$ Algorithm: Creating Datalog Rules Directly}\label{app:FullDR}

The algorithms presented in the body of the paper all create the Datalog rules
needed for the final rewriting as well as intermediate non-full TGDs or rules
that are discarded after all inferences are performed. We now present an
algorithm that produces \emph{only} Datalog rules. Similar algorithms have
appeared in the prior literature \cite{DBLP:journals/lmcs/AmarilliB22}. After
presenting such an algorithm, we explain the shortcomings of this approach.

\begin{definition}\label{def:FullDR}
    The \emph{Full Datalog Rewriting} inference rule $\FullDR$ can be applied
    in two ways, depending on the types of TGDs it takes.
    \begin{itemize}
        \item The \compose variant of the $\FullDR$ inference rule takes full
        TGDs
        \begin{displaymath}
            \tau = \forall \vec x [\body \rightarrow A] \qquad \text{and} \qquad \tau' = \forall \vec z [A' \wedge \body' \rightarrow H']
        \end{displaymath}
        and a substitution $\theta$ such that
        \begin{itemize}
            \item ${\theta(A) = \theta(A')}$,

            \item ${\dom(\theta) = \vec x \cup \vec z}$, and

            \item ${\rng(\theta) \subseteq \vec w \cup \consts(\tau) \cup
            \consts(\tau')}$ where $\vec w$ is a vector of ${\hwidth(\Sigma) +
            |\consts(\Sigma)|}$ variables different from ${\vec x \cup \vec z}$,
        \end{itemize}
        and it derives
        \begin{displaymath}
             \theta(\body) \wedge \theta(\body') \rightarrow \theta(H').
        \end{displaymath}

        \item The \propagate variant of the $\FullDR$ inference rule takes TGDs
        \begin{displaymath}
            \tau  = \forall \vec x [\body \rightarrow \exists \vec y~\head \wedge A_1 \wedge \dots \wedge A_n] \qquad \text{and} \qquad \tau' = \forall \vec z [A_1' \wedge \dots \wedge A_n' \wedge \body' \rightarrow H']
        \end{displaymath}
        and a substitution $\theta$ such that
        \begin{itemize}
            \item ${\theta(A_i) = \theta(A_i')}$ for each $i$ with ${1 \leq i \leq n}$,

            \item ${\dom(\theta) = \vec x \cup \vec z}$,

            \item ${\rng(\theta) \subseteq \vec w\cup \vec y \cup \consts(\tau)
            \cup \consts(\tau') \text{ where } \vec w \text{ is a vector of }
            \hwidth(\Sigma) + |\consts(\Sigma)| \text{ variables different from
            } \vec x \cup \vec y \cup \vec z}$,

            \item ${\theta(\vec x) \cap \vec y = \emptyset}$, and

            \item ${\vars(\theta(\body')) \cap \vec y = \emptyset}$ and ${\vars(\theta(H')) \cap \vec y = \emptyset}$,
        \end{itemize}
        and it derives
        \begin{displaymath}
            \theta(\body) \wedge \theta(\body') \rightarrow \theta(H').
        \end{displaymath}
    \end{itemize}
\end{definition}

\begin{theorem}\label{thm:FullDR-correctness-complexity}
    Program $\FullDR(\Sigma)$ is a Datalog rewriting of a finite set of GTGDs
    $\Sigma$. Moreover, the rewriting can be computed in time ${O(b^{r^d \cdot
    (w + c)^{d a}})}$ for $r$ the number of relations in $\Sigma$, $a$ the
    maximum relation arity in $\Sigma$, ${w = \width(\Sigma)}$, ${c =
    |\consts(\Sigma)|}$, and some $b$ and $d$.
\end{theorem}

\begin{proof}[Proof of Correctness]
The proof follows the same structure as the correctness proof of
Theorem~\ref{thm:ExbDR-correctness-complexity}: we show that property
($\blacklozenge$) holds for each loop on a one-pass tree-like chase sequence
for $I$ and $\Sigma$; a minor difference is that the TGD whose existence is
implied by ($\blacklozenge$) is not necessarily guarded, but has width bounded
by ${\hwidth(\sigma)}$. To this end, we consider an
arbitrary loop ${T_i, T_{i+1}, \dots, T_{j-1}, T_j}$ at vertex $v$ in the
sequence, and assume that the claim holds for all shorter loops in the
sequence. By the definition of the loop, chase tree $T_{i+1}$ is obtained from
$T_i$ by applying a chase step to some non-full TGD ${\forall \vec x[\body_0
\rightarrow \exists \vec y ~ \head_0] \in \Sigma}$. Let $\sigma_0$ and
$\sigma_0'$ be the substitutions used in this chase step, and let $v'$ be the
child of $v$ introduced in $T_{i+1}$. Note that $T_{i+1}(v')$ contains at most
${\hwidth(\Sigma) + |\consts(\Sigma)|}$ distinct terms. We show by another
induction on $k$ that the following property holds for each $k$ with ${i < k
\leq j-1}$:
\begin{quote}
    ($\lozenge$) for each fact ${G \in T_k(v') \setminus T_{i+1}(v')}$, there
    exist a full TGD ${\forall \vec x[\body \rightarrow H] \in \Sigma'}$ of
    width at most $\width(\Sigma)$ and a substitution $\sigma$ such that
    ${\sigma(\body) \subseteq T_{i+1}(v')}$ and ${\sigma(H) = G}$.
\end{quote}

For the base case ${k = i + 1}$, property ($\lozenge$) holds vacuously because
${T_k(v') \setminus T_{i+1}(v') = \emptyset}$. For the inductive step, assume
that ($\lozenge$) holds for some $k$ and consider the possible ways to obtain
$T_{k+1}$ from $T_k$. Property ($\lozenge$) holds by the inductive hypothesis
if $T_{k+1}(v') = T_k(v')$---that is, if the step involves a descendant of
$v'$. Otherwise, ${T_{k+1}(v') = T_k(v') \cup \{ G \}}$ where fact $G$ is
obtained in one of the following two ways.
\begin{itemize}
    \item A full TGD in $\Sigma$ derives $G$ from $T_k(v')$. Set $\Sigma'$
    contains this TGD up to redundancy, so by Definition~\ref{def:redundancy}
    there exist a full TGD ${\body'' \rightarrow H' \in \Sigma'}$ and a
    substitution $\sigma$ such that ${\sigma(\body'') \subseteq T_k(v')}$ and
    ${\sigma(H') = G}$.

    \item Fact $G$ is the output of a loop at vertex $v'$. But then, this loop
    is shorter than ${T_i, \dots, T_j}$ so, by property ($\blacklozenge$),
    there exists a full TGD ${\body'' \rightarrow H' \in \Sigma'}$ and a
    substitution $\sigma$ such that ${\sigma(\body'') \subseteq T_k(v')}$ and
    ${\sigma(H') = G}$.
\end{itemize}
Either way, the width of rule ${\body'' \rightarrow H'}$ is bounded by
$\width(\Sigma)$, and we can assume that ${\body'' \rightarrow H'}$ is of the
form ${A_1' \wedge \dots \wedge A_n' \wedge \body' \rightarrow H'}$ where
${\sigma(A_\ell') \in T_k(v') \setminus T_{i+1}(v')}$ for each ${1 \leq \ell
\leq n}$, and ${\sigma(\body') \subseteq T_{i+1}(v')}$. By property
($\lozenge$), for each ${1 \leq \ell \leq n}$ there exist a full TGD
${\body_\ell \rightarrow H_\ell \in \Sigma'}$ and a substitution $\sigma_\ell$
such that ${\sigma_\ell(\body_\ell) \subseteq T_{i+1}(v')}$ and
${\sigma_\ell(H_\ell) = \gamma(A_\ell')}$. Moreover, set $\rng(\sigma_\ell)$
clearly contains at most ${\hwidth(\Sigma) + |\consts(\Sigma)|}$ distinct
terms. But then, there exist substitutions ${\theta_1, \dots, \theta_n}$ that
allow us to iteratively compose each ${\body_\ell \rightarrow H_\ell}$ with
${A_1' \wedge \dots \wedge A_n' \wedge \body' \rightarrow H'}$ to obtain a full
TGD subsumed by some ${\tau \in \Sigma'}$ and substitution $\sigma_1$ such that
$\tau$ and $\sigma_1$ satisfy property ($\lozenge$).

To complete the proof, consider an arbitrary fact ${F \in T_{j-1}(v') \setminus
T_{i+1}(v')}$ that is propagated from $v'$ to $v$ in $T_j$, and let ${\body''
\rightarrow H' \in \Sigma'}$ and $\sigma$ be the TGD and substitution whose
existence is guaranteed by property ($\lozenge$). Now if ${\sigma(\body'')
\subseteq T_{i+1}(v') \setminus \sigma_0'(\head_0)}$, then TGD ${\body''
\rightarrow H'}$ satisfies property ($\blacklozenge$). Otherwise, we can assume
that the rule is of the form ${A_1' \wedge \dots \wedge A_n' \wedge \body'
\rightarrow H'}$ where ${\sigma(A_i') \in \sigma_0'(\head_0)}$ for each ${1
\leq i \leq n}$, and ${\sigma(\body') \subseteq T_{i+1}(v') \setminus
\sigma_0'(\head_0)}$. Moreover, $\rng(\sigma)$ clearly contains at most
${\hwidth(\Sigma) + |\consts(\Sigma)|}$ distinct terms. But then, there exists
a substitution $\theta$ that allows us to apply the \propagate variant of the
$\FullDR$ inference rule to ${\body_0 \rightarrow \exists \vec y ~ \head_0}$
and ${A_1' \wedge \dots \wedge A_n' \wedge \body' \rightarrow H'}$ to obtain a
full TGD subsumed by some TGD ${\tau \in \Sigma'}$ and substitution $\sigma_1$
such that $\tau$ and $\sigma_1$ satisfy property ($\blacklozenge$).
\end{proof}

\begin{proof}[Proof of Complexity]
The proof is analogous to the proof of complexity of
Theorem~\ref{thm:ExbDR-correctness-complexity}. In particular, the \propagate
variant of the $\FullDR$ inference rule is analogous to the $\ExbDR$ inference
rule, so we can bound in the same way the number of candidate rule pairs and
possible ways to match body atoms of $\tau'$ to head atoms of $\tau$ by
$32^{\ell^2}$, where ${\ell = r \cdot (w + c)^a}$. Once a candidate pair of
$\tau$ and $\tau'$ has been selected, we need to consider all possible
substitutions $\theta$. Each such $\theta$ is defined on at most $2w$ variables
${\vec x \cup \vec z}$. Moreover, each variable is mapped to one of the ${w +
c}$ variables or to one of the $c$ constants in $\consts(\Sigma)$. Hence, there
are at most ${(w + 2c)^{2w} \leq 4^{(w + 2c) \cdot w} \leq 4^{(w + c)^2}}$
different substitutions $\theta$. Consequently, the \propagate variant of the
$\FullDR$ inference rule can be applied at most ${32^{\ell^2} \cdot 4^{(w +
c)^2} < 32^{n \cdot (w + c)^{2a+1}}}$ times. Applications of the \compose
variant can be bounded analogously. Finally, the times needed for subsumption
checking, unification, and all other steps can be bounded analogously as in the
complexity proof of Theorem~\ref{thm:ExbDR-correctness-complexity}, with a
minor difference that only body atoms need to be matched in the subsumption
checks.
\end{proof}

The $\FullDR$ algorithm has several obvious weak points. First, it considers
all possible ways to compose Datalog rules as long as this produces a rule with
at most ${\hwidth(\Sigma) + |\consts(\Sigma)|}$ variables. This may seem
unnecessary, but the \compose variant of the $\FullDR$ inference rule cannot be
simply dropped while retaining completeness. To understand why, consider an
arbitrary loop ${T_i, \dots, T_j}$ at vertex $v$ with child $v'$ and output
fact $F$ in a one-pass chase proof. The \propagate variant of the $\FullDR$
inference reflects only the chase step that derives the loop's output $F$, but
the derivation of $F$ in $v'$ may depend on the prior derivation of another
fact $F'$ in $v'$. The \compose variant allows us to produce $F$ in $v'$
without $F'$, rendering it eligible for \propagate again. Second, it is not
clear how to efficiently select the atoms ${A_1, \dots, A_n}$ and ${A_1',
\dots, A_n'}$ participating in the \propagate variant. Third, the number of
substitutions $\theta$ in the \compose and \propagate variants of the $\FullDR$
inference rule can be very large. Example~\ref{ex:FullDR-problem} illustrates
this problem for the \compose variant, but one can show analogously that the
\propagate variant suffers from the same issues.

\begin{example}\label{ex:FullDR-problem}
Consider the steps of the $\FullDR$ algorithm on GTGDs
\eqref{ex:FullDR-problem:1}--\eqref{ex:FullDR-problem:3}.
\begin{align}
    R(x_1,x_2)                      & \rightarrow \exists y_1,y_2 ~ S(x_1,x_2,y_1,y_2) \wedge T(x_1,x_2,y_2) \label{ex:FullDR-problem:1} \\
    S(x_1,x_2,x_3,x_4)              & \rightarrow U(x_4)                                                     \label{ex:FullDR-problem:2} \\
    T(z_1,z_2,z_3) \wedge U(z_3)    & \rightarrow P(z_1)                                                     \label{ex:FullDR-problem:3}
\end{align}
The \compose variant of the $\FullDR$ inference rule should be applied to GTGDs
\eqref{ex:FullDR-problem:2} and \eqref{ex:FullDR-problem:3}, but it is not
clear which unifier $\theta$, identifying variables $z_i$ in the latter with
variables $x_i$ in the former, one should use. The standard resolution
inference rule from first-order theorem proving would consider only the MGU
$\theta$ that maps $z_3$ to $x_4$; however, this would produce the resolvent
${S(x_1,x_2,x_3,x_4) \wedge T(z_1,z_2,x_4) \rightarrow P(z_1)}$ containing more
than ${\hwidth(\Sigma) = 4}$ variables, so this rule is not derived by the
\compose variant. Eliminating the upper bound on the number of variables is not
a solution: doing so would allow the derivation of full TGDs with an unbounded
number of variables, which would prevent termination. Instead, the \compose
variant requires us to consider every possible substitution $\theta$ that maps
variables ${x_1, \dots, x_4, z_1, \dots, z_3}$ to at most ${\hwidth(\Sigma)}$
variables. Consequently, ${7^4 = 2401}$ substitutions deriving rules such as
\begin{align}
    S(x_1,x_2,x_3,x_4) \wedge T(x_1,x_2,x_4)    & \rightarrow P(x_1), \\
    S(x_1,x_2,x_3,x_4) \wedge T(x_2,x_1,x_4)    & \rightarrow P(x_2), \\
    S(x_1,x_2,x_3,x_4) \wedge T(x_1,x_3,x_4)    & \rightarrow P(x_1), \\
    S(x_1,x_2,x_3,x_4) \wedge T(x_3,x_1,x_4)    & \rightarrow P(x_3),
\end{align}
need to be considered, which is clearly infeasible in practice.
\end{example}

Nevertheless, we implemented $\FullDR$ using the subsumption and indexing
techniques described in Section~\ref{sec:implementation}. Unsurprisingly, we
did not find $\FullDR$ competitive in our experiments. In fact, $\FullDR$ timed
out on 173 ontologies, and there are only three ontologies where another
algorithm reached the timeout but $\FullDR$ did not. For this reason, we do not
discuss the results with $\FullDR$ in Section~\ref{sec:experiments}.

}{}

\end{document}